\documentclass[fleqn]{2023SCGE}
\usepackage{mwe,tikz,ulem}
\usepackage[percent]{overpic}
\usepackage{graphicx}
\usepackage{xspace}
\usepackage[numbers]{natbib}
\usepackage[utf8x]{inputenc}
\usepackage{float}

\usepackage[gen]{eurosym}
\usepackage{microtype}
\usepackage{color}
\usepackage{xcolor}
\usepackage{txfonts}
\usepackage{makeidx}
\usepackage[columns=1]{idxlayout}
\usepackage{textcomp}

\newcommand{\aapr}{A\&AR\xspace}
\newcommand{\apj}{ApJ\xspace}
\newcommand{\apjl}{ApJL\xspace}

\newcommand{\mnras}{MNRAS\xspace}
\newcommand{\aap}{A\&A\xspace}
\newcommand{\aj}{AJ\xspace}

\newcommand{\nat}{Nature\xspace}

\newcommand{\araa}{ARAA\xspace}

\newcommand{\pr}{Phys. Rev. \xspace}
\newcommand{\prl}{Phys. Rev. Lett. \xspace}

\newcommand{\nar}{New Astron.\ Rev.\xspace}
\newcommand{\pasj}{PASJ\xspace}

\begin{document}
\ensubject{subject}

\ArticleType{Article}
\SpecialTopic{SPECIAL TOPIC: }
\Year{ }
\Month{ }
\Vol{ }
\No{ }
\DOI{ }
\ArtNo{ }

\title{Probing the Strong Gravity Region of Black Holes with eXTP}{extp-WG2}


\author[1]{Qingcui Bu}{bu@ccnu.edu.cn}%
\author[2,3]{Cosimo Bambi}{}
\author[4]{Lijun Gou}{}
\author[5]{Yanjun Xu}{}
\author[6]{Phil Uttley}{}
\author[7]{Alessandra De Rosa}{}
\author[8,2,5]{\\Andrea Santangelo}{}
\author[9]{Silvia Zane}{}
\author[5]{Hua Feng}{}
\author[5]{Shuang-Nan Zhang}{}
\author[4]{Chichuan Jin}{}
\author[4]{Haiwu Pan}{}
\author[10]{\\Xinwen Shu}{}
\author[11]{Francesco Ursini}{}
\author[4]{Yanan Wang}{}
\author[12]{Jianfeng Wu}{}%
\author[13]{Bei You}{}
\author[14]{Yefei Yuan}{}
\author[4]{\\Wenda Zhang}{}
\author[11]{Stefano Bianchi}{}
\author[15]{Lixin Dai}{}
\author[16]{Tiziana Di Salvo}{}
\author[17]{Michal Dov\v{c}iak}{}
\author[14]{Yuan Feng}{}
\author[18]{\\Hengxiao Guo}{}
\author[19]{Adam Ingram}{}
\author[20]{Jiachen Jiang}{}
\author[17]{Vladim\'{\i}r Karas}{}
\author[4]{Dongyue Li}{}
\author[8]{Honghui Liu}{}
\author[21]{\\Guglielmo Masteroserio}{}
\author[11]{Giorgio Matt}{}
\author[22]{Sara Motta}{}
\author[23]{Guobin Mou}{}
\author[2]{Abdurakhmon Nosirov}{}
\author[24,25]{\\ Zhen Pan}{}
\author[4]{Erlin Qiao}{}
\author[26,27]{Rongfeng Shen}{}
\author[5,28]{Qingcang Shui}{}
\author[4,28]{Yujia Song}{}
\author[17]{Ji\v{r}\'{\i} Svoboda}{}
\author[5]{Lian Tao}{}
\author[29,30]{\\Alexandra Veledina}{}
\author[18]{Zhen Yan}{}
\author[4,28]{Tong Zhao}{}

\address[1]{Institute of Astrophysics, Central China Normal University, Wuhan 430079, China}
\address[2]{Center for Astronomy and Astrophysics, Center for Field Theory and Particle Physics, and Department of Physics, Fudan University, Shanghai 200438, China}
\address[3]{School of Natural Sciences and Humanities, New Uzbekistan University, Tashkent 100007, Uzbekistan}
\address[4]{National Astronomical Observatories, Chinese Academy of Sciences, Datun Road A20, Beijing 100012, China}
\address[5]{Key Laboratory of Particle Astrophysics, Institute of High Energy Physics, Chinese Academy of Sciences, Beijing 100049, China}
\address[6]{Anton Pannekoek Institute for Astronomy, University of
Amsterdam, Science Park 904, 1098 XH, Amsterdam, the Netherlands}
\address[7]{INAF -- Istituto di Astrofisica e Planetologie Spaziali, Via Fosso del Cavaliere, 00133 Rome, Italy}
\address[8]{Institut f\"ur Astronomie und Astrophysik, Kepler Center for Astro and Particle Physics, Eberhard Karls Universit\"at, Sand 1, 72076 T\"ubingen, Germany}
\address[9]{Mullard Space Science Laboratory, University College London, Holmbury St Mary, Dorking, Surrey RH5 6NT, UK}
\address[10]{Department of Physics, Anhui Normal University, Wuhu, Anhui 241002, China}
\address[11]{Dipartimento di Matematica e Fisica, Universit\'a degli Studi Roma Tre, Via della Vasca Navale 84, I-00146 Roma, Italy}
\address[12]{Department of Astronomy, Xiamen University, Xiamen 361005, China}
\address[13]{Department of Astronomy, School of Physics and Technology, Wuhan University, Wuhan 430072, China}
\address[14]{Department of Astronomy, University of Science and Technology of China, Hefei 230026, China}
\address[15]{Department of Physics, University of Hong Kong, Pokfulam Road, Hong Kong 999077, China}
\address[16]{Dipartimento di Fisica e Chimica, Universit\'a degli Studi di Palermo, via Archirafi 36, 90123 Palermo, Italy}
\address[17]{Astronomical Institute, Czech Academy of Sciences, Bo\v{c}n\'{\i} II 1401, 14100 Prague, Czech Republic}
\address[18]{Shanghai Astronomical Observatory, Chinese Academy of Sciences, Shanghai 200030, China}
\address[19]{School of Mathematics, Statistics, and Physics, Newcastle University, Newcastle upon Tyne NE1 7RU, UK}
\address[20]{Department of Physics, University of Warwick, Coventry CV4 7AL, UK}
\address[21]{Dipartimento di Fisica, Universit\'a degli Studi di Milano, Via Celoria 16, I-20133 Milano, Italy}
\address[22]{Istituto Nazionale di Astrofisica, Osservatorio Astronomico di Brera, via E. Bianchi 46, 23807 Merate (LC), Italy}
\address[23]{Department of Physics and Institute of Theoretical Physics, Nanjing Normal University, Nanjing 210023, China}
\address[24]{Tsung-Dao Lee Institute, Shanghai Jiao Tong University, Shanghai, 1 Lisuo Road, 201210, China}
\address[25]{School of Physics \& Astronomy, Shanghai Jiao-Tong University, Shanghai, 800 Dongchuan Road, 200240, China}
\address[26]{School of Physics and Astronomy, Sun Yat-Sen University, Zhuhai, 519000, China}
\address[27]{CSST Science Center for the Guangdong-Hongkong-Macau Greater Bay Area, Sun Yat-Sen University, Zhuhai, 519082, China}
\address[28]{School of Astronomy and Space Sciences, University of Chinese Academy of Sciences, Yuquan Road 19A, Beijing 100049, China}
\address[29]{Department of Physics and Astronomy, University of Turku, Turku 20014, Finland}
\address[30]{Nordita, KTH Royal Institute of Technology and Stockholm University, Hannes Alfv\'ens v\"ag 12, SE-10691 Stockholm, Sweden}

\AuthorMark{Bu Q.~C., Bambi C., Gou L.~J., Xu Y.~J., Uttley P., De Rosa A.}


\AuthorCitation{Bu Q.~C., Bambi C., Gou L.~J., Xu Y.~J., Uttley P., De Rosa A.,  et al.}





\abstract{We present the novel capabilities of the enhanced X-ray Timing and Polarimetry (eXTP) mission to study the strong gravity region around stellar-mass black holes in X-ray binary systems and supermassive black holes in active galactic nuclei. eXTP can combine X-ray spectral, timing, and polarimetric techniques to study the accretion process near black holes, measure black hole masses and spins, and test Einstein's theory of General Relativity in the strong field regime. We show how eXTP can improve the current measurements of black holes of existing X-ray missions and we discuss the scientific questions that can be addressed.}



\maketitle


\begin{multicols}{2}

\section{Introduction}\label{sec:Introduction}


Black holes are one of the most intriguing predictions of General Relativity. The first black hole solution was discovered by Karl Schwarzschild in 1916~\cite{Schwarzschild:1916uq}, less than a few months after the announcement by Albert Einstein of his theory. However, the actual properties of these solutions were not immediately understood. David Finkelstein was the first, in 1958, to realize that these solutions had an event horizon acting as a one-way membrane~\cite{Finkelstein:1958zz}: if something crosses the horizon, it cannot influence the exterior region any longer. For a long time, the astronomy community was very skeptical of the possibility of the existence of black holes in the Universe. The situation changed in the early 1970s, when Thomas Bolton and, independently, Louise Webster and Paul Murdin identified the X-ray source Cygnus X-1 as the first stellar-mass black hole candidate~\cite{cyg1,cyg2}. Since then, an increasing number of astronomical observations have pointed out the existence of stellar-mass and supermassive black holes. Thanks to technological progress and new observational facilities, in the past 10-20~years there have been tremendous advancements in the understanding of these objects and of their environments.

\begin{figure*}[t]
\centering
\includegraphics[width=\textwidth]{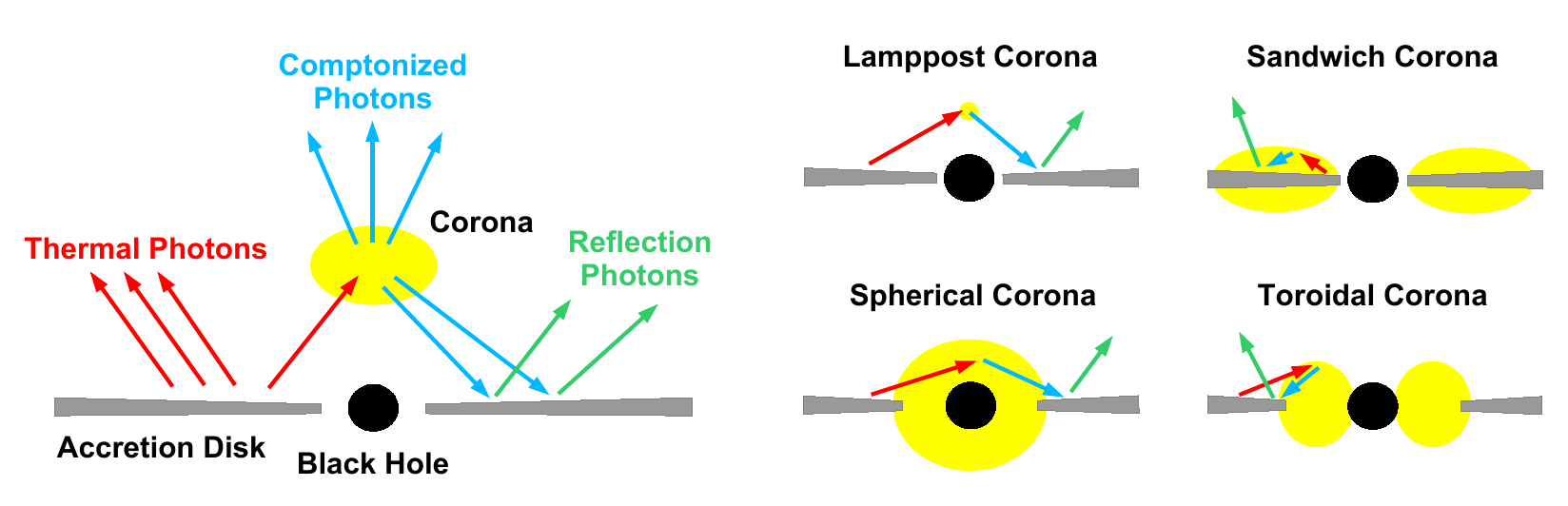}
\caption{Left panel: Disk-corona model. Right panel: Examples of possible coronal geometries. Figure from Ref.~\cite{Bambi:2024hhi}. }
\label{fig:1}
\end{figure*}

The analysis of the X-ray radiation emitted from the strong gravity region around black holes is a powerful tool to study the accretion process of the innermost region of the accretion disk, measure black hole masses and spins, and test Einstein's theory of General Relativity (GR) in the strong field regime. The prototype of astrophysical system for X-ray observations is shown in the left panel in Fig.~\ref{fig:1} and is normally referred to as the \textit{disk-corona model}~\cite{corona1,corona2}. The central black hole can be either a stellar-mass black hole in an X-ray binary (BHXRB) system or a supermassive black hole in an active galactic nucleus (AGN). The crucial ingredient is that the accretion disk is Keplerian, geometrically thin or `slim', and optically thick. Such an accretion disk is \textit{cold} because it can efficiently emit radiation\footnote{Thin accretion disks are cold with respect to other accretion disk models (e.g., advection-dominated accretion flow or ADAF~\cite{Narayan:1994xi}), as well as with respect to the corona.} The thermal spectrum of the accretion disk turns out to be peaked in the soft X-ray band for stellar-mass black holes in X-ray binary systems and in the UV band for supermassive black holes in active galactic nuclei~\cite{Shakura:1972te,Page:1974he}. The ``corona'' is some \textit{hot} plasma ($\sim 100$~keV) near the black hole and the inner part of the accretion disk, but its exact morphology is not yet completely understood~\cite{Bisnovatyi-Kogan:1977xxx,Haardt:1991tp,Titarchuk:1994rz,Dove:1997ei,Liu:2003yg,Markoff:2005ht,Sironi:2019sxv}. The corona may be the base of the jet, the hot atmosphere above the accretion disk, an inner, advection-dominated accretion flow when the inner edge of the disk is truncated at large radii, etc. (see the right panel in Fig.~\ref{fig:1}). 

Since the accretion disk is cold and the corona is hot, thermal photons from the disk can inverse Compton scatter off free electrons in the corona. The X-ray spectrum of the Comptonized photons can be normally approximated by a power law with a high-energy cutoff of the order of the electron temperature of the corona~\cite{Haardt:1991tp}. A fraction of the Comptonized photons can illuminate the accretion disk: Compton scattering and absorption followed by fluorescent emission generate the reflection spectrum (green arrows in Fig.~\ref{fig:1}). The \textit{non-relativistic} reflection spectrum, namely the reflection spectrum in the rest-frame of the material of the disk, is characterized by narrow emission lines in the soft X-ray band and a Compton hump with a peak around 20-40~keV~\cite{Ross:2005dm,Garcia:2010iz}. The most prominent emission line is often the iron K$\alpha$ complex, which is a narrow feature around 6.4~keV in the case of neutral or weakly ionized iron atoms and shifts up to 6.97~keV in the case of H-like iron ions. The \textit{relativistic} reflection spectrum, namely the reflection spectrum of the whole disk as observed far from the source, is blurred because of relativistic effects in the strong gravity region around the black hole: Doppler boosting because of the relativistic motion of the material in the accretion disk and gravitational redshift because of the gravitational well of the black hole~\cite{Fabian:1989ej,Laor:1991nc}. Such relativistically blurred reflection features are commonly observed in the X-ray spectra of both stellar-mass black holes in X-ray binaries and supermassive black holes in active galactic nuclei~\cite{Fabian:1989ej,Tanaka:1995en,Nandra:2007rp}.

The variability of the X-ray spectrum of an accreting black hole also brings useful information about the source. Quasi-periodic oscillations (QPOs) are narrow features at characteristic frequencies in the X-ray power density spectra of accreting black holes and neutron stars~\cite{Ingram:2019mna}. In the case of stellar-mass black holes in X-ray binary systems, QPOs can be grouped into two classes: low-frequency QPOs (0.1-30~Hz) and high-frequency QPOs (40-450~Hz). There are three different types of low-frequency QPOs, which are called, respectively, type~A, type~B, and type~C QPOs. QPOs are observed even in the X-ray power density spectra of intermediate-mass and supermassive black holes~\cite{Pasham:2014ybe,Gierlinski:2008yz}, where we see that the QPO frequency scales as the inverse of the black hole mass~\cite{Zhou:2014cma}. However, the detection of QPO in intermediate-mass and supermassive black holes is more difficult due to insufficient observation lengths and modeling problems.

Despite the remarkable advancements in the past 10-20~years in the understanding of astrophysical black holes and of their environment, there are still many open questions in the physics and astrophysics of these objects:
\begin{enumerate}
\item Are astrophysical black holes the Kerr black holes predicted by GR? Do astrophysical black holes have an event horizon or are they instead compact horizonless objects as predicted by a number of scenarios beyond GR~\cite{Mathur:2024ify}?
\item What is the spin distribution of stellar-mass black holes in X-ray binary systems? Is such a spin distribution consistent with that observed in binary black holes with gravitational waves by the LIGO-Virgo-KAGRA Collaboration or are they two different object classes?
\item What is the spin distribution of supermassive black holes in active galactic nuclei? How do the spins of these objects evolve with time and what is the main mechanism responsible for such an evolution?
\item What is exactly the \textit{corona}? How does it form and evolve with time? How is the corona connected to the accretion disk and outflows?
\item What is the mechanism responsible for QPOs? Are they generated in the accretion disk or in the corona? Can we use QPO observations for measuring black hole masses and spins and for testing Einstein's theory of GR?
\item What is the exact mechanism responsible for the formation of powerful jets, which are commonly observed both in stellar-mass and supermassive black hole systems? Are jets powered by the rotational energy of the black hole or by that of the disk?
\end{enumerate}

The \textsl{enhanced X-ray Timing and Polarimetry} (eXTP) mission has been proposed by a consortium led by the Institute of High-Energy Physics of the Chinese Academy of Sciences to address these questions. Its launch is currently scheduled in 2030. The primary goals of eXTP are: $i)$~the study of matter at very high density~\cite{WP-WG1}, $ii)$~the study of strong gravity with black holes (this paper), $iii)$~the study of matter in strong magnetic fields~\cite{WP-WG3}, $iv)$~the study of violent high-energy astrophysical phenomena with X-ray in the era of multi-messanger astrophysics~\cite{WP-WG4} and $v$~the observatory science~\cite{WP-WG5}.

In the new baseline design, the scientific payload of eXTP consists of three main instruments: the Spectroscopic Focusing Array (SFA), the Polarimetry Focusing Array (PFA) and the Wide-band and Wide-field Camera (W2C). Here, we provide a brief introduction to the scientific payloads and refer to~\cite{SFA-PFA} for a detailed description.

The SFA consists of five SFA-T (where T denotes Timing) X-ray focusing telescopes covering the energy range $0.5$--$10\, \mathrm{keV}$, featuring a total effective area of $2750\,{\rm cm^2}$ at $1.5\, \mathrm{keV}$ and $1670\,{\rm cm^2}$ at $6\,\mathrm{keV}$. The designed angular resolution of the SFA is $\le 1^\prime$ (HPD) with a $18^{\prime}$ field of view (FoV). The SFA-T are equipped with silicon-drift detectors (SDDs), which combine good spectral resolution ($\sim$ 180~eV at 1.5~keV) with very short dead time and a high time resolution of $10\,{\mu\mathrm {s}}$. They are therefore well-suited for studies of X-ray emitting compact objects at the shortest time scales. The SFA array also includes one unit of the SFA-I (where I signifies Imaging) telescope equipped with pn-CCD detectors (p-n junction charged coupled device), to enhance imaging capabilities, which would supply strong upper limits to detect weak and extended sources. The expected FoV of SFA-I is $18^\prime \times 18^\prime$. Therefore, the overall sensitivity of SFA could reach around $3.3\times 10^{-15}\,{\rm erg\,cm^{-2}\,s^{-1}}$ for an exposure time of $1\, \mathrm{Ms}$. Since it is not excluded that the SFA might in the end include six SFA-T units, simulations presented here have taken this possibility into consideration. 

The PFA features three identical telescopes, with an angular resolution better than $30^{\prime\prime}$ (HPD) in a $9.8^{\prime} \times 9.8^{\prime}$ FoV, and a total effective area of $250\,{\mathrm{ cm^{2}}}$ at $3\, \mathrm{keV}$ (considering the detector efficiency). Polarization measurements are carried out by gas pixel detectors (GPDs) working at 2-- 8\,keV with an expected energy resolution of 20\% at 6\,keV and a time resolution better than $10\,{\mathrm {\mu{s}}}$ \cite{2001Natur.411..662C, 2003NIMPA.510..176B,2007NIMPA.579..853B,2013NIMPA.720..173B,Zhang_2019}. The instrument reaches an expected minimum detectable polarization (MDP) at $99\%$ confidence level ($\mathrm{MDP}_{99}$) of about $2\%$ in $1\,\mathrm {Ms}$ for a milliCrab-like source.

The W2C is a secondary instrument of the science payload, featuring a coded mask camera with a FoV of approximately {1500} {square degrees} (Full-Width Zero Response, FWZR). The instrument achieves a sensitivity of $4\times 10^{-7}\,{\rm erg\, cm^{-2}\,s^{-1}}$ (1\,s exposure) across the 10–600\,keV energy range, with an angular resolution of $20^{\prime}$, a position accuracy of 5', and an energy resolution better than $30\%$ at 60\,keV.



\begin{figure*}[t]
    \centering
    \includegraphics[width=\textwidth]{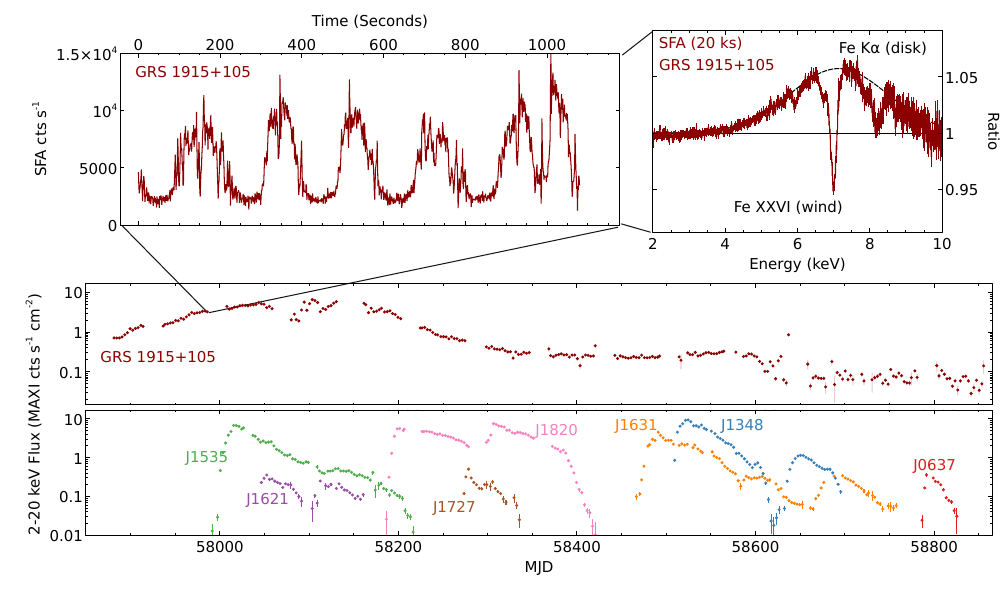}
    \caption{The top two panels show a simulated SFA light curve and spectrum of GRS 1915$+$105, demonstrating eXTP’s capabilities in studying strong-gravity accretion physics. The rapid, periodic variability reveals instabilities of the inner accretion disk of this black hole; the relativistic Fe K emission line provides insights into strong-gravity effects and accretion disk properties; and the absorption features from disk winds enable kinematic studies and an investigation of potential driving mechanisms. These topics will be further explored throughout this paper.  The bottom two panels illustrate the expected number of black hole X-ray novae eXTP will observe over the first three-year period, based on historical events. These include short outbursts lasting a few months and longer outbursts extending over several years. See text for further details.}
    \label{fig:2}
\end{figure*}

To illustrate eXTP’s overall observing capabilities, we present historical X-ray novae in Fig.~\ref{fig:2}, where the bottom panel displays short-lived outbursts lasting for months, while the middle panel highlights long-duration outbursts spanning years or persistent X-ray sources. These light curves, measured in flux units by MAXI \citep{maxi09} between 2017 and 2020, demonstrate the variability observed in the X-ray sky.

Beyond all-sky monitoring, eXTP will conduct detailed studies of X-ray novae in the future, with a particular focus on black hole X-ray binaries in this paper. The top panels of Fig.~\ref{fig:2} showcase a simulated SFA light curve and spectrum of the black hole X-ray binary GRS 1915+105 \citep{liu22}, based on archival observations from Insight-HXMT. eXTP will reveal rapid and periodic variability on short timescales associated with disk instability \citep[see Section~\ref{ss-spin} or reference][]{janiuk00}. A 20\,ks eXTP observation will also demonstrate the spectral capabilities of SFA, detecting relativistic Fe K disk emission as part of disk reflection (discussed in Section~\ref{ss-XRS}), which will provide insight into the properties of the strong gravity region and the accretion disk. Additionally, eXTP will resolve the absorption features of the disk winds, allowing measurements of their velocities and helping to determine their kinematic properties and potential launching mechanisms (see Section 3.1.2).

The paper is organized as follows. In Section~\ref{s-gravity}, we show how eXTP can improve current measurements of black hole masses and spins as well as put stronger constraints on possible deviations from the predictions of GR. In Section~\ref{s-matter}, we show how eXTP can help us to understand better the accretion process in the strong gravity region of black holes. Summary and conclusions are reported in Section~\ref{sec:Summary}.

\section{Measuring the Basic Parameters of Black Holes and Testing GR}\label{s-gravity}

Black holes are relatively simple objects, as they are completely characterized by a few parameters: their mass, their spin angular momentum, and their electric charge. This is the celebrated result of the {\it No-Hair Theorem}, which is actually a family of theorems, and holds under specific assumptions~\cite{Carter:1971zc,Robinson:1975bv,Chrusciel:2012jk}. For astrophysical macroscopic objects, the electric charge is normally negligible, so astrophysical black holes should be completely characterized by their mass $M$ and spin angular momentum $J$ and the spacetime geometry around these objects should be described by the {\it Kerr solution}~\cite{Kerr:1963ud}.

For a Kerr black hole, the mass $M$ is a free parameter, and therefore we may have black holes with arbitrarily low as well as arbitrarily high masses. However, the black holes in our Universe must form through some astrophysical process and therefore it is natural to expect black holes with masses only in specific ranges. This is exactly what we find. Astrophysical observations have discovered at least two classes of black holes: stellar-mass black holes ($M \approx 3-100$~$M_\odot$)~\cite{Remillard:2006fc} and supermassive black holes ($M \sim 10^5 - 10^{10}$~$M_\odot$)~\cite{Kormendy:1995er}. There is likely a third class of objects, the so-called intermediate mass black hole candidates~\cite{Miller:2003sc}, which are black hole candidates with a mass filling the gap between the stellar-mass and the supermassive ones.

Unlike the mass, the spin angular momentum $J$ of a Kerr black hole is subject to certain constraints. If we define the dimensionless spin parameter $a_* = J/M^2$ (in units in which $c = G_{\rm N} = 1$), the condition for the existence of a Kerr black hole is $|a_*| \le 1$ (Kerr bound). For $|a_*| > 1$, there is no event horizon and the Kerr solution describes the spacetime of a naked singularity, which is thought to be impossible to create in the Universe~\cite{Penrose:1969pc}. Spin measurements are crucial to understand how black holes form, evolve, and interact with the host environment~\cite{Reynolds:2020jwt}. 
In the case of stellar-mass black holes in binary systems, it is often thought that the value of the black hole spin parameter is determined by the formation mechanism (and the natal spin is expected to be low~\cite{Woosley:2006fn,Yoon:2006fr,Fuller:2019sxi}), because the amount of mass that can be transferred from the companion star to the black hole is modest and cannot appreciably change the black hole mass and spin angular momentum~\cite{King:1999aq,Valsecchi:2010cw,Wong:2011eg}. However, such a conclusion is not necessarily true~\cite{Podsiadlowski:2002ww}. In the case of low-mass XRBs, the black hole may be spun up immediately after its formation~\cite{Fragos:2014cva}. In the case of high-mass XRBs, the black hole spin may be directly related to the angular momentum of the progenitor star, which can transfer part of its envelope to the companion star and get a core with high angular momentum~\cite{Qin:2018sxk}. See Ref.~\cite{Zdziarski:2025ozs} for a recent review on this topic.
In the case of supermassive black holes in galactic nuclei, the black hole spin parameter should be determined by the mass transferred from the host environment to the compact object and spin measurements can test different scenarios of galaxy evolution and
merger~\cite{Barausse:2012fy,Sesana:2014bea}.

\begin{figure}[H]
\centering
\includegraphics[width=\columnwidth]{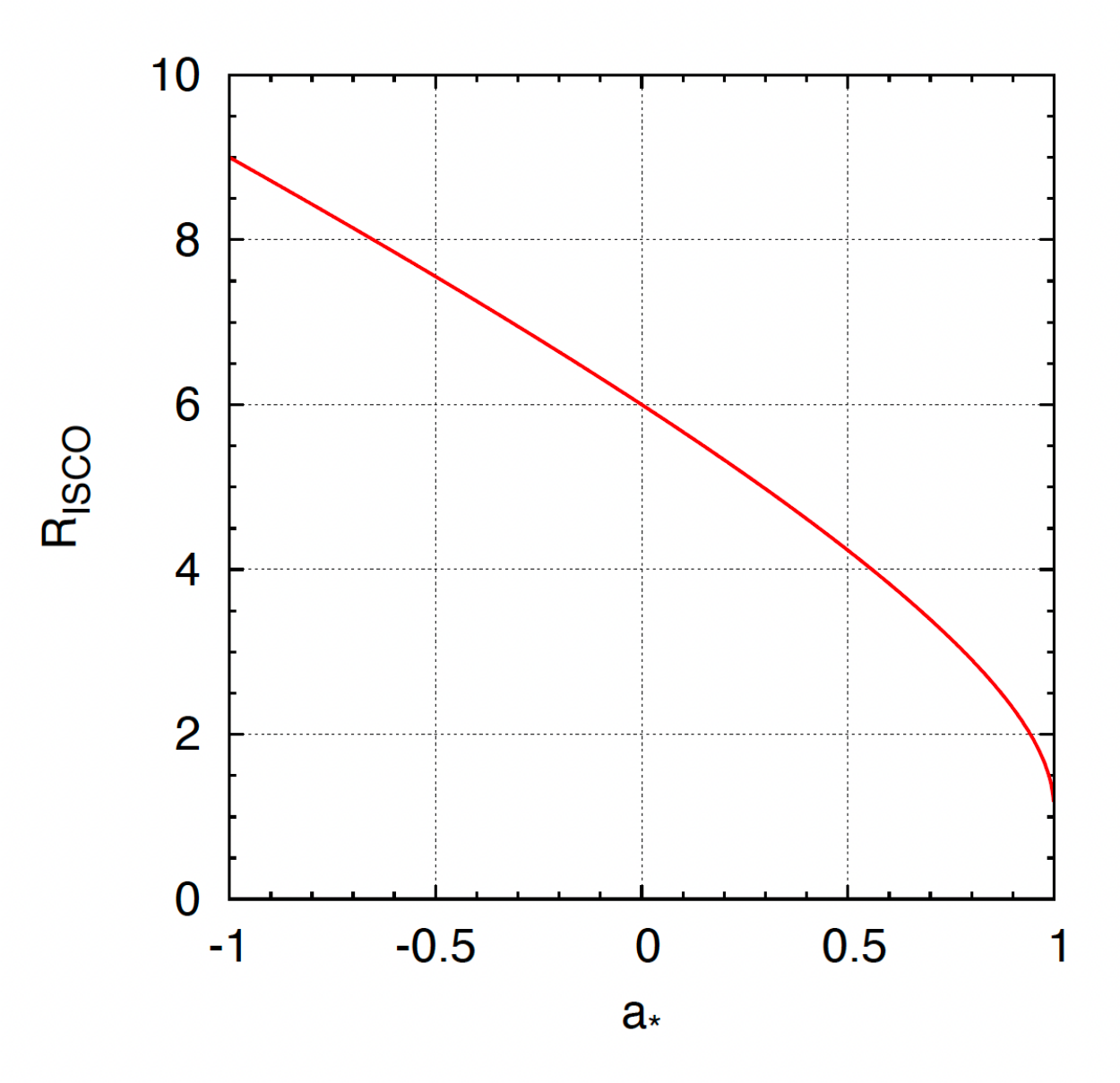}
\caption{ISCO radius $R_{\rm ISCO}$ in the Kerr spacetime in Boyer-Lindquist coordinates as a function of the black hole spin parameter $a_*$. $a_* > 0$ ($a_* < 0$) is for co-rotating (counter-rotating) orbits, namely for orbits with angular momentum parallel (anti-parallel) to the black hole spin. $R_{\rm ISCO}$ in units $M=1$.}
\label{fig:3}
\end{figure}

A crucial ingredient in most black hole spin measurements with X-ray observations is the assumption that the inner edge of the accretion disk is at the innermost stable circular orbit (ISCO) of the Kerr spacetime~\cite{Bardeen:1972fi}. There is indeed a one-to-one correspondence between the ISCO radius $R_{\rm ISCO}$ and the black hole spin parameter $a_*$ (see Fig.~\ref{fig:3}), so a precise and accurate measurement of the location of the inner edge of the accretion disk can be converted to a measurement of the black hole spin parameter. The assumption that the inner edge of the accretion disk is at the ISCO radius is well supported by observational and theoretical studies for black holes with an Eddington luminosity between a few percent and about 30\% and a spectrum dominated by the thermal component of the accretion disk (soft state)~\cite{Steiner:2010kd,Penna:2010hu}. It is still a controversial issue if this is also the case for sources in which the spectrum is dominated by the Comptonized spectrum from the corona (hard state)~\cite{Zdziarski:2020arg}\footnote{We note that both hard and soft states are observed at luminosities $>1\%$ of the Eddington luminosity $L_{\rm Edd}$, implying that both corresponding physical solutions can exist at similar accretion rates.}: the disk is probably truncated at a radius larger than the ISCO at low luminosities and the inner edge of the disk moves to the ISCO radius as the luminosity increases~\cite{Wang-Ji:2017oly,Liu:2023ovm} (see, however, Refs.~\cite{Zdziarski:2021jrw,Zdziarski:2021nem}).

Although the predictions of GR have been extensively tested in weak gravitational fields with Solar System experiments and radio observations of binary pulsars~\cite{Will:2014kxa}, the strong-field regime is still largely unexplored. Black holes are ideal laboratories for testing GR in the strong field regime as they are the sources of the strongest gravitational fields that can be found today in the Universe~\cite{Bambi:2015kza,Karas:2022}. The past 10~years have seen significant advancements in our capability of testing strong gravity and today we can test the nature of astrophysical black holes with X-ray, gravitational waves, and black hole imaging~\cite{Tripathi:2018lhx,Tripathi:2020yts,LIGOScientific:2016lio,Vagnozzi:2022moj}. Current constraints are still weak (especially if compared to our capability of testing atomic and particle physics) but they can be improved with the next generation of observational facilities.

\subsection{Mass measurements in X-ray binaries}
\label{sec:mass}

Mass is one of the two fundamental parameters for astrophysical black holes. Besides dynamical methods using optical spectroscopy/photometry, the black hole mass can also be measured with X-ray observations, via spectroscopy, variability, QPOs, etc. This will fully utilize the advantages of the eXTP mission, such as large effective area, high spectral and timing resolution. 

\subsubsection{Spectroscopic mass estimates}

X-ray spectral modeling will provide the detailed physical properties of the inner accretion disk, which can be used to derive the black hole mass and spin. Both the X-ray thermal continuum and the relativistic reflection spectrum can be utilized to constrain the black hole mass and spin. The underlying principle is that spectroscopic measurements of relativistic broadening of reflection features (notably Fe~K$\alpha$) can yield the disk inner radius in units of gravitational radius $R_g$\footnote{Where for a black hole of mass $M$ the gravitational radius 1~$R_g=GM/c^{2}$} (see Section~\ref{ss-XRS}). If the spectrum presents both a strong thermal component and prominent reflection features, from the analysis of the iron line we can estimate the dimensionless black hole spin parameter and the inclination angle of the disk, and with the continuum-fitting we can infer the black hole mass.

In black hole X-ray binaries (BHXRBs) the physical estimate of radius in length units can be obtained using thermal continuum spectroscopy, where the accretion disk's multi-temperature blackbody spectrum is fitted, together with an estimate for the inclination of the inner disk and the distance to the source. From the Stefan-Boltzmann law and appropriate modifications for the disk atmosphere, the emitting area and hence inner radius of the disk can be determined (see e.g. \citep{Gou:2009, Gou:2011extreme} and Section~\ref{sec:thermal_continuum}). Alternatively, the measurement of physical radius can be carried out with X-ray polarimetric measurements of the disk thermal spectrum (\cite{svoboda_first_2024,marra_ixpe_2024,steiner_ixpe-led_2024} and Section~\ref{sec:spin_polarimetry}). 

To obtain a spectroscopic mass estimate the methods used to estimate the radius in physical dimensions and gravitational radii should ideally be applied to the same data. This likely requires a strong disk thermal component to be present for the physical radius measurement, i.e. the source must lie in the soft spectral state. If the observations are carried out at different times, one must make the assumption that both methods sample the same inner disk radius, which would be justified if the disk is assumed to extend down to the ISCO for both observations. eXTP will be uniquely capable of these measurements because it has excellent spectroscopic capabilities thanks to the large SFA collecting area and can make use of both thermal continuum {\it and} polarization methods to estimate the disk physical dimension.
 


\subsubsection{X-ray reverberation mapping estimates of black hole mass}

X-ray reverberation mapping explores the disk-corona geometry by measuring the time lags between the coronal X-ray emission which travels directly to the observer and the component reflected by the accretion disk. The Fe K$\alpha$ emission line is one of the main features to conduct X-ray reverberation mapping. The time lag between X-ray continuum and the Fe K$\alpha$ line provides the light-traveling distance between the corona and line emitting region on the accretion disk, which, combined with the gas velocity measurement, will enable the constraints on black hole mass. 

\begin{figure*}
\centering
    \includegraphics[width=\textwidth]{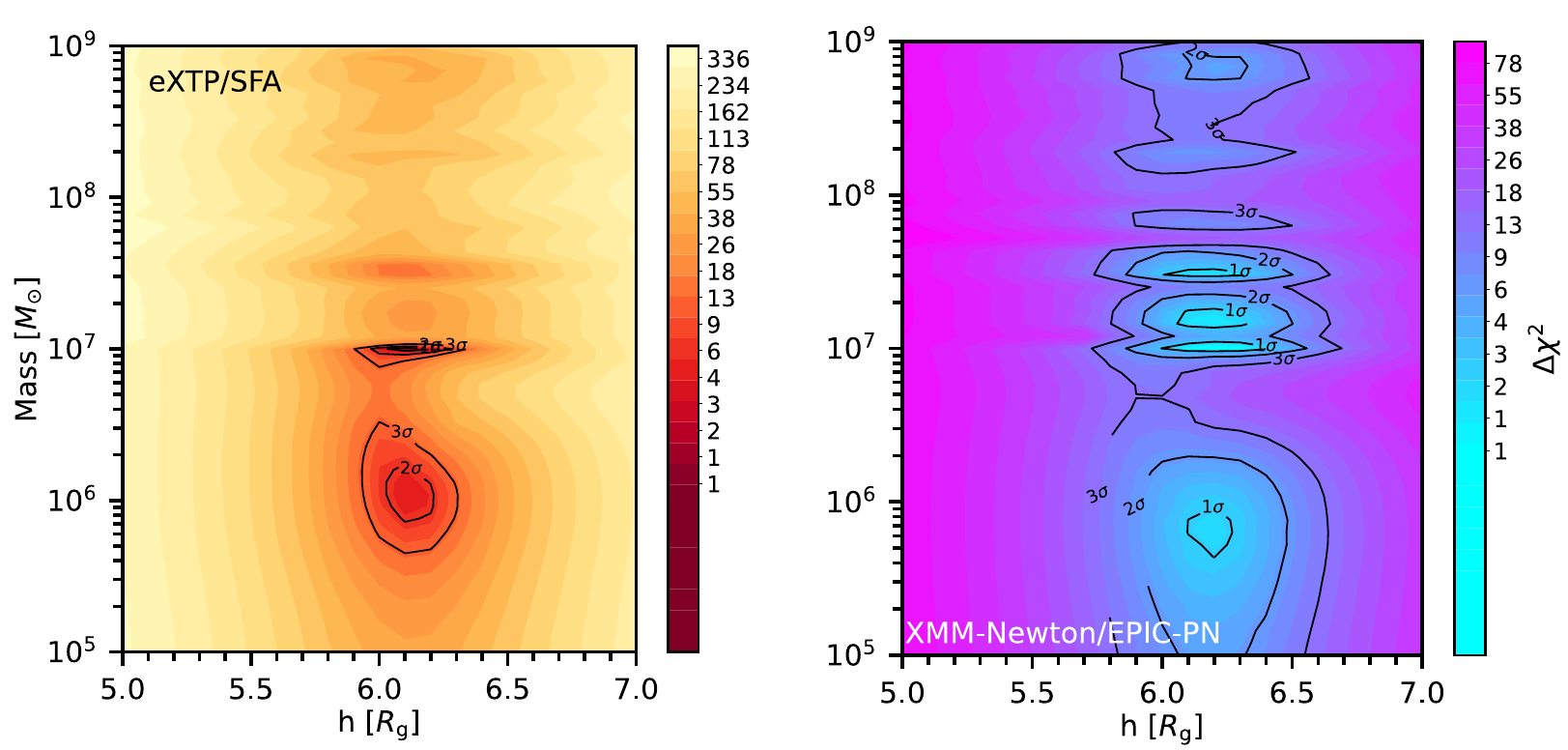}
\caption{Comparison between eXTP-SFA (left panel) and XMM-Newton epic-PN (right panel) of the $\Delta \chi^2$ 2D contour plots between the black hole mass and the height of the lamp-post corona. Both panels are the result of fitting simultaneously the time-averaged and lag energy spectra accounting for hard continuum lags and X-ray reverberation lags. The simulations are performed considering an AGN of $10^7~M_\odot$ and an illuminating corona placed at 6~$R_g$ above the black hole axis, the target is observed for 500~ks at 1~mCrab.} 
\label{fig:4}
\end{figure*}

In particular, Fig.~\ref{fig:4} shows the 2D contour plot of black hole mass \textit{vs} the coronal height obtained from the spectral-timing fit of AGN X-ray reverberation simulations performed with the \texttt{reltrans} model \cite{Ingram2019_reltrans, Mastroserio2021}. 
We simulated and fitted simultaneously the time-averaged energy spectrum and the lag energy spectrum for two distinct timescales: [$10^{-5} - 10^{-4}$]~Hz which is dominated by the hard lags and [$2\times10^{-4} - 10^{-3}$]~Hz where the main contribution to the signal is the reverberation lag (see \cite{Mastroserio2020} as an example). 
The simulation is performed considering a lamp-post coronal geometry placed at 6~$R_g$ above the black hole, that illuminates through a Comptonization-like spectrum ($\Gamma = 2$ and kT$_{\rm e}$ = 60~keV) the accretion disk which extends down to the 
ISCO.\footnote{The parameters of the accretion disk are standard: Fe abundance 1, ionization parameter log ($\xi$ / erg cm s$^{-1}$) = 3, disk number density log${\rm n}_{\rm e}$/cm$^{3}$ = 15 .}.
The simulated AGN holds a $10^7~M_\odot$, maximally spinning black hole, which is observed at 1~mCrab flux for 500~ks exposure.
Fig.~\ref{fig:4} shows clearly how much better eXTP-SFA will perform in constraining the AGN mass compared to the XMM-Newton observations. 
The right panel (XMM-Newton simulations) shows that the black hole mass is unconstrained despite the simultaneous fit of time-averaged spectrum and lag-energy spectrum  (see e.g. \cite{Cackett2014} and \cite{Ingram2019_reltrans}). This happens because we allowed the hard lags to contribute to the reverberation lags at high Fourier frequencies. 
If the reverberation features are not particularly significant in the high-frequency lag-energy spectrum, the continuum variability, which is included in the fitting model through a pivoting Comptonization \cite{Mastroserio2018}, can account for the reverberation lags causing the mass parameter to be unconstrained. On the other hand, in the case of eXTP-SFA (left panel), the correct black hole mass is recovered at 1~$\sigma$ confidence level, and only a second solution is found if 3~$\sigma$ contour level is considered. 
The improvement is due to the much higher signal to noise at the iron line energy range of eXTP-SFA which allows a more accurate constraint of the mass. 
The small degeneracy will be solved when we will also fit for the amplitude of the correlated variability, and not only for the variability phase-lag.

\subsubsection{Mass estimates from X-ray variability and QPOs}

Measurements of the X-ray variability of AGNs can provide constraints on the central black hole mass, since the amplitude of short-term ($<1$~day) X-ray flux variability anti-correlates with the black hole mass. As the metric of flux variability, the ``normalized excess variance" $\sigma^2_{\text{rms}}$ is defined as the following equation, 
\begin{equation}
\sigma^2_{\text{rms}} = \frac{1}{N \mu^2} \sum_{i=1}^{N} \left( (X_i - \mu)^2 - \sigma_{i,\text{err}}^2 \right),
\end{equation}
where $N$ is the number of observational epochs, $X_i$ is the individual X-ray flux measurement with $\sigma_{i, \textrm{err}}$ as its uncertainty, and $\mu$ is the mean flux value. 

The empirical relation between $\sigma^2_{\text{rms}}$ and black hole mass has been calibrated (e.g. \cite{Paolillo2025}). With $\sigma^2_{\text{rms}}$ calculated, the black hole mass can be constrained. With its long ($>100$~ks) uninterrupted exposures and excellent SFA collecting area and sensitivity, eXTP will extend the measurement of masses using the excess variance to much fainter AGN than were previously accessible. In particular, the method can be applied to many AGN with relatively low black hole masses and modest accretion rates, to check the scaling of the relation from higher-mass black holes observed in more luminous AGN.

The quasi-periodic oscillations in the X-ray light curves can often be associated with the orbital motion of the accretion disk. The high-frequency QPOs observed above 100~Hz are believed to be the manifestation of the gas movement in the accretion disk. If these high-frequency QPOs correspond to the orbital frequency at the inner accretion flow, the black hole mass can be constrained. The low-frequency QPOs in the $\sim$Hz range are associated with the Lense-Thirring precession of the accretion disk. The disk precession frequency depends on the black hole mass and spin. The black hole mass can be constrained if independent spin measurement can break the mass-spin degeneracy.



\subsection{Spin measurements}\label{ss-spin}
\label{sec:spin}
\subsubsection{X-ray reflection spectroscopy}\label{ss-XRS}

X-ray reflection spectroscopy refers to the analysis of the relativistically blurred reflection features in the X-ray spectra of accreting black holes and is a powerful technique to probe the strong gravity region around black holes. So far X-ray reflection spectroscopy has provided a spin measurement of about 40~stellar-mass black holes in X-ray binary systems and about 40~supermassive black holes in active galactic nuclei~\cite{Bambi:2020jpe,Draghis:2023vzj}, and is currently the only available technique to measure the spins of supermassive black holes. 

\begin{figure*}[t]
\centering
\includegraphics[width=\textwidth]{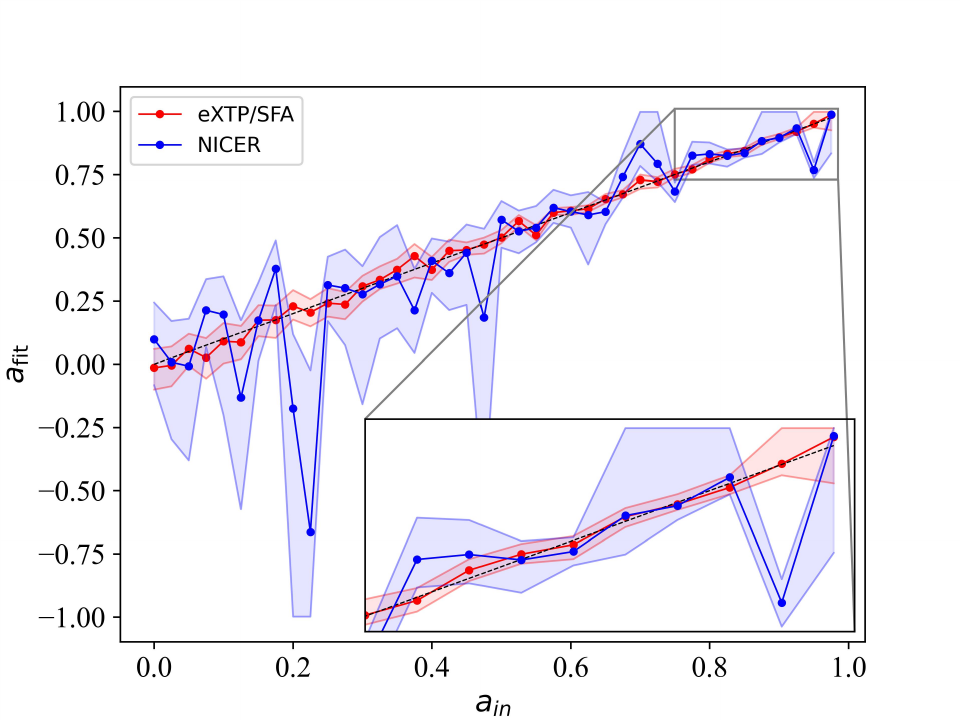}
\caption{The recoverability of the black hole spin parameter for eXTP/SFA and NICER/XTI with simulated BHXRB reflection spectra. The {\tt relxill} model is used assuming a photon index $\Gamma=1.7$ and cutoff energy $E_{\rm cut}=300$~keV. The accretion disk extends to the ISCO and is viewed at an inclination angle of 45 degrees. Other parameters are the reflection fraction ($R_f$=1), the emissivity profile ($q=3$), the iron abundance ($A_{\rm Fe}=1$), the ionization parameter ($\log\xi=3$) and the flux ($2\times 10^{-8}$~erg~cm$^{-2}$~s$^{-1}$ in the 2--10 keV band). We fit the simulated data in the range 4-10~keV. The dashed line represents the condition where the input spin for the simulation is equal to the best-fit spin ($a_{\rm in}=a_{\rm fit}$). The blue and red data points correspond to the best-fit values for NICER/XTI and eXTP/SFA, respectively. The shaded areas represent the 90\% confidence intervals.}
\label{fig:5}
\end{figure*}

The recoverability of the black hole spin parameter using X-ray reflection spectroscopy is shown in Fig.~\ref{fig:5} for eXTP/SFA, in comparison to NICER/XTI. We simulated 30~ks observations of bright black hole X-ray binaries (flux $2\times 10^{-8}$~erg~cm$^{-2}$~s$^{-1}$ in the 2-10~keV energy band) with a black hole spin parameter $a_*$ ranging from 0 to 0.975. For the simulations, we used the publicly available model {\tt relxill}~\cite{Dauser:2013xv,Garcia:2013lxa}, assuming that the coronal spectrum has photon index $\Gamma = 1.7$ and high-energy cutoff $E_{\rm cut} = 300$~keV, the reflection fraction is $R_{\rm f} = 1$, the inclination angle of the disk is $i = 45$~deg, the inner edge of the accretion disk is at the ISCO radius of the spacetime, the emissivity profile of the disk is described by a simple power law with emissivity index $q = 3$, the iron abundance of the disk is the Solar iron abundance ($A_{\rm Fe} = 1$), and the ionization parameter of the disk is $\log\xi = 3$ ($\xi$ in units erg~cm~s$^{-1}$). We fit the simulated data in the range 4-10~keV, as current reflection models are unsuitable to fit the data of black hole X-ray binaries below 4~keV because of the high temperatures of their accretion disks. In general, the spin is better constrained if the black hole is rotating faster. From Fig.~\ref{fig:5}, it is also clear that eXTP/SFA can measure the black hole spin parameter with smaller uncertainties than NICER.

\begin{figure*}[t]
    \centering
    \includegraphics[width=\textwidth]{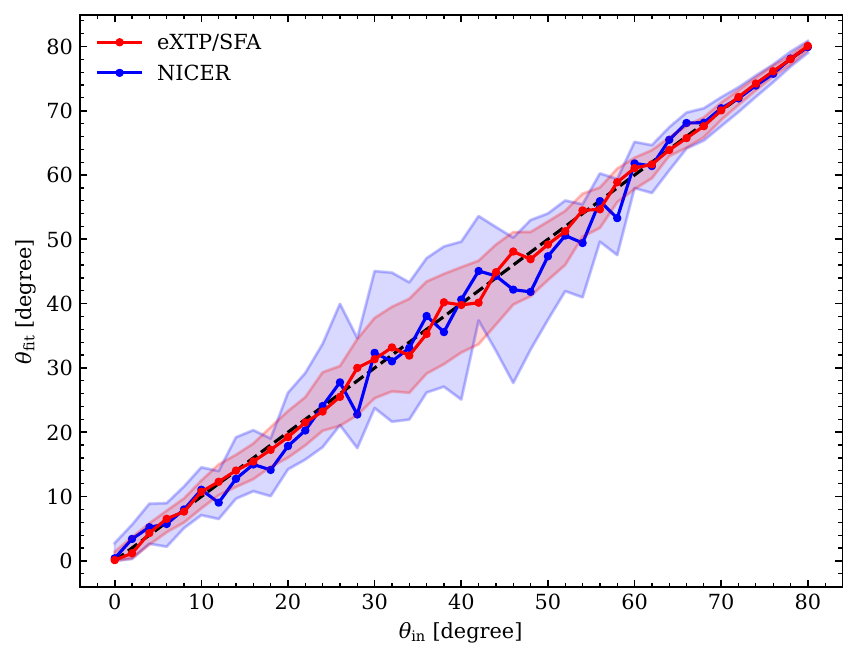}
    \caption{The recoverability of corona inclinations for eXTP/SFA (red) and NICER/XTI (blue) with simulated black hole X-ray binary reflection spectra. The black hole spin is assumed to $a_* = 0.998$. The radius and azimuth angle of the corona are assumed to $r_c = 6\, R_g$ and $\phi_c = 0^\circ$, respectively, and the corona moves radially outward at a speed $v_r = 0.5\, c$. The values of other parameters are the same as those in Fig.~\ref{fig:5}. The dashed line represents that the input inclination for the simulation is equal to its best-fit value. The blue and red data points correspond to the best-fit values for NICER/XTI and eXTP/SFA, respectively. The shaded areas represent the 90\% confidence intervals.}
    \label{fig:6}
\end{figure*}

For simplicity we initially assumed a lamp-post coronal geometry which is axisymmetric. Deviations from axisymmetry might be expected for precessing coronae (e.g. due to differential disk illumination in the precession cycle or by causing a misalignment of a coronal 'jet base' with the thin disk), or in case parts of the disk are obscured relative to the observer (e.g. by the corona itself). Such situations are too complex to model directly for our purposes here, so we show the SFA sensitivity to such effects by considering non-axisymmetric illumination of the accretion disk from an off-axis lamp-post corona.
We simulated the spectrum provided by a corona away from the black hole spin axis using \texttt{offaxxill} model \citep{2023ApJ...955...53F,Feng:2025}, which is an extension of \texttt{relxill} and allows the corona to be located off-axis and to move with arbitrary velocity. 
Fig.~\ref{fig:6} shows the recoverability of corona inclinations using X-ray reflection spectroscopy for eXTP/SFA. 
The black hole spin is assumed to $a_* = 0.998$ in simulations. 
The radius and azimuth angle of the corona are assumed to $r_c = 6\, R_g$ and $\phi_c = 0^\circ$, respectively, and the corona moves radially outward at a relativistic speed $v_r = 0.5\, c$. 
The corona inclination is better constrained if the corona has a low or high inclination.
It is constrained worse if the corona has a moderate inclination.
Fig.~\ref{fig:6} shows that eXTP/SFA provides better constraints on corona inclinations by contrast with NICER.

The presented simulations include the statistical uncertainty due to stochastic flux variability. However, larger uncertainties can be due to imperfect modelling and/or deviations of the real accretion flow from the model assumptions. In such a case, independent ways to test the accretion physics and measure the black hole spin are essential. The X-ray corona geometry (often assumed in the relativistic reflection models as a lamp post) as well the relativistic reflection component itself 
will also leave their signature in the X-ray polarization properties~\cite{Podgorny:2023egq}. eXTP black hole spin measurements can thus be further improved if we combine spectroscopic and polarimetric techniques by analyzing simultaneously SFA and PFA data (see the next subsection).

\subsubsection{X-ray Polarimetry}
\label{sec:spin_polarimetry}

X-ray polarimetry represents an independent tool for measuring the geometry of the innermost accretion flow, the X-ray corona, and the black hole spin. While for the coronal emission the X-ray polarization is expected to be sensitive to the geometry and motion of the corona~\cite{schnittman_x-ray_2010,tamborra_moca:_2018,zhang_constraining_2019,zhang_investigating_2022,ursini_prospects_2022,poutanen_polarized_2023}, for the purpose of measuring the black hole spin the X-ray polarization of emission from the thin disc is more relevant as the relativistic signatures are imprinted in the level and energy dependence of the polarization degree and angle \cite{dovciak_thermal_2008,li_inferring_2009,schnittman_x-ray_2009}. In the case of strongly polarized accretion disks seen at high (almost edge-on) inclination, the polarization signal of returning blackbody radiation from the disk is expected to be significant \cite{schnittman_x-ray_2009,taverna_towards_2020,taverna_spectral_2021,mikusincova_x-ray_2023}.

The recently launched X-ray polarimetry explorer IXPE \cite{weisskopf_imaging_2022} has measured polarization spectra for a dozen black holes in X-ray binaries as well as in active galactic nuclei. The first observation of an accreting black hole in Cyg X-1 revealed that the geometry of the X-ray corona is rather horizontally extended than vertically \cite{krawczynski_polarized_2022}. Similar results were found in other sources with a measured orientation of the radio jet, the X-ray polarization angle being always aligned with it.

The first black hole spin measurement based solely on the IXPE polarization data was reported for LMC X-3 \cite{svoboda_first_2024} and found to be consistent with the spectral continuum fitting method. The high significance of the returning radiation was proposed for 4U 1957+115 \cite{marra_ixpe_2024} and Cyg X-1 \cite{steiner_ixpe-led_2024} in the soft state. These results indicate the presence of an accreting material very close to the black hole, which is only possible around a highly spinning black hole. The eXTP/PFA will have in total $\sim 5$ times larger collecting area than IXPE, which will allow to make more accurate spin measurements for more sources.

\begin{figure*}[t]
\centering
\includegraphics[width=\textwidth]{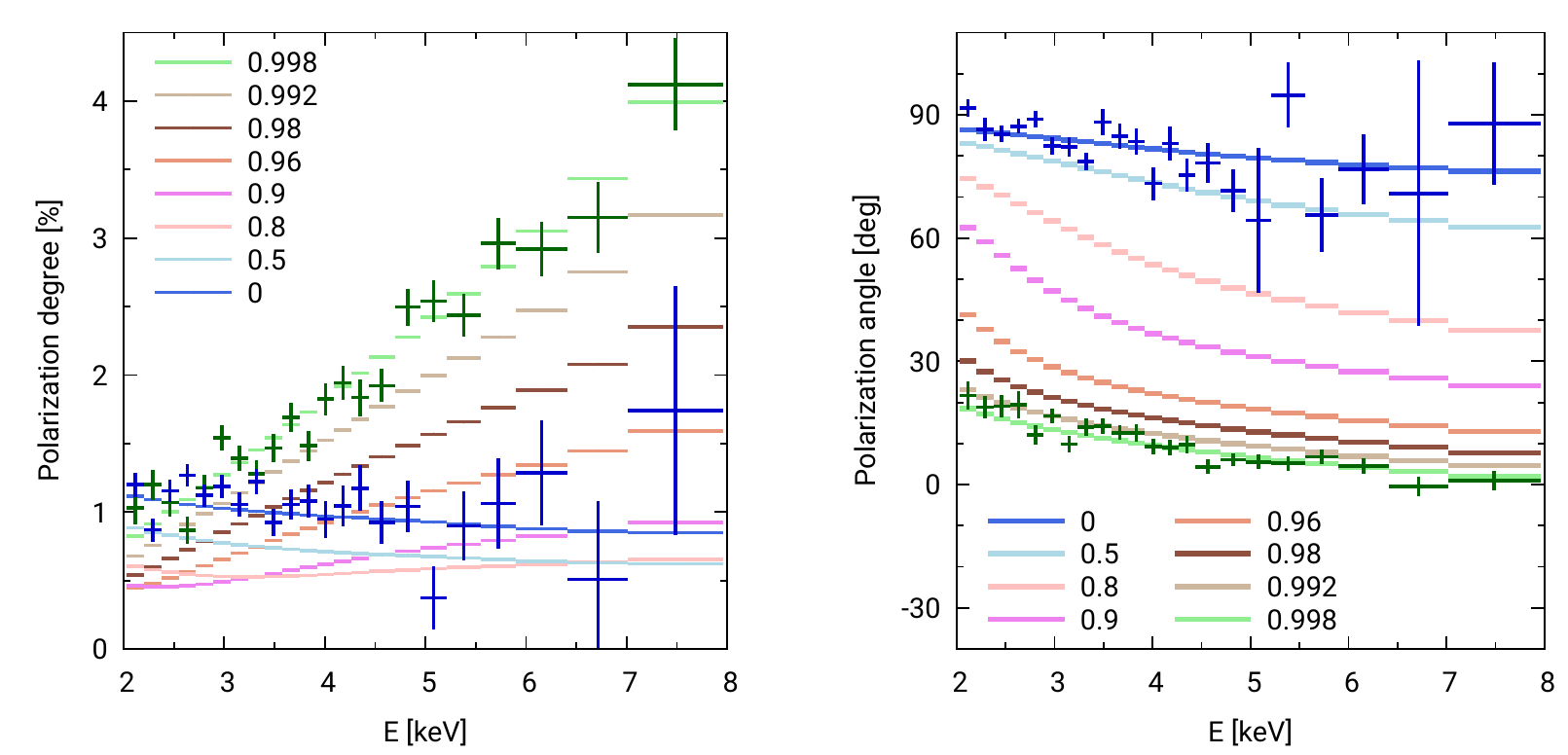}
\caption{The simulated eXTP/PFA observations for polarization degree (PD) and polarization angle (PA) for extremely rotating (green crosses, $a_*=0.998$) and non-rotating (blue crosses, $a_*=0$) black holes. The simulations of 500\,ks observations of a generic 1 Crab black hole X-ray binary (in 2-8 keV) in the soft state, have been prepared with the ixpeobssim v31.0.3 updated for the eXTP/PFA responses. Black hole mass of 7 $M_\odot$, inclination of $60^\circ$, distance of 3 kpc and high disk albedo of 1 was assumed. The model predictions with Stokes parameters binned in 22 energy bins for different values of the black hole spin is also shown. Both the simulated observations and the model predictions were created with {\tt plot polfrac} and {\tt plot polang} commands in XSPEC v12.14.1d.}
\label{fig:7}
\end{figure*}

To demonstrate the capability of eXTP in constraining the black hole spin by X-ray polarimetry, we perform simulations of eXTP/PFA polarisation observations of a bright black hole X-ray binary in the soft (thermal-dominated) state. For the simulations, we assume a black hole X-ray binary with a flux of 1\, Crab in 2--8\,keV and at a distance of 3\,kpc. The mass of the black hole is $7M_\odot$. The observer is located at an inclination of $60^\circ$. The energy spectra and polarization is computed with the {\tt kynbbrr} model \cite{dovciak_thermal_2008,taverna_towards_2020,taverna_spectral_2021,mikusincova_x-ray_2023} and a disk albedo of 1 is assumed. The results are shown in Fig.~\ref{fig:7} (polarization degree and angle in left and right panels, respectively). By looking at the simulated data for an extremely rotating black hole ($a_*=0.998$; green crosses) and a non-rotating ($a_*=0$; blue crosses) black hole, as well as the model predictions for different values of black hole spins, it is evident that eXTP observations will be able to put stringent constraints on black hole spin.

\subsubsection{Thermal continuum fitting}
\label{sec:thermal_continuum}
The X-ray thermal continuum fitting method offers an effective way to measure black hole spin by analyzing thermal emission in the X-ray band from a multi-temperature accretion disk surrounding accreting black holes. This method requires the source to be in the high-soft state where the thin-disk model assumptions hold, with disk emission contributing dominantly ($>$75\%~\citep{Remillard:2006fc}) to the total flux and weak scattering/reflection components. The temperature distribution in the accretion disk and various gravitational effects are highly sensitive to the mass and spin of the central black hole, allowing the spectral energy distribution to provide sufficient details to constrain the black hole's properties. This approach requires precise measurements of the black hole mass $M$, the inclination angle of the system $i$, and the distance $D$. To date, the spins of more than 20 black holes have been successfully measured using this method~\citep{Gou:2009, Gou:2011extreme}.

The black hole spectra for three configurations: eXTP/SFA(full array of 5 SDD modules + 1 pnCCD module), eXTP/SFA-I(one pnCCD only), and NICER/XTI were simulated using the relativistic continuum model {\tt kerrbb}, assuming a 300~ks exposure time. A total of 40 simulated spectra were generated, covering the full range of black hole spin from 0.000 to 0.998. A disk inclination of $i=57^{\circ}$, a source distance of $D=7.47$~kpc, and a black hole mass of $M=8.46M_\odot$ were adopted. The inclination corresponds to the expectation value ($\bar{i} = 1$ radian or $57^\circ$) for randomly oriented orbital planes, computed from the probability distribution $P(i) \propto \sin i$ on a sphere. The mass and distance were chosen to correspond to the median values in the observed distribution of black hole X-ray binaries~\citep{Abdulghani:2024}. The mass accretion rate $M_{\rm {dd}}$ was adjusted to yield a luminosity corresponding to 5\% of the Eddington limit. By default, a canonical spectral hardening factor of 1.7 was assumed. To account for self-irradiation and limb-darkening effects, both corresponding flags in {\tt kerrbb} were set to 1. The self-irradiation, i.e. due to returning blackbody radiation, is assumed to  be fully-absorbed (i.e. albedo 0) and thus reprocessed at the local blackbody temperature. This is inconsistent with the assumed fully reflecting disk in the polarization modelling of the soft state shown in the previous section. Modelling of combined SFA and PFA data from {\it eXTP} will require the development and application of fully self-consistent spectro-polarimetric models, including more realistic disk atmospheres. The power-law photon index was set to 2.0, and the power-law normalization was set to ensure that the disk emission contribution excced 75\% of the total unabsorbed flux in the 2-10 keV energy band. In the actual synthetic spectra, the disk component contributed 80-85\% of the total unabsorbed flux. The absorbed flux ranging from $3.4 \times 10^{-9}$ erg cm $^{-2}$ s$^{-1}$ to $8.7 \times 10^{-9}$ erg cm $^{-2}$ s$^{-1}$ after accouting for the specified column density of $N_{\rm{H}}=1.0 \times 10^{22} cm^{-2}$. No systematic uncertainties were added to the simulated spectra for any of the instruments.

\begin{figure*}[t]
\centering
\includegraphics[width=\textwidth]{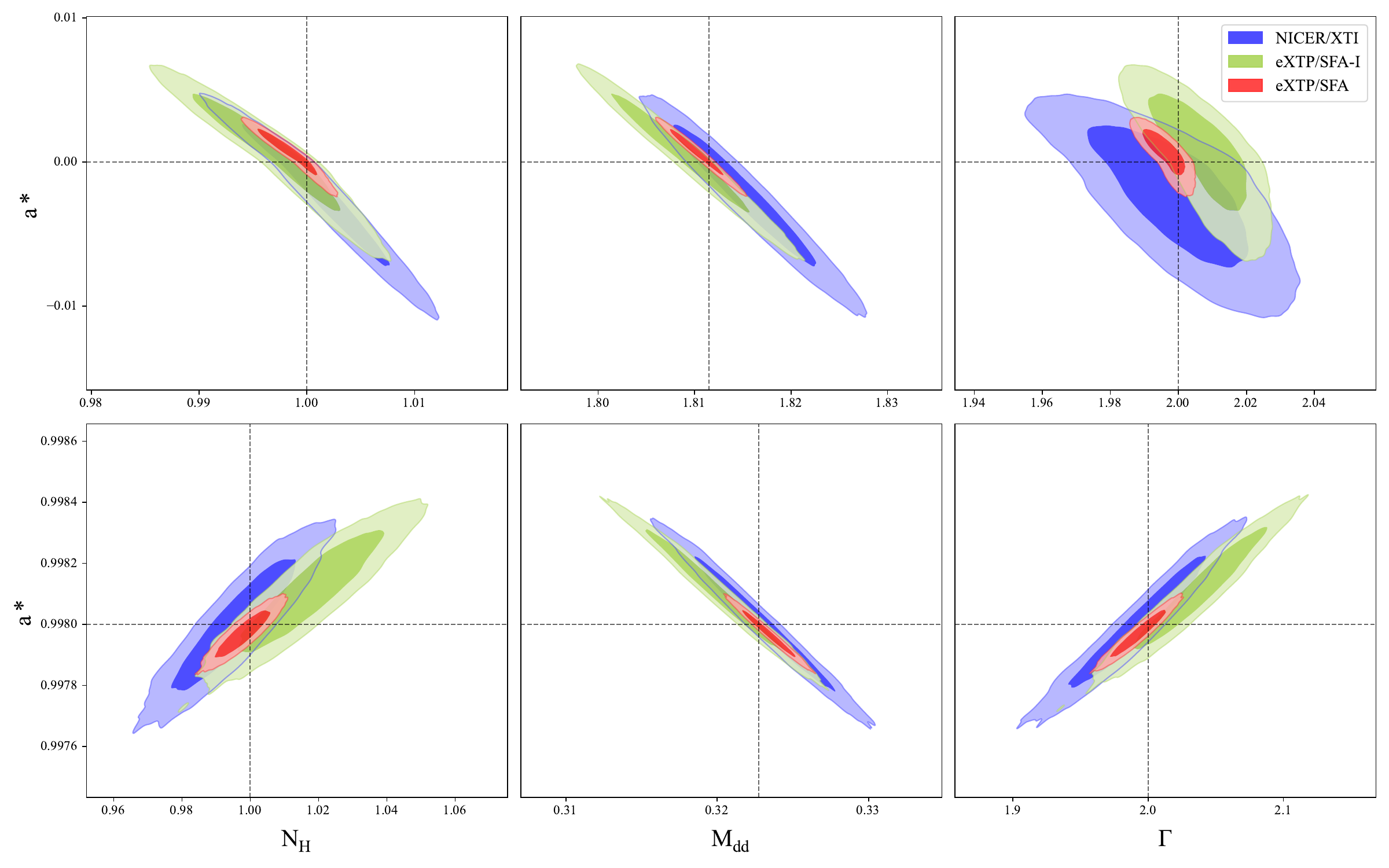}
\caption{
MCMC sampling results for simulated spectra with spin assumed at extreme values of 0.000 and 0.998. The simulation paramters were set as $M = 8.46 M_{\odot}$, $D = 7.47$ kpc, $i = 57.0^{\circ}$, while other parameters were adjusted to achieve a 5\% Eddington luminosity and the disk component contributing 80–85\% of the unabsorbed flux. The flux of the brightest spectrum corresponds to 0.07 Crab in the 2-10~keV energy band. No systematic uncertainties were added. Confidence contours are presented at the 68\% and 95\% confidence levels, with red, green and blue denoting the confidence intervals for eXTP/SFA, eXTP/SFA-I, and NICER/XTI, respectively. The tightest constraints were obtained for eXTP/SFA.}
\label{fig:8}
\end{figure*}

To estimate confidence intervals for the freely fitted parameters ($N_{\rm{H}}$, $a_*$, $M_{\rm{dd}}$, $\Gamma$, and $Norm_{\rm{PL}}$), Markov Chain Monte Carlo (MCMC) sampling was performed using the Goodman-Weare algorithm. Each chain was sampled for $10^6$ steps, and convergence was evaluated using the $\hat{R}$ diagnostic. All model parameters were successfully recovered from the simulated spectra in all configurations. For each spin value investigated, the eXTP / SFA configuration yielded the tightest constraints, with confidence intervals reduced by a factor of 3 to 4 compared to those obtained with NICER/XTI. While the SFA-I produced slightly broader intervals than NICER/XTI, both remained within the same order of magnitude, indicating overall comparable performance. Representative results are shown in Fig.~\ref{fig:8} for spin values of 0.000 and 0.998, with red, green, and blue denoting the confidence intervals for eXTP/SFA, eXTP/SFA-I, and NICER/XTI, respectively.

Near the surface of the black hole, the gravitational field is extremely intense. The final region of spacetime, known as the plunging region, is where infalling matter cannot be halted and inevitably descends into the black hole. It has been found that for black holes with high spin, the combined emission from the plunging region can produce a weak power-law tail in the high-energy spectrum, which may be the origin of corona emission \cite{zhu2012eye}. Fitting synthetic spectra from three-dimensional general relativistic magnetohydrodynamic (GRMHD) simulations with current relativistic thermal emission models (e.g., \texttt{kerrbb} for thin disks and \texttt{slimbh} for slim disks) could help to understand systematics in spin measurements arising from the neglected plunging region emission and other modeling simplifications. Studies of this kind have not converged to a consistent understanding yet, with some simulations shown negligible systematic errors \cite{kulkarni2011measuring} while others indicate a significant overestimation of the black hole spin \cite{Wielgus2022MNRAS.514..780W, Lancova2023AN....34430023L}. Semi-analytical models that include plunging region emission have been developed \cite{Mummery2024MNRAS.531..366M}, and their application to data has shown that black hole spin measurements can be lower than those obtained using standard models in some cases (e.g., MAXI J0637-430 \cite{mummery2024plunging}, MAXI J1820+070 \cite{Mummery2024MNRAS.531..366M} and M33 X-7 \cite{Mummery2025arXiv250513119M}).

Another aspect concerning the reliability of standard disk models is the stability. While the standard disk models predict the disk to be viscously and thermally
unstable in the radiation pressure-dominated regime \cite{Lightman1974ApJ...187L...1L}, observations of BHXRBs in the corresponding soft state show very stable X-ray emission. A possible solution to this discrepancy is the inclusion of magnetic fields \cite{Begelman2007MNRAS.375.1070B}, which could alter the disk structure and temperature profile \cite{Wielgus2022MNRAS.514..780W, Lancova2019ApJ...884L..37L}.




Observations by eXTP will help us explore the effects of radiation from the plunging region and gain deeper insights into the fundamental physical processes at work.
Pressure effects, magnetic fields, and turbulence affect the structure of gaseous flows; nevertheless, the ISCO radius of particle motion plays a crucial role in shaping the inner regions of accretion disks, where strong gravitational effects dominate the structure. While it is known to influence standard geometrically thin, equatorial gaseous flows, its role in vertically extended and/or non-planar (e.g., twisted or warped) accretion flows remains insufficiently explored. This raises a long-standing question about a boundary between stable (energetically bound), plunging, and escaping (unbound) trajectories for non-equatorial motion \cite{2024ApJ...966..226K}.

The signal from the plunging region, albeit weak, needs to be modelled in order to discriminate between different viable explanations \cite{2025MNRAS.537.1963M}. The study of analytical models that include the emission from the plunging region to soft-state data was reported in~\cite{mummery2024plunging}. The deviation from a standard ISCO stress-free disk model is very significant in the data, when observed in a certain range of mass accretion rate. The plunging region is only optically thick when the mass accretion rate is sufficiently high.  



\subsubsection{X-ray Quasi Periodic Oscillations}
\label{sec:qpos}

X-ray Quasi Periodic Oscillations (QPOs) have been routinely observed in stellar mass black holes at low (0.1–10\, Hz) and high (few hundred Hz) frequencies, within 10-20\% of the relativistic precession, orbital, and epicyclic frequencies in the inner disk. These frequencies correspond to the fundamental frequencies of orbital motion in strong-field GR. The low and high-frequency QPOs have sometimes been observed together, in combinations of frequencies consistent with those expected from the multiple relativistic signals associated. These observations suggest that QPOs may be indicative of oscillations occurring within specific radii, thereby offering a potential means to accurately  probe relativistic dynamic of particles in the vicinity of a black hole event horizon. Furthermore, QPOs can provide measurements of black hole spin, a crucial parameter that influences the observed frequencies. 

The Relativistic Precession Model (RPM) has been pivotal in linking QPOs to black hole parameters. It explains black hole QPOs as motions of accreting matter in the strong gravitational field of a Kerr black hole, matching observed QPO frequencies to general-relativistic orbital time-scales. 
In this model, a low-frequency QPO (typically a few Hz) is produced by Lense–Thirring frame dragging that causes the nodal precession of matter around the black hole, while high-frequency QPOs (tens to hundreds of Hz) arise from the Keplerian orbital frequency and the relativistic periastron precession of matter. Such types of motion are assumed to occur at the same characteristic radius, often associated with the inner truncation radius of the geometrically thin accretion disk. These fundamental frequencies depend on the black hole’s mass and spin, so detecting a “triplet” of QPOs (one low-frequency plus two high-frequency signals) allows one to infer the black hole mass and dimensionless spin by solving the GR frequency relations \citep{Motta2014,Ingram2014,Motta2024,2023CoSka..53d.175K}. 

The leading model for low-frequency QPOs assumes that the corona corresponds to an inner hot accretion flow, with orbital motion misaligned with the black hole spin. The misalignment causes the corona to undergo solid-body nodal (Lense-Thirring) precession around the black hole spin axis due to the relativistic frame dragging effect, whilst the disk remains stationary due to viscous forces \citep{Ingram2009}. eXTP is ideally suited to two distinctive tests of this model.

\begin{figure*}
    \centering
    \includegraphics[width=\textwidth]{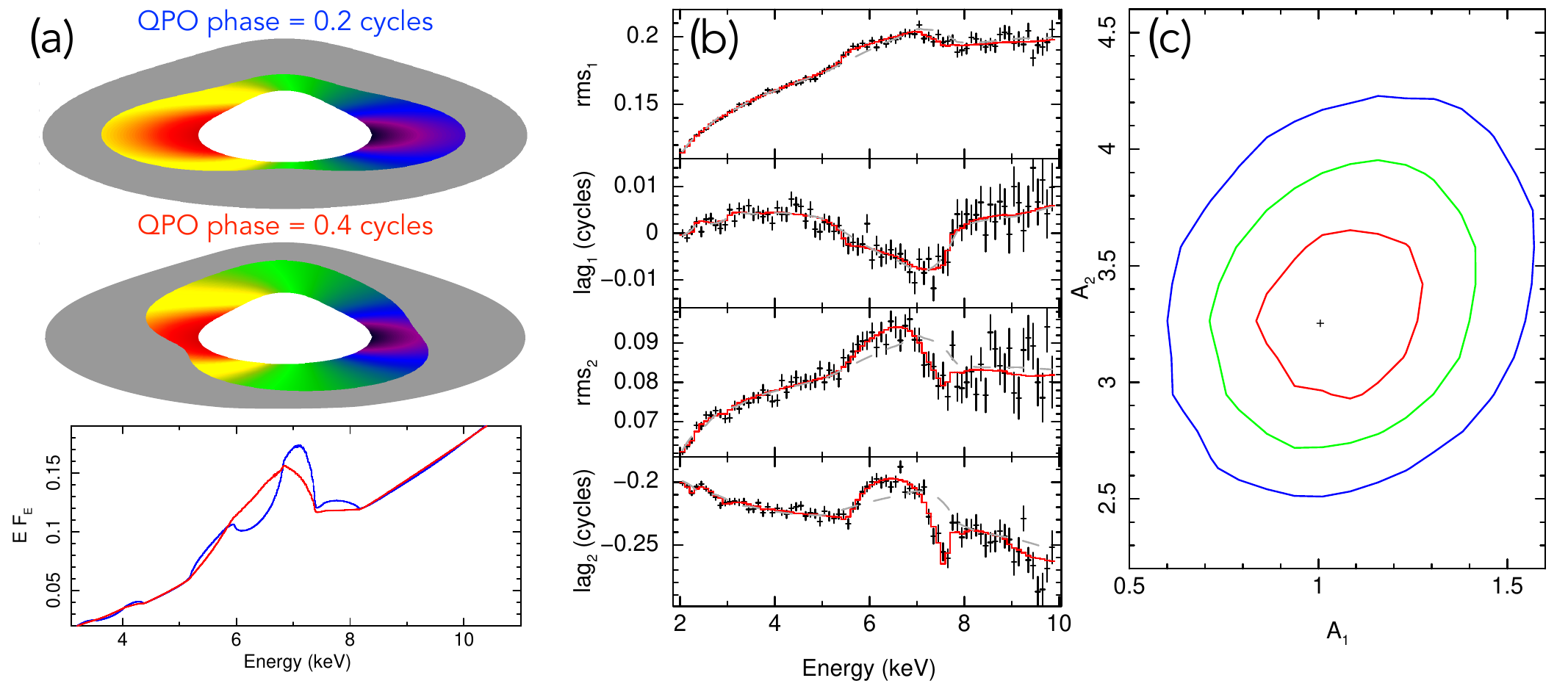}
    \caption{(a) Disk images representing asymmetric illumination by a precessing corona at two precession phases (top and middle) and the corresponding iron line profiles (bottom). The QPO phase-dependent illumination of the disk is parameterized by an analytic function with asymmetry parameters that we set to the best fitting values for H 1743-322 \cite{Ingram2017} of $A_1=0.93$ and $A_2=3.50$. (b)
    Simulated 50 ks eXTP observation of a 250 mCrab black hole X-ray binary with a QPO at 0.4 Hz. We show the amplitude and phase lag of the first and second QPO harmonics as a function of energy (as labeled). The simulated data are unfolded around the best fitting model (red stepped lines). The gray dashed lines represent the null-hypothesis model. (c) Statistical confidence contours (red: $68.27\%$, green: $95.45\%$ and blue: $99.73\%$) on the measurement of the asymmetry parameters from fitting the input model back to the simulated data.}
    \label{fig:9}
\end{figure*}

First, the model predicts that the iron line should rock between Doppler red shifted and Doppler blue shifted as the corona preferentially illuminates different disk azimuths throughout its precession cycle \citep{Ingram2012}. Fig.~\ref{fig:9}a shows an example for two QPO phases. The disk images (top and middle) demonstrate the illumination profile at the two phases, where the rainbow color scheme is used for disk patches illuminated by a flux above a threshold value. The corresponding line profiles for the same two phases are shown in the bottom panel. For axi-symmetric illumination, the line profile would instead be independent of QPO phase. Since the QPO frequency varies stochastically, it is very difficult to phase-fold. Instead, a QPO phase-dependence in spectral shape is optimally inferred in the Fourier domain \cite{Ingram2016}, by measuring the amplitude and phase lag with respect to a reference band as a function of photon energy for the first and second QPO harmonics. Fig.~\ref{fig:9}b demonstrates that a phase-dependence of the iron line profile creates distinctive features in the iron line region of the amplitude and lag spectra.

This method has been used to detect an iron line centroid energy modulation for the black hole X-ray binaries H 1743-322 and GRS 1915+105 \citep{Ingram2016,Nathan2022}. These observations enabled tomographic mapping, whereby the QPO phase-dependent illumination profile of the disk is represented by an analytic function that mimics illumination by a precessing corona \citep{Ingram2017}. In these studies, the best fitting model consists of an asymmetric illumination profile consistent with the illuminator being a precessing corona. However, the null-hypothesis of axi-symmetric illumination (as would result from a stationary corona) has thus far never been ruled out with a statistical confidence greater than $2.5~\sigma$.

The high throughput and spectral resolution of the SFA will enable the first firm detections of asymmetric disk illumination. To demonstrate this, we simulate a 50 ks eXTP observation of a 250 mCrab black hole X-ray binary with a QPO at 0.4 Hz with quality factor $Q=8$. We assume that the photon index $\Gamma$ and the reflection fraction $f_R$ oscillate with QPO phase around their mean values of $\Gamma=1.8$ and $f_R=0.4$, and we model asymmetric illumination with the asymmetry parameters $A_1$ and $A_2$ (using the best fitting values for H 1743-322 \cite{Ingram2017}). We also set $\log\xi=3$, $r_{\rm in}=30~R_g$, $i=70^\circ$ for the disk ionization parameter, inner radius and inclination angle, respectively. We simulate using the cross spectral error formulae of ref. \cite{Ingram2019}. Fig.~\ref{fig:9}b shows the resulting amplitude and lag spectra for the first two QPO harmonics. The features in the $\sim 5-8$ keV range indicative of a QPO phase-dependent iron line profile can be clearly seen in the simulated data. The dashed gray lines represent the best fitting null hypothesis model with $A_1$ and $A_2$ set to zero. We see that this model can clearly be ruled out by the synthetic eXTP observation. Fig.~\ref{fig:9}c shows the statistical confidence with which the asymmetry parameters can be constrained by fitting the input model back to the simulated data\footnote{We fit to the real and imaginary parts of the cross spectrum instead of directly to the amplitude and phase lag, which is statistically favorable.}. The cross represents the best fitting values and the contours represent $68.27\%$, $95.45\%$ and $99.73\%$ confidence. The null hypothesis of axi-symmetric illumination ($A_1=A_2=0$) is ruled out with $\gg 5~\sigma$ significance.

The second test, enabled by the PFA, is to search for modulations of the polarization degree (PD) and polarization angle (PA) on the QPO period. Such polarization QPOs are expected from a precessing corona in analogy with the rotating vector model for radio pulsars. eXTP holds many key advantages over IXPE concerning this test. One is the larger effective area, not just of the PFA, but also of the SFA, which can be utilized by cross correlating the PFA signal with the SFA signal \citep{Ingram2017a}. Another is the superior telemetry. Telemetry limitations restrict each IXPE observation to  a maximum of $\sim 15$ million counts, which will not be the case for eXTP. eXTP will also be better able to slew quickly to outbursting BHXRBs, enabling it to observe when the QPO properties are optimal, and the higher orbit will enable longer uninterrupted observations.

\begin{figure}[H]
    \centering
    \includegraphics[width=0.5\textwidth]{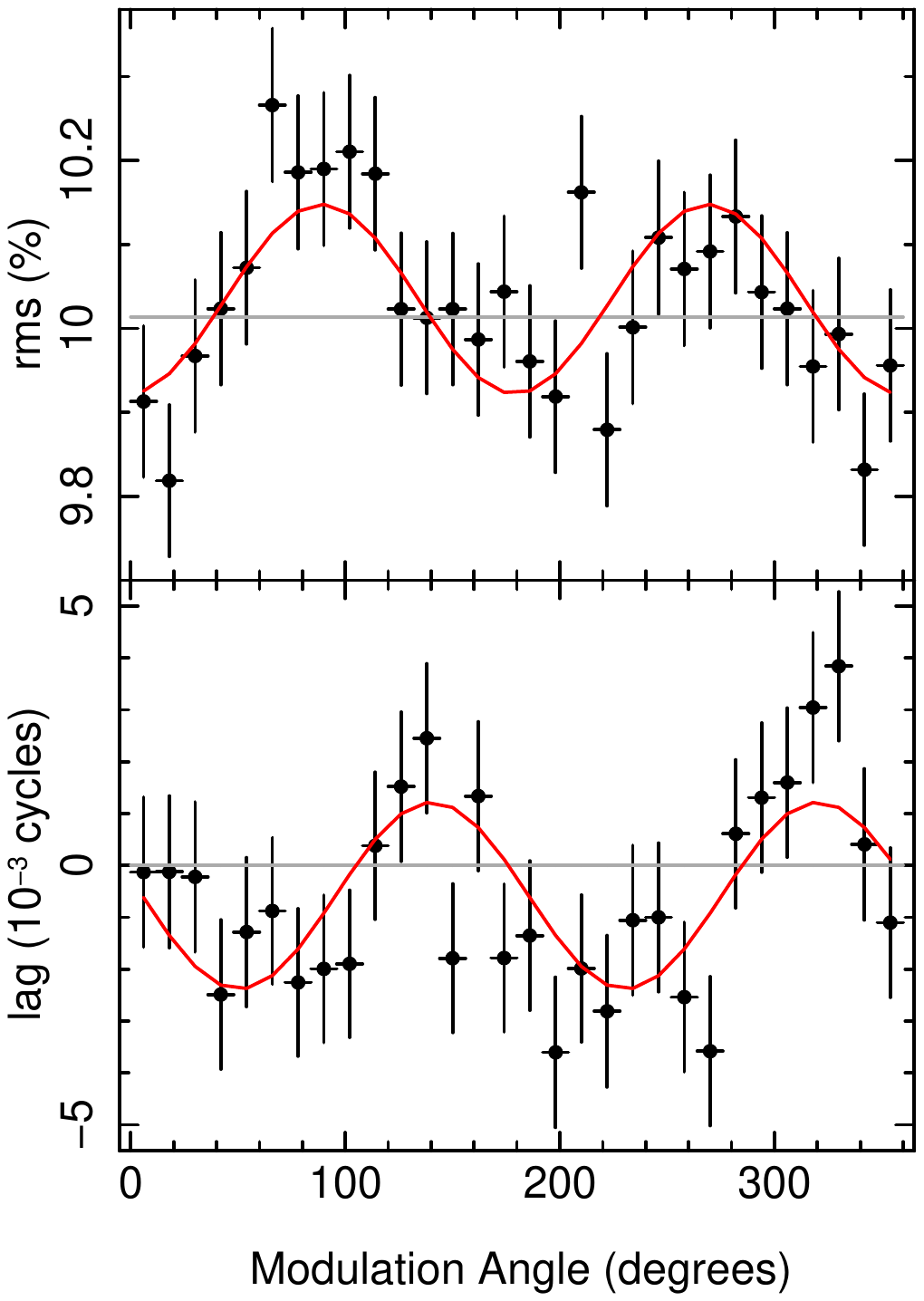}
    \vspace{-1cm}
    \caption{Simulation of a polarization QPO due to relativistic precession of the corona. We use the Fourier detection method of ref. \cite{Ingram2017a}, which measures the QPO amplitude (top) and phase lag with respect to the SFA (bottom) as a function of the modulation angle measured by the PFA. We simulate a 150 ks eXTP observation of a 300 mCrab BHXRB. The best fitting sinusoidal model (red line) is preferred over the null-hypothesis of constant PD and PA (gray line) with $5.2~\sigma$ confidence.}
    \label{fig:10}
\end{figure}

Detection of a polarization QPO requires a long exposure of a bright source, therefore we simulate a 150 ks eXTP observation of a 300 mCrab black hole X-ray binary in the hard state exhibiting a QPO with centroid frequency $0.4$ Hz and quality factor $Q=8$. We calculate the QPO phase dependence of flux, PD and PA assuming that the corona is a torus of outer radius $20~R_g$, misaligned with the disk by an angle $\beta=10^\circ$, which is viewed from inclination $i=70^\circ$ (i.e. Fig 6 of \cite{Ingram2015}). We assume that the time-averaged PD is $4\%$. We simulate using the Fourier detection method of \cite{Ingram2017a}, using an improved treatment of statistical errors \citep{Ingram2019}. This method consists of extracting light curves selected for different modulation angle bins and, for each of these light curves, measuring the QPO amplitude and phase lag with respect to a reference light curve. This is in direct analogy to spectral-timing techniques, except the different time series considered are selected on modulation angle rather than on energy channel. We simulate for 30 modulation angle bins, and we use the SFA for the reference band. We simulate $10^4$ realizations of the same observation. Fig.~\ref{fig:10} shows the results of one representative realization of the simulation. The points are the simulated eXTP data, the gray lines represent the null hypothesis of constant PD and PA, and the red lines the best fitting model of variable PD and PA. We use an F-test for each realization of the simulation to find that the best fitting model is preferred over the null-hypothesis with a statistical confidence of $5.2~\sigma$. We calculate this significance from the median F-statistic of all $10^4$ realizations.

IXPE has thus far made one observation of a strong low-frequency QPO from a black hole X-ray binary, at a frequency of $\sim 1.4$ Hz from the bright source Swift J1727.8-1613. Polarization QPOs are not detected in this observation, with $1 \sigma$ upper limits on the amplitude of any PD and PA oscillations of $0.4\%$ and $3^\circ$ respectively \citep{Zhao2024}. The precession model predicts that the amplitude of PD and PA modulations depends sensitively on source geometry and viewing angle, and thus it is perfectly feasible that future observations will feature much stronger polarization modulations. However, eXTP will even be able to detect very small polarization modulations consistent with the existing Swift J1727.8-1613 observation, for a sufficiently bright source. We simulate a 350 ks eXTP observation of a 500 mCrab source. The QPO frequency and quality factor are the same as before, but now we input polarization modulations set by the best fitting model of \cite{Zhao2024} (PD amplitude of $0.2\%$ and PA amplitude of $1.4^\circ$) and the rms of the QPO in the flux is set to $14\%$. We find that eXTP can detect even these small modulations with a significance of $3.6~\sigma$.

\subsection{Testing GR}
\label{sec:testing_gr}

\begin{figure*}[t]
\centering
\includegraphics[width=\textwidth]{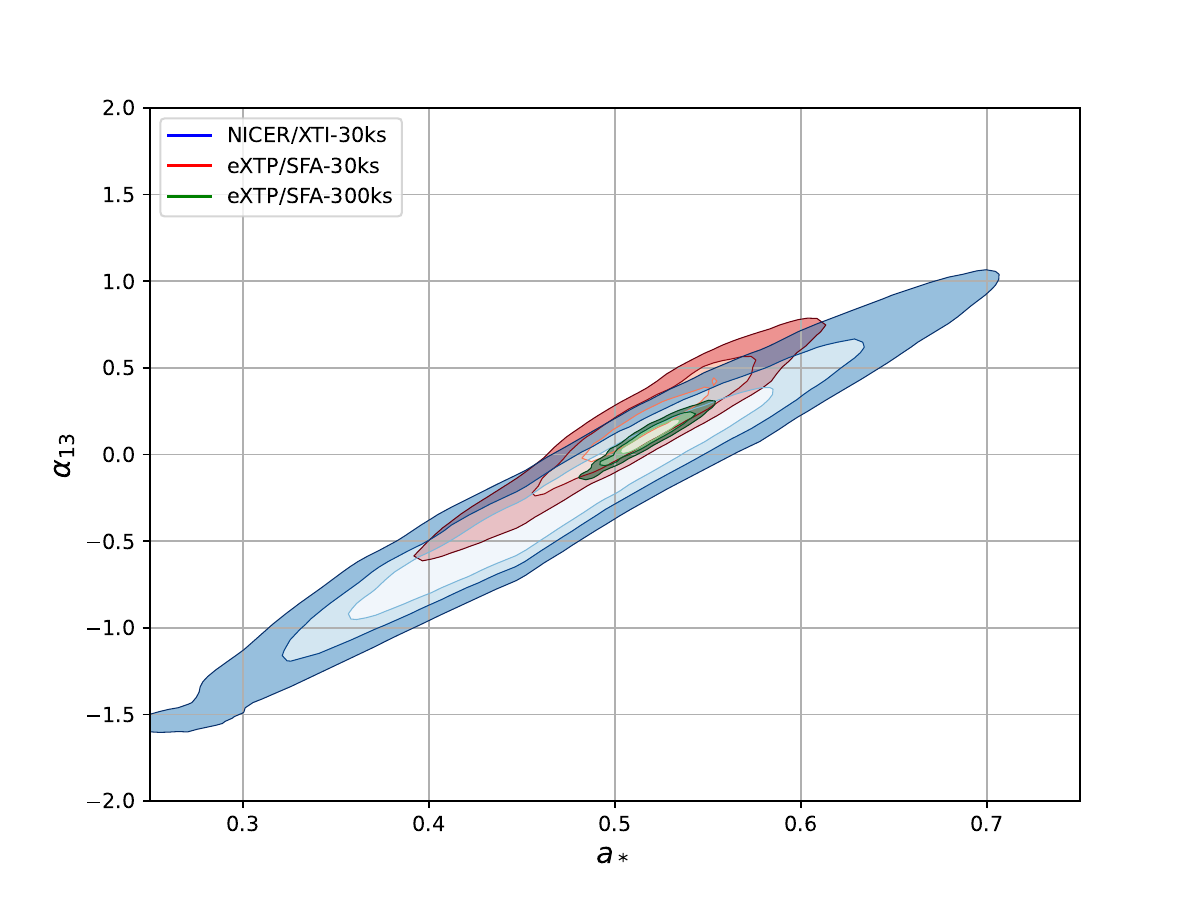}
\vspace{-0.4cm}
\caption{Constraints on the black hole spin parameter $a_*$ and the Johannsen deformation parameter $\alpha_{13}$ ($\alpha_{13} = 0$ corresponds to the Kerr solution of GR and $\alpha_{13} \neq 0$ implies deviations from GR) from a simulated 30~ks NICER observation (in blue), a simulated 30~ks eXTP observation (in red), and a simulated 300~ks eXTP observation (in green) of a bright black hole X-ray binary. In the simulations, we assume that the source is a Kerr black hole with spin parameter $a_* = 0.5$ and the inclination angle of the disk is 70~deg. The other parameters have the values used in the simulations in Fig.~\ref{fig:5}. For every measurement, we show the 68\%, 90\%, and 99\% confidence level curves for two parameters.}
\label{fig:11}
\end{figure*}

The spacetime metric around an astrophysical black hole is normally thought to be approximated well by the Kerr solution. However, there are models beyond GR allowing for macroscopic deviations from the Kerr geometry. Testing the Kerr nature of astrophysical black holes is thus a crucial test of GR in the strong field regime~\cite{Bambi:2015kza}. 

The Kerr nature of astrophysical black holes can be tested with the techniques discussed in Subsection~\ref{ss-spin} for measuring black hole spins: X-ray reflection spectroscopy (see, e.g.,~\cite{Tripathi:2018lhx, Tripathi:2020yts}), X-ray polarimetry (see, e.g.,~\cite{Krawczynski:2012ac, Liu:2015ibq}), continuum-fitting method (see, e.g.,~\cite{Zhou:2019fcg, Tripathi:2020qco}), and QPOs (see, e.g., Refs.~\cite{Bambi:2012pa, Bambi:2013fea}).

X-ray reflection spectroscopy is currently the technique that can provide the most stringent test of the Kerr hypothesis. We simulated a 30~ks observation of a bright black hole X-ray binary with eXTP/SFA and NICER/XTI assuming that the compact object is a Kerr black hole with spin parameter $a_* = 0.5$ and that the inclination angle of the disk is $i = 70$~deg. We fit the data with the Johannsen metric allowing for a non-vanishing deformation parameter $\alpha_{13}$. For the simulations and the fits, we used the model {\tt relxill\_nk}~\cite{Bambi:2016sac,Abdikamalov:2019yrr}, which is an extension of the {\tt relxill} package to non-Kerr spacetimes. The constraints on the black hole spin parameter $a_*$ and the deformation parameter $\alpha_{13}$ (where $\alpha_{13} = 0$ corresponds to the Kerr metric) are shown in Fig.~\ref{fig:11} (after marginalizing over all other parameters): the eXTP/SFA measurements are in red and the NICER/XTI measurements are in blue. eXTP can roughly halve the uncertainty of the NICER measurement with the same exposure time. Fig.~\ref{fig:11} also shows the eXTP/SFA measurements of a putative 300~ks observation of the same source (in green). A longer exposure time can clearly improve the constraints on possible deviations from the Kerr geometry. Unlike spin measurements, where we want to measure the spin of as many black holes as possible, in the case of tests of GR we can properly select the most suitable sources, where systematic uncertainties are under control and subdominant with respect to the statistical ones. It is indeed plausible to image that either GR works or does not, regardless of the source.

With eXTP, the spectral analysis of the reflection spectrum could be combined with the polarization analysis of the source. As of now, there are no models suitable to analyze the X-ray polarimetric data and test the Kerr nature of a black hole, but such models will be surely available within a couple of years from now. 

From the analysis of the thermal spectrum of the disk, it is normally very difficult to constrain possible deviations from the Kerr metric, because of a strong degeneracy between the estimate of the black hole spin and the deformation parameter~\cite{Tripathi:2020qco}. However, the analysis of the thermal spectrum can be combined with that of the reflection spectrum, improving the constraints that can be obtained from the sole analysis of the reflection spectrum~\cite{Tripathi:2020dni,Zhang:2021ymo}. 

The measurement of the QPO frequencies can {\it potentially} provide very stringent constraints on the spacetime geometry around black holes, especially if combined with the constraints inferred with other techniques~\cite{Bambi:2012pa,Bambi:2013fea}. However, the constraints inferred from the QPO frequencies are strongly model-dependent and, at the moment, we do not know the exact mechanism responsible for the observed QPOs, so this technique cannot currently provide reliable tests of the Kerr nature of astrophysical black holes. 

\section{Exploring Accretion Physics in Extreme Gravity}\label{s-matter}


Since the pioneering work by \cite[][]{Shakura:1972te}, accretion power has been understood to be the energy source for emission from accreting stellar-mass black holes in X-ray binaries and supermassive black holes in active galactic nuclei (AGN).

According to their spectral features and timing properties, black hole X-ray binaries can be divided into three spectral states, i.e., the hard state, intermediate state and soft state (also known as the hard, steep power-law and thermal-dominant states) \cite[][]{Remillard:2006fc,Belloni:2005}. The different spectral states are believed to be linked to the relative strengths of disk and coronal emission components.
Although the basic picture has been established, some key questions are still needed to be clarified, such as the formation, evolution and the radiation of the hot corona, the physical mechanism of the truncation of the accretion disk, the origin of QPOs, and the production mechanism of outflows such as disk winds (observed in the soft state) and relativistic jets (observed in the hard and intermediate states). 

A great number of efforts have been made towards establishing the intrinsic connection between black hole X-ray binaries and AGN, such as the fundamental plane of black hole activity \cite{Merloni:2003,Kang:2025} and the unified scheme for black hole accretion (e.g. \cite{Hagen:2024}), in which it is proposed that different types of AGN are determined by Eddington ratio (Eddington rate scaled mass accretion rate) and the viewing angle. As for BHXRBs, the basic picture is in place but many key questions still need to be settled, such as the structure of the disk and corona, the physical mechanism of the production and evolution of AGN QPOs, the origin of quasi-periodic eruptions (QPEs) and the implications for AGN feedback. The latter question has wide relevance, because the process of accretion on to black holes plays an important role in shaping the Universe. The energy released in a small region (e.g., a few tens of gravitational radii) near the black holes in AGN can cause significant feedback which affects the evolution of their host galaxies. 




\subsection{The geometry and dynamics of the accretion flow in stellar mass black holes}
\label{sec:geom_xrbs}


In the hard state, the X-ray energy spectrum is dominated by the emission of a power law, which is believed to originate from the inverse Comptonization of seed photons by a hot plasma (corona) near the black hole.  
The advection-dominated accretion flow (ADAF) was proposed as the mode of accretion in the hard state~\cite{Narayan:1994xi}. It has been suggested that the Comptonization within the ADAF is associated with the source of X-ray power-law emission~\cite{Remillard:2006fc}. However it is clear that this component can be extremely luminous in the brightest hard states during an outburst, unlike a radiatively inefficient ADAF. Therefore we can generalise to consider a hot accretion flow which is not very optically thick, but is at least geometrically thick, so that the presence of disk reflection can be explained. Alternatively, it has been argued that Comptonization could also occur within the jet \cite{Markoff:2005ht}. However, spectral models based on both the hot-flow and jet scenarios can fit the broadband spectral-energy distributions~\cite{2014ARA&A..52..529Y}, making the origin of the X-ray source uncertain.

Moreover, it is well known that the Comptonization-dominated energy spectrum softens as the source transitions from the hard state to the soft state. During the transition, the spectrum becomes dominated by the thermal blackbody emission from the thin disk~\cite{2007A&ARv..15....1D}. Such spectral evolution indicates the co-evolution of the disk and corona during the outburst. In the truncated disk model, the thin disk is truncated beyond the ISCO, and the inner disk is replaced by a hot inner flow. The hot flow contributes to hard X-ray radiation, and its radial suppression and expansion lead to spectral evolution~\cite{1997ApJ...489..865E}. However, some studies, based on X-ray time-lag measurements, have proposed that the corona may vertically expand and then contract as an X-ray binary undergoes the state transition to the soft state~\cite{2019Natur.565..198K}. Thus there is significant uncertainty over the inner structure during state transition, which limits our understanding of the origin of the hard X-ray emission and the accretion physics behind the state transition.

To answer these questions, determining the geometry of the X-ray emitting regions is essential. eXTP can measure and constrain this geometry using several independent techniques, which make use of its advanced instrument capabilities: the excellent SFA soft X-ray response to measure delays between disk-thermal and coronal emission; the combination of SFA effective area and spectral resolution at iron~K line energies for Fe K reverberation studies and QPO tomography of the disk-corona system; PFA's high sensitivity to polarisation signals, which can constrain the coronal geometry even in the early stages of black hole transient outbursts.  

\subsubsection{Disk-corona time delays}
In addition to spectral analysis, timing analysis has been employed to study the accretion geometry and its evolution. Low- and high-frequency noise in the power spectrum of the X-ray fast variations carries information about the accretion flow~\cite{2011MNRAS.414L..60U}. In some X-ray binaries, it has been observed that at low frequencies, the hard X-ray variation lags behind the soft X-ray (hard delay), and the time delay decreases with decreasing optical variation time scale. These `hard lags' are thought to arise from perturbations in the accretion rate at different radii of the accretion disk propagating inward~\cite{2006MNRAS.367..801A}. In contrast, at high variability frequencies (typically $>1$~Hz), soft X-ray variations are observed to lag behind those in hard X-rays (soft delay or `soft lags'). This effect may originate from the time difference corresponding to the irradiation of the accretion disk by the thermal accretion flow, i.e., the X-ray reverberation signal.

To demonstrate the power of eXTP observations, we consider whether the combination of spectral and timing analysis could strongly constrain the accretion geometry. More specifically, we will investigate whether the timing differences between various geometries can be disentangled from additional observations while maintaining an identical time-averaged energy spectrum. To achieve this, theoretical modeling of the X-ray fast variability, considering the variability mechanisms mentioned above, is required.

Phenomenological models which incorporate the disk-corona propagation and reverberation show that complex time-lag behaviour can result \cite{2025MNRAS.536.3284U}. The X-ray radiative process in the disk-corona system was simulated in \cite{2018ApJ...858...82Y} and \cite{2020ApJ...897...27Y}, which includes the thermal emission from the disk, the Comptonization within the corona, and the reflection of coronal photons by the disk. Recently, \cite{2025arXiv250203995Z} incorporated propagating fluctuations through the disk and the light-traveling effect between the disk and corona into the simulation, in order to model the fast variability of an accreting black hole. Here, we utilize Zhan's simulation code to model the energy- and time-dependent variability, demonstrating the capability of additional spectral and timing observations to constrain the inner accretion geometry.

\begin{figure}[H]
\centering
\includegraphics[width=\columnwidth]{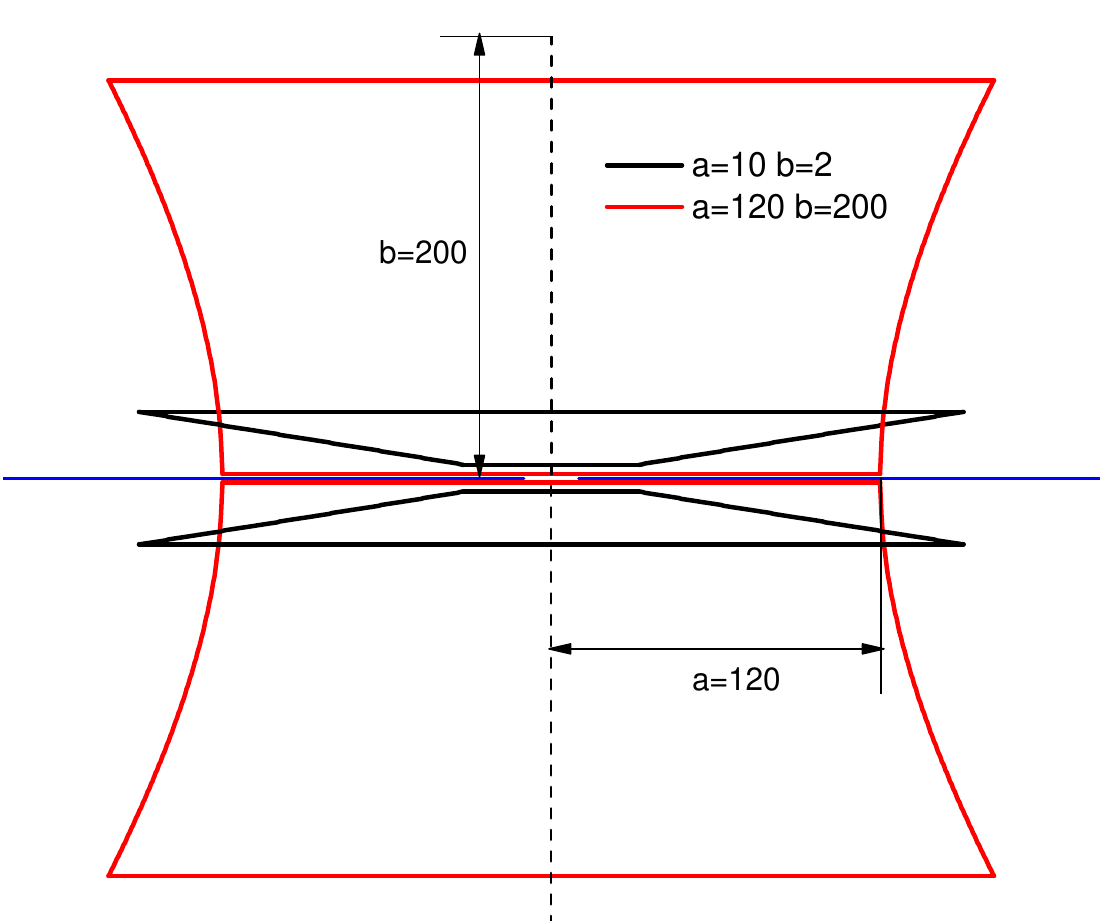}
\caption{Cross-section of the hyperbolic corona in the black hole frame. \textit{a} and \textit{b} are the semi-major and semi-minor axis of the hyperboloid. For disk-like corona (black solid line): \textit{a} = 10, \textit{b} = 2, upper boundary height $\textit{h}_{\rm t} = 30$, lower boundary height $\textit{h}_{\rm b} = 6$. For jet-like corona (red solid line): \textit{a} = 120, \textit{b} = 200, $\textit{h}_{\rm t} = 180$, $\textit{h}_{\rm b} = 2$. The units of all numbers are $R_g$.}
\label{fig:12}
\end{figure}

We model the corona as a hyperboloid with upper and lower surfaces (see Fig.~\ref{fig:12}) and can parameterize it using the following equation:
\begin{equation}
\left\{ 
\begin{array}{cl}
 \frac{x^{2}+y^{2}}{a^{2}}-\frac{z^{2}}{b^{2}}=1  \\
h_{\rm b} \leq |z| \leq h_{\rm t} \\
\end{array} \right.
\end{equation}
where \textit{a} and \textit{b} are the semi-major and semi-minor axis of the hyperboloid. $\textit{h}_{\rm t}$ and $\textit{h}_{\rm b}$ represent the heights of the top and bottom boundaries, respectively. The reference coordinate system is Cartesian, with the black hole at zero, where a geometrically thin and optically thick disk is situated in the \textit{xy} plane, depicted by the blue solid line in Fig.~\ref{fig:12}. The disk has a truncation radius of $10 R_g$ and an external radius of $500 R_g$. Two distinct geometric types (disk-like and jet-like) correspond to varying parameters of the hyperboloid (see Fig.~\ref{fig:12}). The disk-like corona is wider than it is tall, resembling a radially extended disk, while the jet-like corona appears the opposite, resembling a jet.

In the two geometric models described above, we simulated the count rates over a duration of 200 s using a 1 ms time bin. The energy range extends from 0.1 to 12.0 keV, divided into 0.1 keV bins. The count rates were normalized so that the flux in the 2–10 keV band corresponds to 1 Crab. The code is based on the Monte Carlo simulations and tracks only one photon in a single cycle. For simplicity, the trajectories of photons do not take into account relativistic effects. Photons originate from a thin disc, which is described by the Novikov-Thorne model ~\cite{1973blho.conf..343N}. Their emission positions are uniformly distributed as determined by the inverse cumulative distribution function (iCDF), which is based on the radial emission profile of the disc. Photons will obtain randomly assigned momentum and shoot towards infinity or the corona, and the initial energy follow the blackbody radiation function. Photons in the corona will undergo inverse-Compton scattering and then be shot to infinity or return to the disc, that is, reflection occurs. Reflection is a complicated process. We have to take into account all atomic processes, including atomic absorption and excitation, scattering, ionization, and so on. To simplify the program, we adopt the reflection model developed by Garc{\'\i}a et al.~\cite{2013ApJ...768..146G} and Dauser et al. ~\cite{2013MNRAS.430.1694D}, which can generate normalized reflection spectra, including continuum and line emission. We add the propagation fluctuations of accretion rate to generate variability and obtain the signal in the time dimension by recording the number of photons received in different time bins. 

\begin{figure}[H]
\centering
\includegraphics[width=\columnwidth]{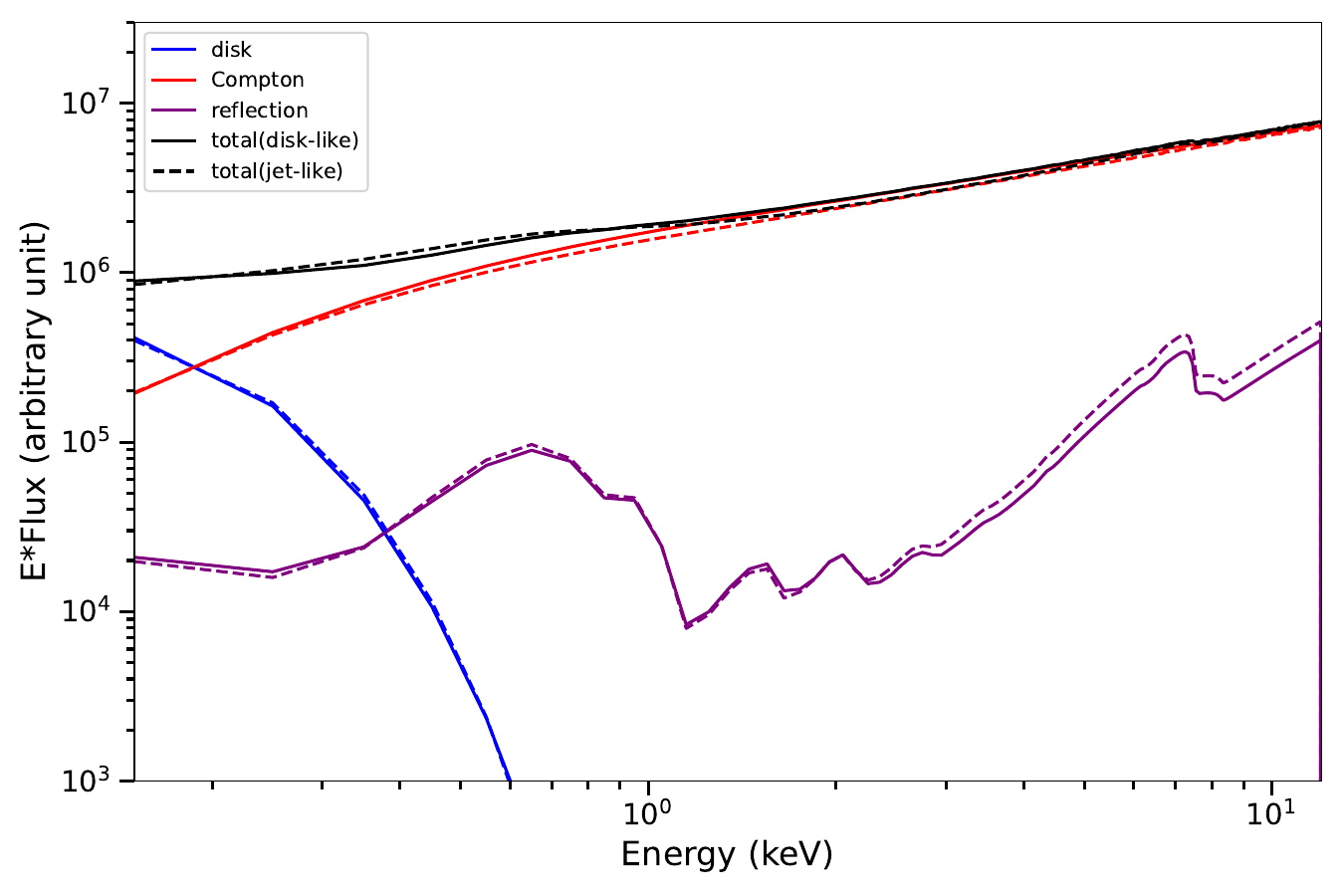}
\caption{Time-averaged energy spectra of the three components and total for both disk-like (solid line) and jet-like models (dashed line). The Y-axis is labeled by the energy multiplied by the flux. }
\label{fig:13}
\end{figure}

Fig.~\ref{fig:13} shows the time-averaged energy spectrum in the simulation results. The solid and dashed lines represent the energy spectra of the disk-like and jet-like models respectively. Although the two geometric models shown in Fig.~\ref{fig:12} show significant differences, we can still ensure that the spectral shapes are similar by adjusting the optical depth of corona. Blue, red and purple represent disk, Compton and reflection components respectively in Fig.~\ref{fig:13}. Compton component exhibits a power law with photon index $\sim 1.4$. Reflection component shows distinct emission lines, such as iron line located in 6-7 keV. The black lines represent the total of all the components. As expected, they are dominated by the Compton component.

\begin{figure}[H]
\centering
\includegraphics[width=\columnwidth]{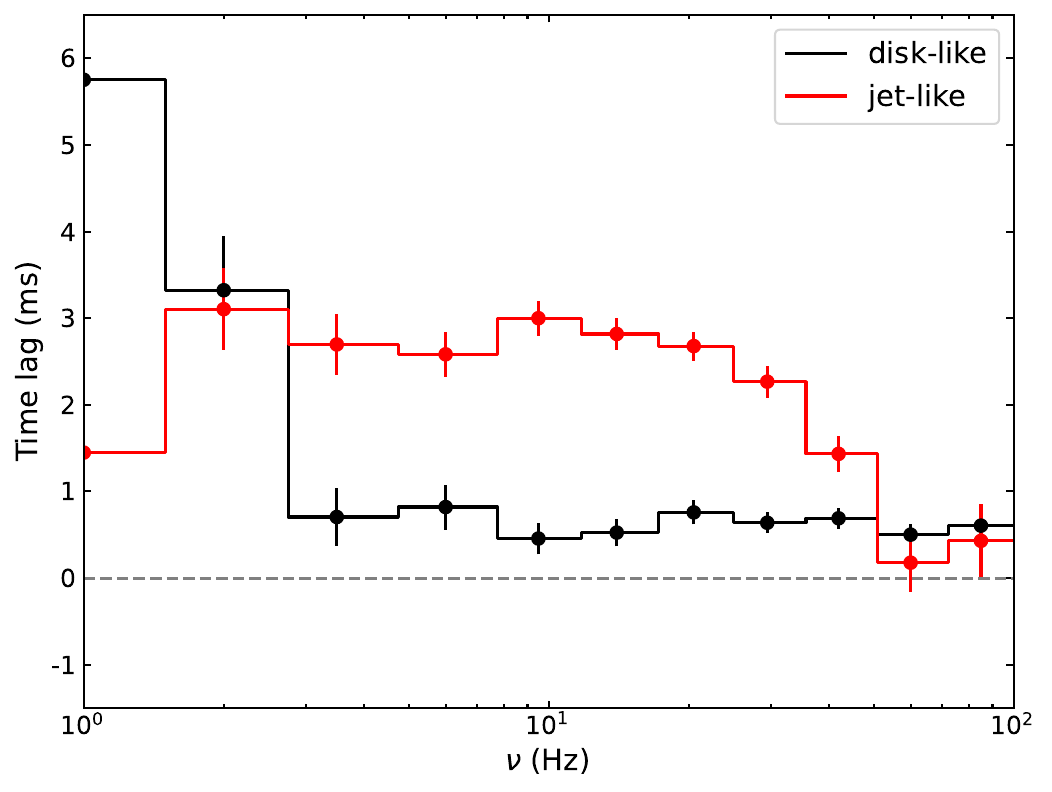}
\caption{The time lag as a function of frequency between 0.5-1.0 keV and 2.0-5.0 keV. The disk-like model is depicted in black, while the jet-like model is shown in red.}
\label{fig:14}
\end{figure}

For the light curve, the count rates in each energy bin were multiplied by the mean effective area for the corresponding energy bin. The count rates within the specified energy bands (0.5-1.0 keV and 2.0-5.0 keV) were summed to produce the light curve. Using the Python library {\tt Stingray v2.0.0}~\cite{matteo_bachetti_2024_10813181,2019ApJ...881...39H,bachettiStingrayFastModern2024}, the time lag-frequency between the 0.5-1.0 keV and 2.0-5.0 keV bands was computed. Fig.~\ref{fig:14} presents the results, where a positive lag indicates a hard lag, meaning the hard photons are delayed relative to the soft photons. The simulation results in Fig.~\ref{fig:13} and Fig.~\ref{fig:14} clearly indicate that, while the energy spectra of the disk-like and jet-like corona are indistinguishable, the frequency-dependent lag reveals significant differences.

\subsubsection{Reflection reverberation}

X-ray reverberation mapping is a powerful technique to probe the structures of the innermost regions of accreting black hole systems, allowing for constraining the properties of black hole independent of the direct spectral-fitting methods (Section 2.2.1) as well as mapping disk-corona geometry and dynamics. 
In the simplest picture of X-ray reverberation mapping, a compact corona that emits Compton-scattered X-ray photons is placed above the black hole. The incident coronal photons irradiate the cold accretion disk and are reprocessed to give rise to a
reflection spectrum, with the most prominent feature being the Fe K$\alpha$ fluorescent line. 
Temporal changes in the reflection spectrum are expected to lag
behind the coronal continuum variations 
because of the extra light-travel time from the corona to the disk. 
More sophisticated reverberation models include the effects of more extended coronae, either in the radial or vertical directions.
So far, X-ray reverberation lags were discovered in both supermassive and stellar-mass black hole accreting systems, 
including AGNs and X-ray binaries \cite[][for a detailed review]{Uttley2014}, however Fe~K reverberation signatures have only been systematically observed in AGN (\cite{Kara2016a}, but see also \cite{2019Natur.565..198K}).

\begin{figure}[H]
\centering
\includegraphics[width=\columnwidth]{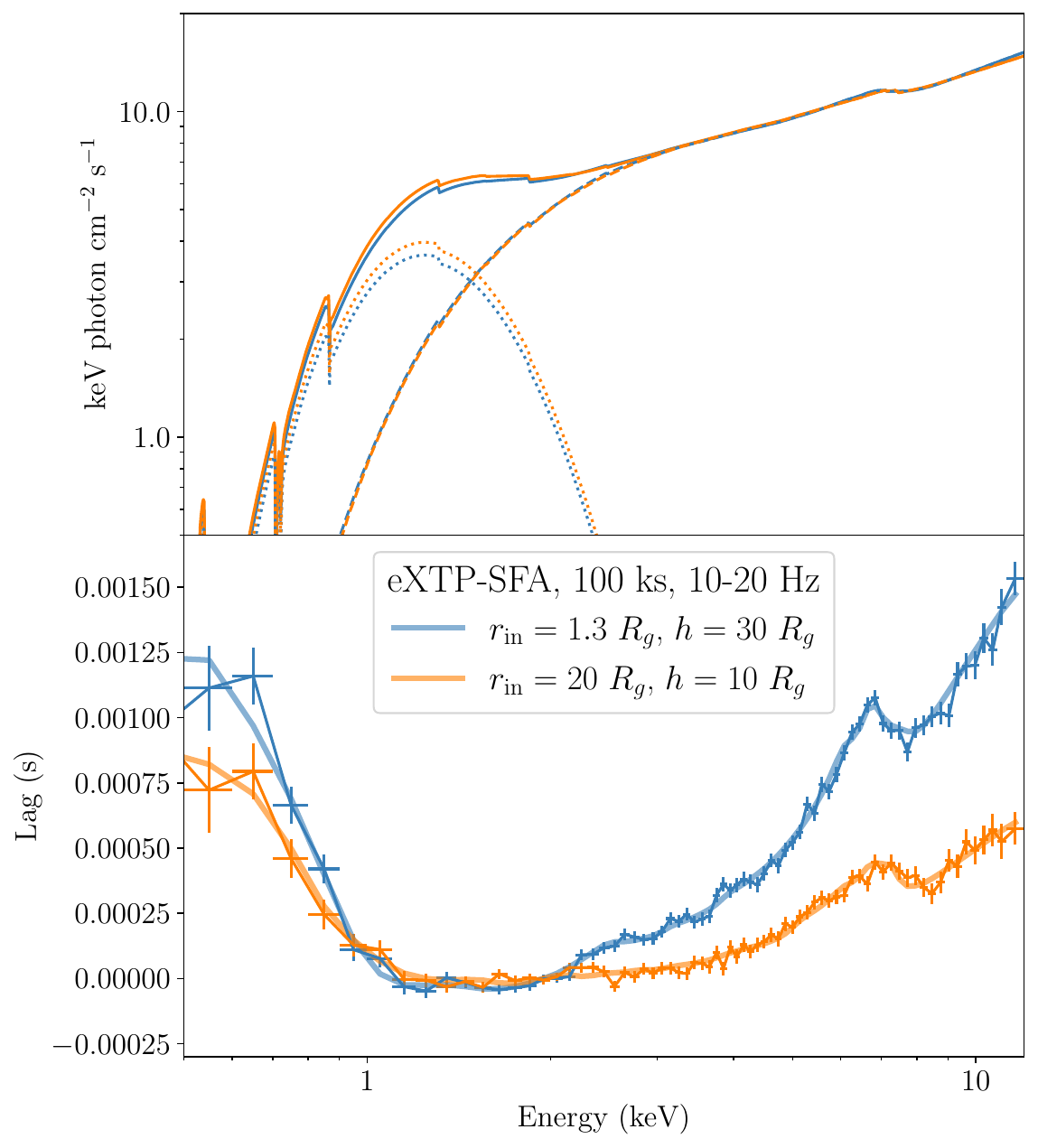}
\caption{Spectral model and simulated eXTP-SFA lag vs. energy spectra for a 1 Crab hard state BHXRB, showing that two distinct inner geometries for disk inner radius ($r_{\rm in}$ and source height ($h$) can be easily distinguished with the lags. The upper panel shows the assumed spectral models with disk emission (dotted lines) set to have similar disk temperature, making the overall spectral energy distributions nearly identical. The lower panel shows the simulated lag data. The simulation assumes 100~ks exposures and shows the lags obtained for the 10-20~Hz frequency range, where evidence for soft reverberation lags has been seen by NICER and where reverberation delays are expected to dominate over propagation delays, which are washed out on these time-scales. The simulation uses the \texttt{reltrans} \cite{Ingram2019_reltrans, Mastroserio2021} lamp-post model for reverberation lags. The lag model curves are shown with the smooth solid lines and data and model are set to zero lag at 2~keV so the shapes can be easily compared.}
\label{fig:15}
\end{figure}

Because of its large collecting area at Fe K energies, eXTP/SFA will open up new possibilities for the use of X-ray reverberation to constrain corona and disk geometries in X-ray binaries. Fig.~\ref{fig:15} (lower panel) shows the simulated lag vs. energy spectrum obtained by eXTP/SFA from a bright (1 Crab) hard state black hole X-ray binary with a 100~ks exposure (which would also be typical to obtain excellent polarization measurements). The upper panel shows the corresponding model spectra including disk and Comptonized plus reflection components. For the simulation we use the \texttt{reltrans} model \cite{Ingram2019_reltrans, Mastroserio2021} and assume different geometries with either a truncated disk and moderately compact source height, or a disk extending to the ISCO of a maximally spinning black hole with a large source height, comparable to what might be expected from a jet-like corona. Given our current limited knowledge of accretion geometries in the hard state, both models could show similar disk temperatures due to a combination of intrinsic heating and coronal reprocessing. The reflection spectral shapes are also similar on account of the degeneracy of large source height vs. disk radius, so spectra alone are insufficient to distinguish the geometries. However, the lag-spectra are distinct and clearly show the Fe~K feature which is currently barely detectable with NICER, allowing the lags to be unambiguously associated with reverberation. Thus eXTP can use reflection reverberation signals for the first time as an unambiguous probe of disk-corona geometry, enabling comparison with multiple independent techniques at the same time.

\subsubsection{Polarization measurements of coronal geometry}
X-ray polarimetry provides an independent tool to probe the accretion geometry.
Polarization degree provides the measure for the elongation of the hot medium -- the corona -- while the polarization angle indicates its orientation.
The polarization degree depends on system inclination, with maximum values for edge-on views and zero for a spherical corona.
Previous hard-state polarization measurements find the corona is extended in the direction orthogonal to the jet axis \cite{krawczynski_polarized_2022, Veledina2023, Ingram2024,Podgorny2024}.
The high observed polarization degree ($\sim$4\%), however, challenges standard models of a static corona, suggesting the coronal matter might be outflowing \cite{poutanen_polarized_2023}.
Key questions remain: is the dynamic, outflowing corona always present in these systems? What is the outflow velocity and does it evolve over the course of the outburst?
Additionally, the link between inner coronal outflows to the optical and infrared wind outflows ubiquitously seen at the initial stages of the outburst is presently missing.

\begin{figure}[H]
\centering
\includegraphics[width=\columnwidth]{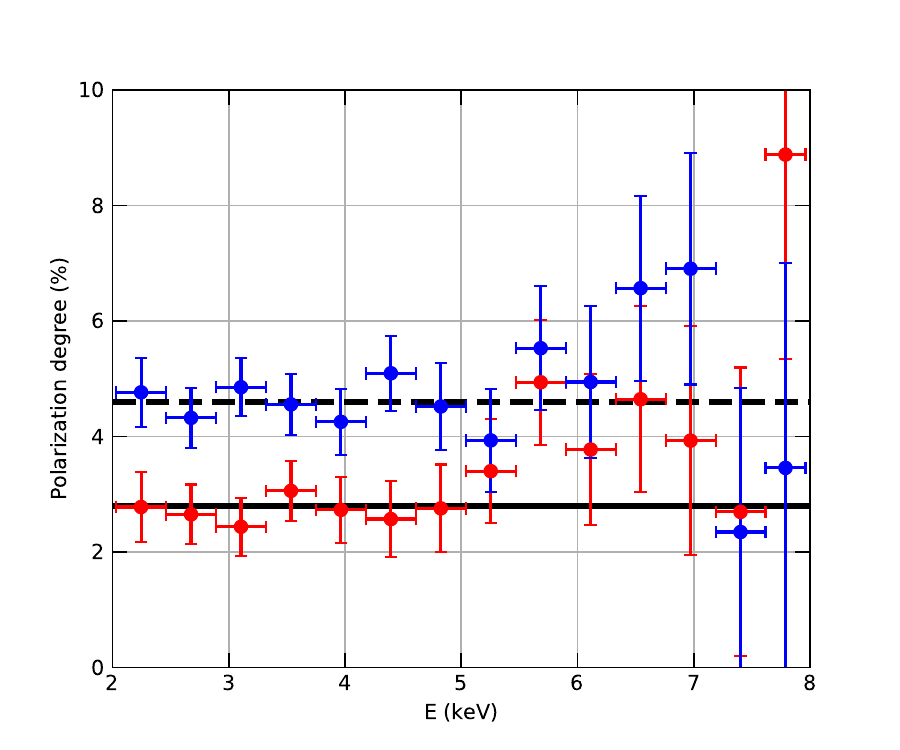}
\caption{Simulated eXTP/PFA data for a generic BHXRB source at the initial stages of an outburst. The binary is assumed to have an inclination of $i=30^{\circ}$, with an incident spectrum characterized by a photon index $\Gamma=1.6$ in the 2--8 keV range and an unabsorbed flux of 100 mCrab. The model PD corresponds to a slab corona geometry with an outflow relative velocity of $\beta=0$ (static, black solid line) and $\beta=0.4$ (mildly relativistic, black dashed line), based on \cite{poutanen_polarized_2023}. An eXTP exposure of 100 ksec is sufficient to distinguish between these alternatives.}
\label{fig:16}
\end{figure}

These questions are not expected to be solved with the IXPE instrument, given the fast evolution of the transient black hole X-ray binaries, their typically dim fluxes in the initial outburst stages and the requirement of polarization measurement with sub-per cent accuracy. To get to a precision of 0.2\% on polarization degree, required to constrain the outflow parameters, a generic 100 mCrab hard state black hole X-ray binary source would need to be observed by IXPE for more than a week (accounting for the $\sim$50\% effective exposure time).
This time is comparable to the characteristic timescales of the outburst evolution, preventing precise measurements of the geometry at distinct times, and a clear understanding of geometric evolution.
In contrast, as shown in Fig.~\ref{fig:16}, eXTP/PFA brings the exposure time down to about a day, hence becoming uniquely suited to track the evolution of accretion geometry for black hole transients.

\subsubsection{Low-frequency QPOs as a disk-corona probe}

The geometry of the corona surrounding black holes is not static during outbursts. Recent studies propose that the coronal geometry undergoes dynamic evolution, particularly during spectral state transitions from the hard to soft state \citep[e.g.][]{Wang2022,Cao2022}. High-cadence observations with the eXTP mission will be uniquely positioned to probe this evolution through its unprecedented combination of simultaneous X-ray spectral, timing, and polarimetric measurements. Furthermore, eXTP’s capacity for prolonged exposure significantly enhances the likelihood of capturing rapid changes of corona during state transitions \citep[e.g.][]{Liu2023,Yang2023,Li2025}.





Low-frequency QPOs are commonly observed in X-ray light curves of black hole X-ray binaries \cite{Ingram:2019mna} and are widely attributed to the relativistic (Lense-Thirring) precession of the inner hot accretion flow \cite{Ingram2009}. Recent studies using NICER and NuSTAR observations have revealed that the disk-blackbody component also exhibits variations on the QPO timescale. Notably, the disk temperature variations were found to precede changes in both the non-thermal and disk fluxes by approximately 0.4--0.5 QPO cycles \cite{2024ApJ...973...59S}. This phenomenon has been proposed to result from the precession of the inner disk. 

It is important to note that the fitted peak temperature ($T_{\rm in}$) may be affected by electron scattering and general relativistic (GR) effects, and can be expressed as: $T_{\rm in}=f_{\rm GR}\left(i,a_*\right)f_{\rm col}T_{\rm eff}$, where $f_{\rm GR}$ represents the fractional change of the fitted color temperature due to the GR effects, $i$ is the inclination angle, $a_*$ is the black hole spin, $f_{\rm col}$ denotes the hardening factor, and $T_{\rm eff}$ is the effective temperature of the inner disk \cite{1997ApJ...482L.155Z}. Numerical simulations suggest a canonical value of $f_{\rm col}\sim1.6-1.8$ \cite{1995ApJ...445..780S,2005ApJ...621..372D,2006ApJ...636L.113S}, whereas $f_{\rm GR}$ strongly depends on the inclination. As demonstrated in previous studies, the GR effects cause the spectrum to be redshifted at small inclinations, but blueshifted at large inclinations \cite{1975ApJ...202..788C,1997ApJ...482L.155Z}. This indicates that the apparent temperature could display a positive correlation with the inclination. However, for the accretion disk flux over the precession period, it is primarily influenced by the projected area, and is therefore expected to be negatively correlated with the inclination. Consequently, a large phase difference (roughly 0.5 QPO cycles) between the temperature variation and the flux variation could be potentially observed. By employing the Hilbert-Huang Transform method \cite{1998RSPSA.454..903H,2023ApJ...957...84S}, phase-resolved spectral analysis with eXTP/SFA observations will be sufficiently sensitive to detect variations in the disk-blackbody component on the QPO timescale, thanks to the significantly larger effective area of eXTP/SFA in the 0.5--10 keV range compared to NICER. Consequently, the origin of these variations can be effectively constrained.



\subsection{Mapping the inner regions of supermassive black holes}
\label{sec:geom-smbh}
\subsubsection{X-ray reverberation mapping of the inner regions}



As for accreting stellar-mass black holes, the measurement of energy-dependent time delays, i.e., lag-energy spectra, are useful to test detailed models for the innermost emitting regions of AGN, such as propagating fluctuations and reverberation \cite[e.g.,][]{Kara2016a}. 
However, the spectral-timing measurements require observations with very high count rates, so can only be done for a handful of bright AGNs with current instruments. Furthermore, for past and current proportional-counter instruments with large collecting area at Fe~K energies (e.g. RXTE, ASTROSAT), the spectral resolution is poor, which prevents detailed investigation of reverberation properties in the Fe K band.  
With its large effective area and CCD-quality spectral-resolution in the 0.5--10 keV band, eXTP/SFA will provide much better signal-to-noise measurements of the lag-energy spectra, at sufficient resolution to resolve Fe~K reverberation signatures. 

The improved energy resolution provided by eXTP/SFA allows for measuring more precisely the changes of the Fe K line profile. With the light-travel lags, the extension of the redshifted Fe K line emission and its location can be simultaneously measured, which are critical 
to test the effects of GR at small radii, and constrain the
accretion disk dynamics. 

The high count rates enable direct time-resolved spectral studies of variable components in the Fe K complex, probing the flaring regions of the accretion disk down to spatial scales of several to tens of gravitational radii from the central black hole \cite{Marinucci2020}. 
The physical connection between the local flare activity of accretion disk and accretion rate can be revealed. 
In some AGNs, the disk reflection component and the broad Fe K line (if present) are found to vary with much smaller amplitude than the continuum \cite{Markowitz2003}, which can be interpreted in the context of the light-bending model. In such a model, physical motion of the X-ray source causes variability in the X-ray continuum but the illumination of the disk (and therefore the Fe K line flux) can remain relatively unaffected in some regimes \cite{Fabian2004, Miniutti2004}. 
eXTP/SFA observations will be able to reveal characteristic properties of 
reflection continuum (including broad Fe K line) in response to the large-amplitude continuum variability \cite{Shu2010, Liang2022}, allow for testing the light-bending model and constraining coronal geometry and dynamics. 

X-ray reflection and reverberation can be used to probe accretion flow and black holes in super-Eddington accretion. The study of super-Eddington accretion is important as such extreme accretion likely has played an important role in the growth of massive black holes in the early universe, as recently observed by JWST \cite{Madau2024, Suh2024}.  It has been theoretically predicted that super-Eddington accretion around black holes produces slim disks instead of conventional thin disks \cite{Abramowicz1988}.
Recent state-of-the-art numerical simulations of super-Eddington accretion further reveal that optically thick winds are launched from the geometrically and optically thick disks \cite{Ohsuga2009, Jiang2014}. Therefore, one expects that X-ray reflection geometry in the super-Eddington regime should be morphologically different from the thin disk picture. 
Ref.\cite{Thomsen2019,Thomsen2022, Zhang2024} conducted theoretical studies on X-ray reflection and reverberation from super-Eddington accretion flows and demonstrated that the reflection geometry in super-Eddington accretion flows leads to unique observable features. They found that the Fe K$\alpha$ line energy spectra, emitted by the fast-moving wind, are broad and blueshifted with a symmetric profile, which is distinct from the skewed line profiles produced from thin disks. Also, the Fe line lag spectra from super-Eddington accretion flows have characteristics different from the thin-disk lag spectra. Not only can these features be used to distinguish super-Eddington accretion systems from sub-Eddington systems, but they are also key for constraining the reflection
geometry and extracting parameters from the observed lags. 
By applying the model to fit the observed lag-energy spectrum of Swift J1644+57, the first jetted TDE \cite{Kara2016b}, it is found that super-Eddington disk geometry is preferred over the thin disk geometry, and key parameters of black hole and corona can be constrained.
eXTP/SFA observations will allow further exploration of how X-ray reverberation signals can be used to probe super-Eddington accretion flow funnel geometry and wind kinematics, which are  linked to fundamental physical parameters like black hole spin and Eddington ratio. 

Given the large effective area of eXTP-SFA, we can perform X-ray reverberation mapping of both the continuum and Fe K line for the brightest nearby intermediate-mass black holes (IMBHs, 10$^{2}$-10$^{6}$ M$_{\odot}$), such as NGC 4395. IMBHs serve as a crucial population bridging stellar-mass black holes and supermassive black holes (SMBHs). Measuring X-ray time lags in IMBHs is essential for filling the gap between these two mass regimes and for establishing a global scaling relation between the continuum/emission-line lags and black hole mass \citep{Kara2016a}. These measurements also provide key insights into the corona-disk-jet connection across different accretion modes \citep{Wang2021}. Furthermore, recent IXPE observations \citep{Gianolli2023} suggest that the corona geometry extends along the same direction as the accretion disk, challenging the simple lamp-post model assumption. eXTP offers a unique opportunity to test this corona illumination and disk reprocessing scenario, providing new constraints on the physics of accretion and black hole environments.

\begin{figure}[H]
    \centering
    \includegraphics[width=\columnwidth]{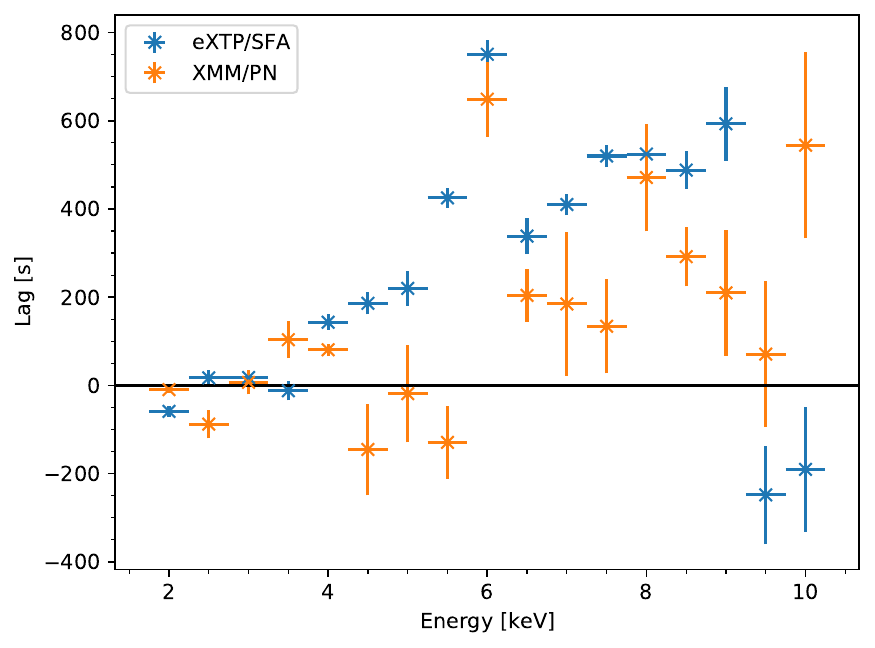}
    \caption{The time-lags at $1.14\times 10^{-4}~\rm Hz$ versus the 2--4 keV band as a function of the photon energy, for a warped disc with an inner edge-on and outer face-on components. The warped disc is around a $10^7~\rm M_\odot$ black hole. The results extracted from simulated eXTP/SFA data with an exposure of 35\,ks are shown in blue symbols. For comparison, we also plot in orange symbols the results extracted from simulated XMM-Newton/pn data, for which larger uncertainties are seen.}
    \label{fig:17}
\end{figure}

Fe K reverberation is an extremely useful tool for probing warped disks. While several mechanisms can lead to disk warping, one particularly interesting process is the Bardeen-Petterson effect \citep{bardeen_lense-thirring_1975}. In this scenario, the inner region of a misaligned disk aligns with the black hole’s rotation axis due to the combined effects of viscous torque and general relativistic Lense-Thirring precession. This effect may manifest in tidal disruption events, where the angular momentum of the disrupted star—and consequently the resulting debris disk—is random \cite{zhang_predictions_2015}. Zhang et al. 2019~(\cite{zhang_probing_2019}) demonstrated that a warped disk can produce distinct emission line reverberation mapping features. Such features can be detected with SFA, as evident in Fig.~\ref{fig:17}.

\subsubsection{AGN non-thermal superflare from an orbiting hotspot at the SMBH inner radii}

Recently, Very Large Telescope Interferometer (VLTI) observed several near-infrared superflares which might be from hot spots moving around Sgr A* at only 7--10 gravitational radii from the central black hole \cite{Gutierrez2020}. It is believed that these hot spots are likely due to relativistic magnetic reconnection events and they might be also the sources of X-ray flares which exhibit a double-peak structure in the in the light curve of the total X-ay flux due to their orbital motion. Here we simulate the X-ray light curve of these hot spots observed by eXTP-SFA.

\begin{figure*}[t]
    \centering
    \includegraphics[width=\textwidth]{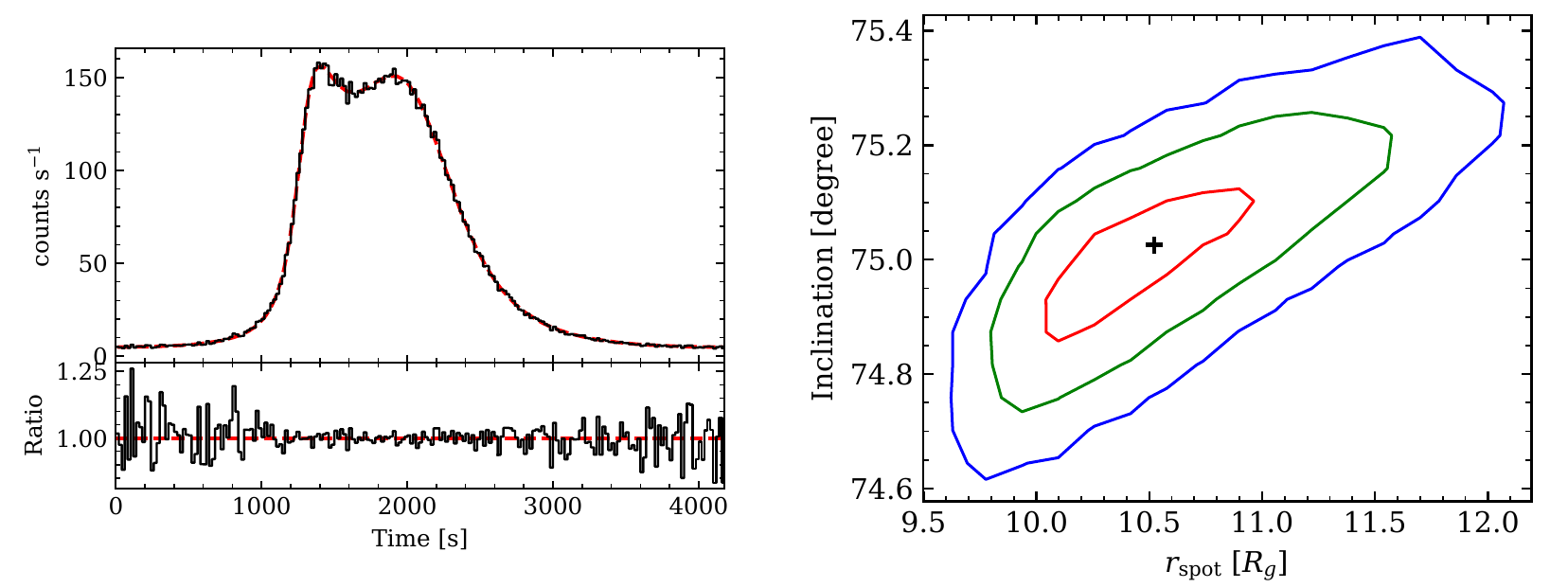}
    \caption{Left: The light curve (black) produced by a hot spot orbiting a black hole, as observed by eXTP/SFA in 2-10 keV energy band. The light curve are binned to 50 s. The black hole mass is assumed to $4.14\times 10^6\, M_\odot$ and black hole spin is 0.998.
    The orbital radius of the hot spot is $r_\mathrm{spot} = 10\, R_g$.
    Its radiation is assumed as a power law spectrum with the index of $\Gamma = 2$.
    The observing inclination is assumed to $75^\circ$.
    The red dashed line represents the count rate of the best fit model. The ratio between the data and the best fit model is shown in lower subpanel. Right: the error contours of hot spot orbital radius and observing inclination. The red, green and blue lines represent 1, 2, and 3 $\sigma$ confidence levels, respectively.}
    \label{fig:18}
\end{figure*}

Fig.~\ref{fig:18} (left) shows a simulation of the light curve produced by a hot spot orbiting a black hole for one period, as observed by eXTP-SFA in the 2--10 keV energy band.
We simulate a 4.2 ks observation with an average flux of $10^{-11}$ erg cm$^{-2}$ s$^{-1}$.
The black hole mass is assumed to $4.14\times 10^6\, M_\odot$ and black hole spin is 0.998.
The orbital radius of the hot spot is $r_\mathrm{spot} = 10\, R_g$.
Its radiation is assumed as a power law spectrum with index $\Gamma = 2$.
The observing inclination is assumed to be $75^\circ$.
There is an obvious double-peak structure in the light curve, with a time interval of 500\,s between the two peaks.
In the lower subpanel, we show the ratio between the data and the best fit model.
We obtain a good fit and recover the input parameters.

Fig.~\ref{fig:18} (right) shows the error contours (1, 2, and 3 $\sigma$ in red, green and blue, respectively) of hot spot orbital radius and observing inclination.
The orbital radius of the hot spot can be measured to a precision $<3\, R_g$ at $3\sigma$ confidence, while the inclination is measured to a precision of $<1^\circ$.
The technique also provides us with a way to measure the black hole spin, if the mass of the black hole is measured independently (e.g. via optical reverberation mapping \citep{2006NewAR..50..796P}).
This would apply to any hot spot radius and hence is independent of whether or not the disk is assumed to extend to the ISCO, as is the case for spin constraints from spectral fitting.
Thanks to its unprecedented throughput, eXTP will be able to perform such measurements for any AGN with X-ray flux above 1 mCrab.

\subsubsection{Quasi Periodic Oscillations from SMBHs}

AGN, powered by accretion on to SMBHs at the center of of galaxies, are thought to be scaled-up versions of BHXRBs \cite{2010LNP...794..203M, 2012A&A...544A..80G, 2015ApJ...812L..25M, 2015SciA....1E0686S}. A compelling line of evidence for this postulation is the striking similarity in the variability of the X-ray radiation between AGN and BHXRBs \cite[e.g.][]{2002MNRAS.332..231U, 2003ApJ...593...96M, 2003MNRAS.345.1271V, 2011MNRAS.413.2489V, 2006Natur.444..730M}. The power spectral densities (PSDs) of both can be well described as red-noise. However, the characteristic QPOs, which have been observed in the X-ray light curves of dozens of BHXRBs \cite[e.g.][]{2001ApJ...552L..49S, Remillard:2006fc}, are rarely detected in AGN. More detections of QPOs in SMBHs will strengthen the connection between AGN and BHXRBs. Also, the properties of QPOs in AGN can provide a powerful probe to diagnose the inner structure of the X-ray emitting region around SMBHs.

A number of models have been proposed to interpret the phenomena of QPOs, e.g., the relativistic precession model \cite{1999ApJ...524L..63S}, the resonance model \cite{2001A&A...374L..19A}, acoustic oscillation modes in pressure-supported accretion tori \cite{2003MNRAS.344L..37R}, instability at disk-magnetosphere interface \cite{2004ApJ...601..414L}, the accretion-ejection instability in magnetized disks \cite{1999A&A...349.1003T}, and the global corotational instability of non-axisymmetric g-mode or p-mode trapped in the innermost region of the accretion disk \cite{2003ApJ...593..980L}, etc. Nevertheless, the origin of QPOs still remains unclear. A sample of QPOs detected in SMBHs can help understand the mechanism of QPOs generated in accreting black hole systems.

So far, only a few possible QPOs have been found in accreting systems with SMBHs: the Narrow-Line Seyfert 1 (NLS1) galaxies RE\,J1034+396 \cite{2008Natur.455..369G}, 1H\,0707-495 \cite{2016ApJ...819L..19P}, ESO\,113-G010 \cite{2020AcASn..61....2Z}, Mrk\,766 \cite{2017ApJ...849....9Z}, MCG-06-30-15 \cite{2018A&A...616L...6G},  MS\,2254.9–3712 \cite{2015MNRAS.449..467A}, the ultrasoft AGN candidate 2XMM\,J123103.2+110648, as well as the tidal disruption events (TDE) Swift\,J164449.3+573451 \cite{2013ApJ...776L..10L}, etc. But the reliability of some cases are still under debate \cite[e.g.][]{2023ApJ...946...52Z}. Considering the the dependence of the timescale on the black hole mass \cite{Remillard:2006fc, Zhou:2014cma}, it is not hard to understand the lack of QPO detections in SMBHs. More QPOs in SMBHs will be found if more high-quality observations with long, continuous, uninterrupted exposure can be taken.

\begin{figure*}[t]
\centering
\includegraphics[width=\textwidth]{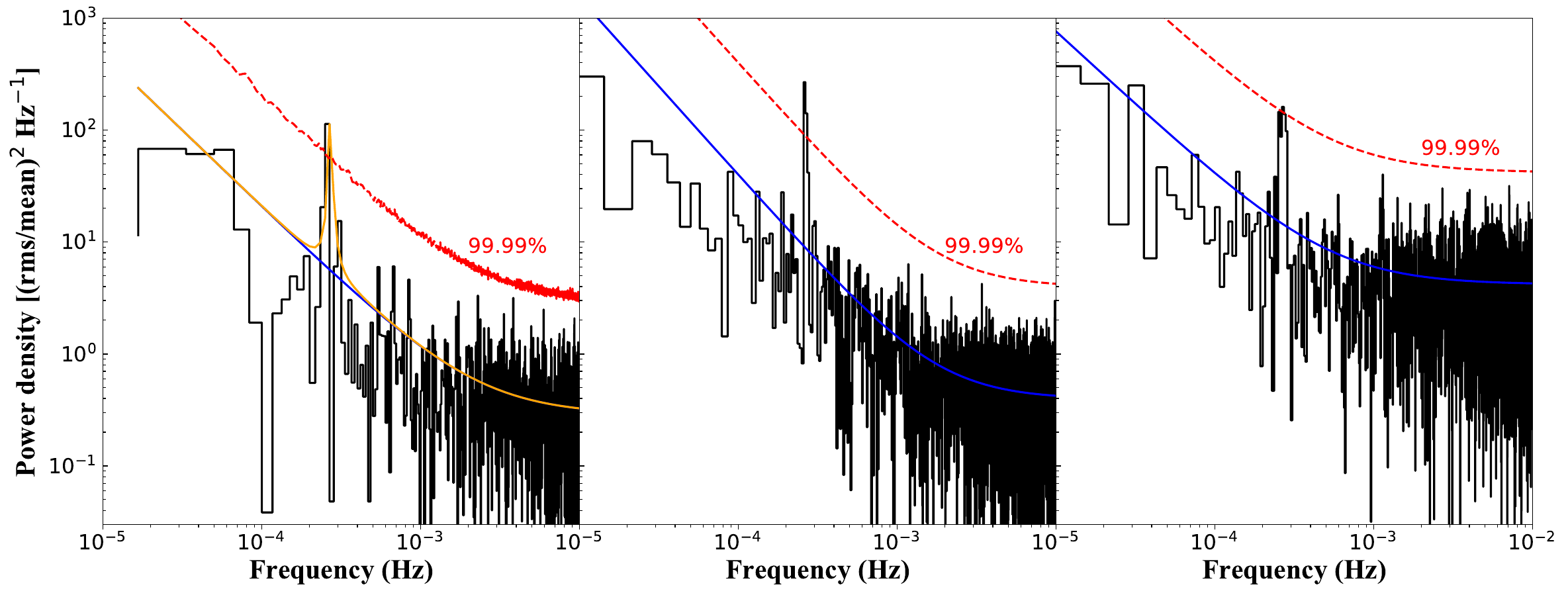}
\caption{
Left panel: the PSD of RE\,J1034+396 observed with {\it XMM-Newton}. The blue line represents the best-fit power law, the orange line represents the PSD model with an added Lorentzian profile, and the red dashed line represents the 99.99\% confidence level. Middle and right panels: the examples of simulated PSDs observed with SFA, with flux of $1\times10^{-11}\,\rm erg\,s^{-1}\,cm^{-2}$ and $2.5\times10^{-13}\,\rm erg\,s^{-1}\,cm^{-2}$, respectively. }  \label{fig:19}
\end{figure*}

Among all the QPOs detected in SMBHs, the most acceptable and significant case is the one in RE\,J1034+396. The continuum of the PSD can be fitted by a powerlaw model, and the QPO signal around $2.7\times10^{-4}$\,Hz can be well described by a Lorentzian profile, as shown in the left panel of Fig.~\ref{fig:19}. The 0.3-10\,keV count rate (PN+MOS1+MOS2) of RE J1034+396 observed with XMM-Newton is about 7.2 counts\,s$^{-1}$, which corresponds to a flux of about$1\times10^{-11}\,\rm erg\,s^{-1}\,cm^{-2}$, by assuming a typical power-law spectrum model with photon index of 2 and hydrogen column density of $3\times10^{20}\,\rm cm^{-2}$. Using the RE\,J1034+396 PSD model as input, and the count rate of about 20~counts\,s$^{-1}$, we simulate 1000 lightcurves of RE\,J1034+396 observed with SFA, with duration of 140\,ks. In over 80\% of the observations, significant QPO signals with confidence level higher than 99.99\% can be detected. The detection probability will decrease to about 50\% when the flux of source is about $2.5\times10^{-13}\,\rm erg\,s^{-1}\,cm^{-2}$. The capability of SFA can expand the sample of QPOs in SMBHs. Or the observations can help answer another question: whether the QPOs are really rarer in accreting systems with SMBHs than those with stellar black holes.

\subsubsection{The disk-corona structure of AGN from X-ray polarimetry}

IXPE has significantly advanced our understanding of the AGN central engine by providing the first X-ray polarization measurements of hot coronae in unobscured Seyferts \cite{Marin2024}. These observations have helped constrain theoretical models of X-ray emission in such extreme environments. Indeed, the X-ray polarization signal strongly depends on the morphology of the corona. Radially extended configurations, such as a slab sandwiching the disk or a wedge structure, produce polarization parallel to the symmetry axis, with a degree increasing with inclination \cite{ursini_prospects_2022}. Vertically extended coronae instead produce polarization perpendicular to the symmetry axis, generally with a lower degree. IXPE results on MCG-5-23-16 \cite{Marinucci2022,Tagliacozzo2023}, IC~4329A \cite{Ingram2023}, and NGC~4151 \cite{Gianolli2023, Gianolli2024} are consistent with the X-ray corona lying along the equatorial plane; however, only one solid detection ($>3 \sigma$) is obtained in the case of NGC~4151. 

\begin{figure}[H]
    \centering
    \includegraphics[width=\columnwidth]{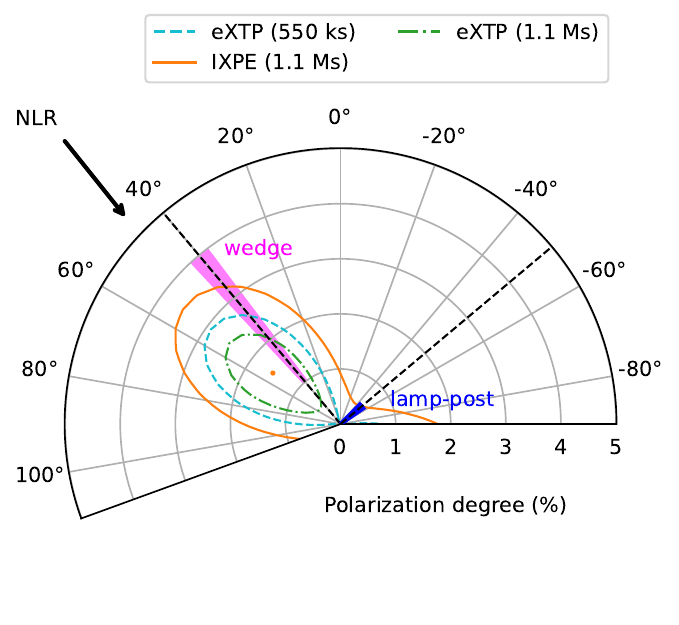}
    \vspace{-1cm}
    \caption{Polarization degree and angle contour plots, at the 99\% confidence level, of MCG-5-23-16 from the IXPE observation \cite{Tagliacozzo2023} (solid orange line) and eXTP simulations with 550 ks (cyan dashed line) and 1.1 Ms (green dash-dotted line). The black dashed line at 40° represents the elongation of the narrow-line region (NLR) and indicates the symmetry axis of the disk, while the black dashed line at -50 deg represents the perpendicular direction.}
    \label{fig:20}
\end{figure}

The case of MCG-5-23-16 well illustrates the issue (see Fig.~\ref{fig:20}). The IXPE upper limit, obtained with a total exposure of 1.1 Ms, rules out certain configurations (such as a conical outflowing corona \cite{Tagliacozzo2023}) but remains consistent with a spherical lamp-post model (see the orange contour in the figure). Thanks to the larger PFA effective area (up to five times greater than IXPE), eXTP will provide a measurement at the 99\% confidence level with an exposure of 550 ks, and a $4 \sigma$ detection with 1.1 Ms (see the cyan and green contours in the figure). This also means that eXTP will be able to investigate the corona-disc interplay with high signal-to-noise ratio for a significantly larger number of sources. Assuming that a total exposure of about 10 Ms is dedicated to non-jetted, unobscured Seyferts with fluxes exceeding 2~mCrab in the eXTP band, and extrapolating the IXPE results, we expect at least 10–12 new measurements. These observations will provide a robust sample for investigating the corona in supermassive black holes, which, together with stellar-mass black holes in X-ray binaries, will offer fundamental insights into the accretion properties in extreme regimes.

\subsubsection{Probing strong gravity effects with QPEs and TDEs}

X-ray quasi-periodic eruptions (QPEs) are intensive, recurring soft X-ray bursts from galaxy nuclei. The first tentative QPE source RX J1301.9+2747 
was reported  in 2013 \cite{Sun2013}, and the first confirmed QPE source was observed in 2019 in GSN 069 with nine-hour cycles of intense flares \cite{Miniutti2019}. This discovery was then followed by more detections in RX J1301.9+2747 (20-hour cycles), eRO-QPE1, eRO-QPE2, and AT2019qiz (48-hour cycles post-TDE) \cite{Giustini2020, Arcodia2021, 2024Natur.634..804N}. QPEs are preferentially found in dwarf galaxies which host SMBHs of $10^{5-7} M_{\odot}$.
The sample, though small, shows diverse recurrence patterns, from regular to erratic \cite{Quintin2023}.

Observationally, QPEs are defined by their dramatic X-ray variability. The eruptions are soft, with spectra peaking below 2 keV, often modeled as thermal emission from a disk with temperatures around 50 eV during quiescent phases \cite{Miniutti2019}. Burst durations range from tens of minutes to a few hours, with peak times and durations shortening at higher energies, indicating complex energy-dependent evolution \cite{Arcodia2022}. Recurrence times vary widely, from GSN 069’s nine hours to AT2019qiz’s two days, and some sources like eRO-QPE1 display irregular intervals, challenging simple periodicity models. Notably, some QPE hosts lack optical signatures of active galactic nuclei (AGN), suggesting these events can occur without pre-existing accretion activity \cite{Arcodia2021}.

\begin{figure*}
    \centering
    \includegraphics[width=\textwidth]{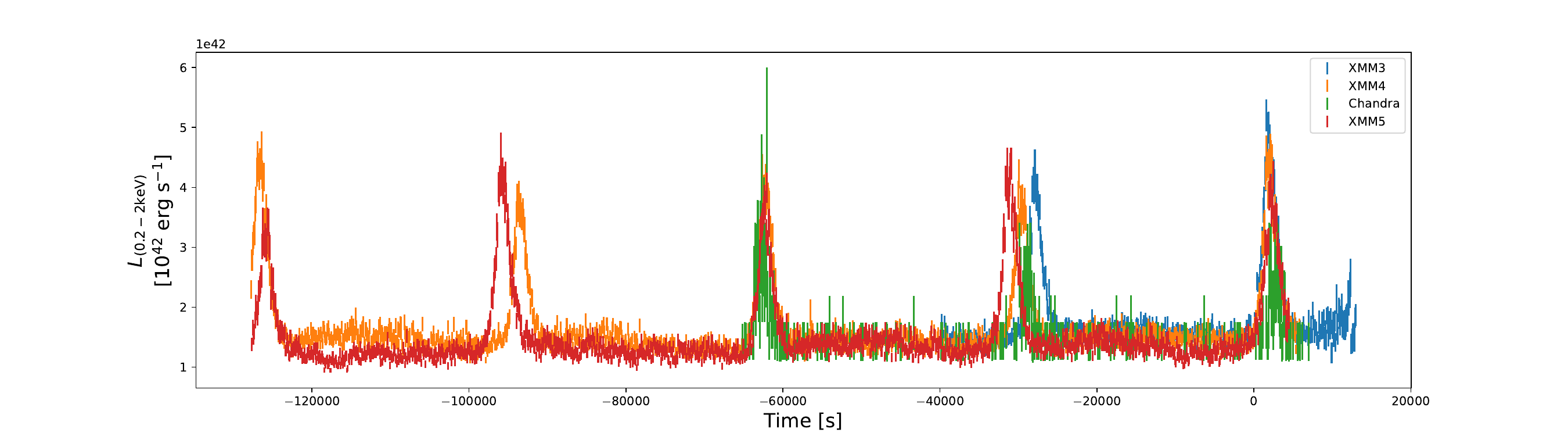}
    \caption{Light curves from 4 observations of GSN 069 QPEs  during Dec. 2018-May 2019 \cite{Miniutti2019}, 
    where the alternating long-short pattern in the recurrence times, and the clear variations in both $P_{\rm long}$ and $P_{\rm short}$ along with the approximately constant $P_{\rm long}+P_{\rm short}$ are exhibited \cite{Zhou2024a,Zhou2025b}.}
    \label{fig:21}
\end{figure*}

Among the diverse phenomena found in QPE observations, two illuminating discoveries reported in 2024 are particularly informative for understanding the origin of QPEs. 
The first discovery is the direct detection of QPEs in light curves of two recent TDEs AT~2019qiz and AT~2019vcb (\cite{2024Natur.634..804N} \cite{Bykov2025}),
confirming that at least some of QPEs are connected to TDEs. The second discovery is about the alternating long-short pattern $P_{\rm long}\ \& P_{\rm short}$ in the recurrence times of QPEs, which have been noticed in several QPE sources sources including the most famous GSN 069 
\cite{Miniutti2019, Giustini2020, Arcodia2021, Arcodia2022}. In a recent work, \citet{Zhou2024a} further pointed out the recurrence times of  GSN 069, both $P_{\rm long}$ and $P_{\rm short}$, show clear variations, while the sum of two consecutive recurrence times $P_{\rm long}+P_{\rm short}$ remains nearly a constant (see Fig.~\ref{fig:21}). This second discovery indicates 
that the fundamental period of QPEs is actually $P_{\rm long}+P_{\rm short}$ instead of either $P_{\rm long}$ or $P_{\rm short}$, and two bursts are produced per period.

Both the TDE-QPE connection and the pattern of constant $P_{\rm long}+P_{\rm short}$ favor the EMRI (extreme mass ratio inspiral)+disk model, 
in which a stellar object orbiting around the SMBH, colliding with the accretion disk to trigger quasi-periodic  bursts \cite{Linial2023,Franchini2023,Tagawa2023}. Though alternative models, including disk instabilities and  repeating partial TDEs have also been proposed as 
possible origins of QPEs, they are incompatible with either the TDE-QPE connection or the constant $P_{\rm long}+P_{\rm short}$ pattern in the recurrence times. Additionally, stellar transits across a magnetized accretion flow that quasi-periodically ejects plasmoids has also been put forward as a promising mechanism \cite{2021ApJ...917...43S}.

In the framework of EMRI+disk model, the variations of $P_{\rm long}$ and $P_{\rm short}$ are the natural consequences of EMRI apsidal precession
in the curved spacetime.
Therefore, two intrinsic periods of the EMRI trajectory are encoded in the QPE timing, the EMRI orbital period $P_{\rm obt}\approx (P_{\rm long}+P_{\rm short})$
and the EMRI apsidal precession period $P_{\rm prec}$ in which the recurrence times are modulated.
Several previous works have attempted to fit the QPE timing data with EMRI geodesics  
(\cite{Xian2021,Franchini2023, Zhou2024a, Zhou2024b}) or EMRI trajectories that are perturbed by the accretion disk \cite{Zhou2025a}.
Taking GSN 069 as an example, \citet{Zhou2025a} found an orbital period $P_{\rm obt}\approx 65$ ks, an apsidal precession period $P_{\rm prec}\approx 76$ days, and an orbital eccentricity $e\approx 0.04$. The two periods are related to the EMRI orbital parameters as 
\begin{equation}
    P_{\rm obt}= 2\pi\left( \frac{A}{M_\bullet}\right)^{3/2}M_\bullet \ , \quad \frac{P_{\rm prec}}{P_{\rm obt}} = \frac{p}{3M_\bullet} = \frac{A}{3M_\bullet}(1-e^2)\ ,
\end{equation}
where geometric units with $G=c=1$ are used, $M_\bullet$ is the gravitational radius of the SMBH, $A$ is the semi-major axis and  $p$ is the semi-latus rectum.
From the two periods inferred from QPE timing data, 
it is straightforward to obtain the value of the SMBH mass $M_\bullet$.
This QPE timing method is promising in accurate measurement of SMBH masses \cite{Zhou2025a,Zhou2025b}. 


Polarization studies for QPEs are under-explored, but theoretical models offer some predictions. In the stellar interaction model, if eruptions arise from localized disk collisions, the X-ray emission could be polarized due to scattering in an asymmetric geometry, such as a hot spot or outflow. The degree of polarization would depend on the viewing angle and disk structure, potentially reaching a few percent in soft X-rays, as seen in some accretion disk simulations \cite{schnittman_x-ray_2009}. For disk instability models, polarization might be lower or absent unless the instability triggers anisotropic outflows or a corona, which could scatter photons and induce polarization \cite{Marinucci2022}. No direct polarization measurements exist yet for QPEs, as current X-ray polarimeters like IXPE lack the sensitivity for such short and faint events.



A tidal disruption event (TDE) occurs when a star ventures within the tidal radius of a supermassive black hole \cite{Hills1975,Rees1988}. These events generate flares lasting months to years, exhibiting extreme variability across multi-wavelengths \cite{Gezari2021} and revealing previously dormant SMBHs. For approximately a year after the disruption, the accretion rate remains high, resulting in a geometrically thick nascent accretion disk. Due to the typical misalignment between the disrupted star’s orbit and the black hole spin axis, strong viscous forces and frame-dragging effects cause the entire disk to undergo rigid-body Lense-Thirring precession around the black hole spin axis (e.g., \cite{Papaloizou1995,Fragile2007}).
For an accretion disk extending from the ISCO to approximately the tidal disruption radius, the precession period varies from one to tens of days, depending on the SMBH mass and spin parameter (\cite{Stone2012,Franchini2016}). In the case of intermediate-mass black holes, the precession period can be significantly shorter, e.g. 85~s in 3XMM~J215022.4--055108 \cite{Zhang2025}.

Unlike in AGNs and BHXRBs, coronae are typically absent in TDEs, at least during the early stages post-disruption, making disk reflection undetectable in most cases. As a result, periodic X-ray variations provide the most viable method for measuring the black hole spin parameter. This approach has been applied to TDEs such as Swift J1644 (\cite{Stone2012,Lei2013} and AT2020ocn \cite{2024SciA...10J8898P}, which exhibited periodicities of 2.7 days and 15 days, respectively, attributed to jet and disk precession. 
GRMHD simulations predict that both alignment and precession should gradually slow down over time due to angular momentum transport and radial disk expansion \cite{Liska2018}. However, these predictions remain unverified, particularly at late evolutionary stages of TDEs, due to the limited statistical constraints from current instruments. 
The large effective area of eXTP/SFA in the soft X-ray band makes it an ideal instrument to thoroughly investigate the evolution of X-ray periodicity and its associated spectral changes. More precise measurements on these quantities could further test the Bardeen--Petterson effect \cite{bardeen_lense-thirring_1975}, which dampens LT precession, thereby providing key constraints on disk dynamics as it transitions from geometrically thick to geometrically thin.

\textit{Extremely relativistic TDEs}

When the stellar orbital periastron $R_p$ is very close to the black hole, i.e., $<~ 6\, R_g$, the strong GR effect causes the orbital dynamics of the stellar debris dramatically different from that in ordinary TDEs, thus rendering the term extremely relativistic TDEs (eTDEs). Such regime dominates those TDEs with black hole mass in the range of $10^7 - 10^8~ M_\odot$. Particularly, the strong apsidal precession is so strong that the debris will wind a few times at $R_p$ around the black hole, before its path extends out to large distances ~\cite{Ryu2023}. Internal shocks occur within the wound debris stream, dissipating energy, and followed by the accretion of the compact, circularized flow, as was shown by simulation in \cite{Ryu2023}. 


Consequently, eTDEs are predicted to have an X-ray light curve that rises rapidly (within hours) to about the Eddington level, and maintains at $L_{\rm Edd}$ for weeks to a year, during which fast and significant variabilities are expected. eXTP has the potential to detect such a new type of transient.

\section{Summary}\label{sec:Summary}

The study of accreting black holes provides insights into some of the most fundamental objects and processes that impact our Universe, while presenting great theoretical and observational challenges. Black holes themselves are a manifestation of gravity at its most extreme and provide a setting which allows us to explore our best theory of gravity - GR - and test the spacetime metric. The physics of accretion and the outflows it produces lead us to couple our understanding of electromagnetism and plasma physics, thermodynamics and radiative processes to this deeper theory. This synthesis allows us to use electromagnetic observations as probes of the behaviour of matter accreting in strong gravity, and gravity itself, but is incredibly demanding in terms of the quality of data required. 

In this work we have shown how eXTP can revolutionize accreting black hole studies by providing, with its large collecting area and unique combination of instruments, order-of-magnitude advances in sensivity to polarization, spectral and timing signatures which are out-of-reach to current X-ray telescopes. To summarize the key advances:
\begin{enumerate}
    \item The collecting area and spectral resolution of SFA enable the direct estimation of black hole masses via X-ray reverberation mapping (Section~\ref{sec:mass}). The high orbit of eXTP extends this technique to supermassive black holes by sampling their variability time-scales with long continuous exposures. These characteristics allow more than an order of magnitude improvement in precision of the mass estimate, compared to current X-ray instruments. 
    \item Black hole spins (Section~\ref{ss-spin}) of stellar mass black holes can be estimated from a variety of techniques that estimate the inner disk radius and hence the spin, in soft spectral states where the disk can be assumed to extend to the ISCO. The eXTP-PFA can constrain spin to the per-cent level due to the spin effect on both polarization angle and degree caused by light-bending and returning radiation. Thermal continuum fitting of the disk spectrum in SFA data provides a spin estimate  using a technique which is completely independent of the polarization approach, but using the same eXTP observations.
    \item Quasi-periodic oscillations (QPOs, Section~\ref{sec:qpos}) could provide a totally novel probe of black hole mass and spin if the physical origin of the QPOs could be confirmed. Both the SFA and PFA can enable this with independent approaches. The SFA will carry out phase-resolved spectroscopy of QPOs to determine for the first time whether the QPOs are produced by coronal precession and the resulting disk-illumination pattern. The PFA combined with SFA will measure polarization changes and delays on the QPO cycle to measure the changing coronal geometry. With this powerful combination eXTP will determine the nature of QPOs and hence enable their use as a spin estimator.
    \item The spectral sensitivity of SFA provides an additional constraint on spin from reflection spectroscopy, while also enabling a probe of deviations from the Kerr metric (Section~\ref{sec:testing_gr}). Such deviations in the fundamental shape of spacetime could also be probed using other techniques listed here, once the appropriate model predictions have been developed.
    \item eXTP-SFA and PFA will enable powerful constraints on the coronal and inner disk geometry in the hard and intermediate spectral states of accreting stellar mass black holes (Section~\ref{sec:geom_xrbs}), to shed light on the nature of these states and the origin of the relativistic jets they produce. Disk-corona geometry can be determined via independent combinations of reverberation time-delays from soft thermal emission and the X-ray reflection signature which are both sampled by SFA. Measurements by PFA can determine the coronal geometry and outflow velocity even in the early stages of an outburst.
    \item The techniques to constrain the disk-corona geometry in stellar mass systems could also be extended to accreting supermassive black holes (Section~\ref{sec:geom-smbh}) thanks to the large collecting area of eXTP-SFA and PFA combined with the long uninterrupted exposures permitted by the high spacecraft orbit. These measurements will allow a detailed test of whether and how the inner geometry depends on black hole mass and whether different classes of AGN show the distinct accretion-state geometries that are suggested for stellar mass systems.
    \item The combination of collecting area, sensitivity, spectral-response and high orbit of eXTP-SFA make it ideal for the detection and study of quasi-periodic signals from supermassive black holes which can result from a variety of physical phenomena (Section~\ref{sec:geom-smbh}). Dynamical-time-scale QPOs can be easily detected from even faint AGN. Strong constraints on orbital dynamics of hotspots orbiting the Sgr A$^{*}$ supermassive black hole in our own galaxy can be obtained by their X-ray flare profiles. The soft emission from accretion disks produced by tidal disruption of stars can be probed for precession signals and the QPEs now associated with repeated disruption of the disk by the stellar orbits.    
\end{enumerate}
It is almost certain that for some measurements, the true physical situation in accreting black holes will present greater challenges than anticipated in our simulations shown here. However the diversity of techniques enabled by eXTP, including spectroscopy, spectral-timing, polarimetry and ultimately polarimetric-spectral-timing, will provide many independent checks that can validate the methods and models used. Critically, measurements of black hole spin and mass can be obtained and cross-checked for the same stellar-mass systems throughout an X-ray binary outburst and in different states. For most supermassive black holes in AGN we likely cannot cross-check results in this way for a given system, due to the long evolutionary time-scales expected. However, rapidly-evolving `changing-look' AGN or transient events driven by tidal stellar disruption provide an opportunity to use multiple techniques as the system evolves, and an additional setting for comparison with the more numerous and predictable AGN.

The richness and diversity of topics described in this work reflect the rapid growth of the field over the last two decades. This growth was largely driven by the breakthroughs in capability of the current generation of X-ray observatories, combined with advances in theory and computation. eXTP represents the next step-change in capability, which should increase the pace of discovery and will inevitably lead to surprises and further puzzles.  Fortunately, the scientific potential of the mission to solve these unanticipated puzzles is clear, thanks to eXTP's unique and complementary combination of instrumentation, large collecting area, powerful dynamic range in source flux and capability for long uninterrupted observations. In combination with other new missions and advances in multiwavelength and multimessenger time-domain astronomy, we can look forward to an exciting future for black hole research in the 2030s.


\emph{Acknowledgements.} 

This work is supported by China's Space Origins Exploration Program. S-NZ is supported by the National Natural Science Foundation of China (No. 12333007), the International Partnership Program of Chinese Academy of Sciences (No.113111KYSB20190020) and the Strategic Priority Research Program of the Chinese Academy of Sciences (No. XDA15020100). AV acknowledges support from the Academy of Finland grant 355672. Nordita is supported in part by NordForsk.

Conflict of Interest

The authors declare that they have no conflict of interest.





\begin{thebibliography}{99}

\bibitem{Schwarzschild:1916uq}
K.~Schwarzschild, 
Sitzungsber. Preuss. Akad. Wiss. Berlin (Math. Phys. ) \textbf{1916}, 189-196 (1916)
[arXiv:physics/9905030 [physics]. doi:10.48550/arXiv.physics/9905030


\bibitem{Finkelstein:1958zz}
D.~Finkelstein. \pr, \textbf{110}, 965-967 (1958)
doi:10.1103/PhysRev.110.965

\bibitem{cyg1}
C.~T.~Bolton. \nat. {\bf 235}, 271 (1972). doi:10.1038/235271b0

\bibitem{cyg2}
B.~L.~Webster and P.~Murdin. \nat, {\bf 235}, 37 (1972). doi:10.1038/235037a0

\bibitem{Bambi:2024hhi}
C.~Bambi. arXiv:2408.12262 [astro-ph.HE]. doi:10.48550/arXiv.2408.12262

\bibitem{corona1}
G.~Matt, G.~C.~Perola and L.~Piro, 
\aap. {\bf 247}, 25 (1991). 

\bibitem{corona2}
A.~Martocchia and G.~Matt. 
\mnras, {\bf 282}, L53 (1996). doi:10.1093/mnras/282.4.L53

\bibitem{Narayan:1994xi}
R.~Narayan and I.~s.~Yi,
Astrophys. J. Lett. \textbf{428}, L13 (1994)
doi:10.1086/187381
[arXiv:astro-ph/9403052 [astro-ph]].

\bibitem{Shakura:1972te}
N.~I.~Shakura and R.~A.~Sunyaev,
Astron. Astrophys. \textbf{24}, 337-355 (1973). 

\bibitem{Page:1974he}
D.~N.~Page and K.~S.~Thorne,
Astrophys. J. \textbf{191}, 499-506 (1974)
doi:10.1086/152990

\bibitem{Bisnovatyi-Kogan:1977xxx}
G.~S.~Bisnovatyi-Kogan and S.~I.~Blinnikov,
Astron. Astrophys. \textbf{59}, 111-125 (1977).

\bibitem{Haardt:1991tp}
F.~Haardt and L.~a.~M.~U.~Maraschi,
Astrophys. J. Lett. \textbf{380}, L51-L54 (1991)
doi:10.1086/186171

\bibitem{Titarchuk:1994rz}
{L.~Titarchuk,
Astrophys. J. \textbf{434}, 570-586 (1994)
doi:10.1086/174760}

\bibitem{Dove:1997ei}
J.~B.~Dove, J.~Wilms, M.~Maisack and M.~C.~Begelman,
Astrophys. J. \textbf{487}, 759 (1997)
doi:10.1086/304647
[arXiv:astro-ph/9705130 [astro-ph]].

\bibitem{Liu:2003yg}
B.~F.~Liu, S.~Mineshige and K.~Ohsuga,
Astrophys. J. \textbf{587}, 571-579 (2003)
doi:10.1086/368282
[arXiv:astro-ph/0301142 [astro-ph]].

\bibitem{Markoff:2005ht}
S.~Markoff, M.~A.~Nowak and J.~Wilms,
Astrophys. J. \textbf{635}, 1203-1216 (2005)
doi:10.1086/497628
[arXiv:astro-ph/0509028 [astro-ph]].

\bibitem{Sironi:2019sxv}
L.~Sironi and A.~M.~Beloborodov,
Astrophys. J. \textbf{899}, 52 (2020)
doi:10.3847/1538-4357/aba622
[arXiv:1908.08138 [astro-ph.HE]].

\bibitem{Ross:2005dm}
R.~R.~Ross and A.~C.~Fabian,
Mon. Not. Roy. Astron. Soc. \textbf{358}, 211-216 (2005)
doi:10.1111/j.1365-2966.2005.08797.x
[arXiv:astro-ph/0501116 [astro-ph]].

\bibitem{Garcia:2010iz}
J.~Garcia and T.~Kallman,
Astrophys. J. \textbf{718}, 695 (2010)
doi:10.1088/0004-637X/718/2/695
[arXiv:1006.0485 [astro-ph.HE]].

\bibitem{Fabian:1989ej}
A.~C.~Fabian, M.~J.~Rees, L.~Stella and N.~E.~White,
\mnras. \textbf{238}, 729-736 (1989)
doi:10.1093/mnras/238.3.729

\bibitem{Laor:1991nc}
A.~Laor,
Astrophys. J. \textbf{376}, 90 (1991)
doi:10.1086/170257

\bibitem{Tanaka:1995en}
Y.~Tanaka, K.~Nandra, A.~C.~Fabian, \textit{et al.}
Nature \textbf{375}, 659 (1995)
doi:10.1038/375659a0

\bibitem{Nandra:2007rp}
K.~Nandra, P.~M.~O'Neill, I.~M.~George and J.~N.~Reeves,
Mon. Not. Roy. Astron. Soc. \textbf{382}, 194 (2007)
doi:10.1111/j.1365-2966.2007.12331.x
[arXiv:0708.1305 [astro-ph]].

\bibitem{Ingram:2019mna}
A.~Ingram and S.~Motta,
New Astron. Rev. \textbf{85}, 101524 (2019)
doi:10.1016/j.newar.2020.101524
[arXiv:2001.08758 [astro-ph.HE]].

\bibitem{Pasham:2014ybe}
D.~R.~Pasham, T.~E.~Strohmayer and R.~F.~Mushotzky,
Nature \textbf{513}, 74 (2014)
doi:10.1038/nature13710
[arXiv:1501.03180 [astro-ph.HE]].

\bibitem{Gierlinski:2008yz}
M.~Gierlinski, M.~Middleton, M.~Ward and C.~Done,
Nature {\bf 455}, 369 (2008)
[arXiv:0807.1899 [astro-ph]].

\bibitem{Zhou:2014cma}
X.~L.~Zhou, W.~Yuan, H.~W.~Pan and Z.~Liu,
Astrophys. J. Lett. \textbf{798}, L5 (2015)
doi:10.1088/2041-8205/798/1/L5
[arXiv:1411.7731 [astro-ph.HE]].

\bibitem{Mathur:2024ify}
S.~D.~Mathur and M.~Mehta,
arXiv:2412.09495 [hep-th].

\bibitem{WP-WG1} A. Li, A. L. Watts, G. Zhang et al. 2025, Science China Physics, Mechanics, and Astronomy, Dense Matter in Neutron Stars with eXTP, this issue

\bibitem{WP-WG3} M.~Y. Ge, L. Ji, R. Taverna, et al. 2025, Science China Physics, Mechanics, and Astronomy, Physics of Strong Magnetism with eXTP, this issue

\bibitem{WP-WG4} S.~X. Yi, W. Zhao, R.~X. Xu et al. 2025, Science China Physics, Mechanics, and Astronomy, Prospects for Time-Domain and Multi-Messenger Science with eXTP, this issue

\bibitem{WP-WG5} P. Zhou, J.~R. Mao, L. Zhang et al. 2025, Science China Physics, Mechanics, and Astronomy, Observatory Science with eXTP, this issue

\bibitem{SFA-PFA} S.~N. Zhang, A. Santangelo, Y.~P. Xu, et al. Science China Physics, Mechanics, and Astronomy, The enhanced X-ray Timing and Polarimetry mission - eXTP for launch in 2030, this issue

\bibitem{2001Natur.411..662C} E. Costa, P. Soffitta, R. Bellazzini et al. Nature, \textbf{411}, 662–665 (2001). doi:10.1038/35079508

\bibitem{2003NIMPA.510..176B} B. Mikulec, M. Campbell, E. Heijne, X. Llopart, L. Tlustos. Nuclear Instruments and Methods in Physics Research Section A: Accelerators, Spectrometers, Detectors and Associated Equipment, \textbf{511}, 282-286(2003), ISSN 0168-9002, doi:10.1016/S0168-9002(03)01807-2.

\bibitem{2007NIMPA.579..853B} R. Bellazzini, G. Spandre, M. Minuti, L. Baldini, A. Brez, L. Latronico, N. Omodei, M. Razzano, M.M. Massai, M. Pesce-Rollins, et al. Nuclear Instruments and Methods in Physics Research Section A: Accelerators, Spectrometers, Detectors and Associated Equipment, \textbf{579}, 853-858 (2007), ISSN 0168-9002, doi:10.1016/j.nima.2007.05.304.

\bibitem{2013NIMPA.720..173B} R. Bellazzini, A. Brez, E. Costa, M. Minuti, F. Muleri, M. Pinchera, A. Rubini, P. Soffitta, G. Spandre, Nuclear Instruments and Methods in Physics Research Section A: Accelerators, Spectrometers, Detectors and Associated Equipment, \textbf{720}, 173-177 (2013), ISSN 0168-9002, doi:10.1016/j.nima.2012.12.006.

\bibitem{Zhang_2019} S. Zhang, A. Santangelo, M. Feroci et al. Sci. China Phys. Mech. Astron. \textbf{62}, 29502 (2019). doi:10.1007/s11433-018-9309-2

\bibitem{maxi09}
M. Matsuoka, {\it et al.},
PASJ, \textbf{61}, 999 (2009)
doi:10.1093/pasj/61.5.999

\bibitem{liu22}
H. Liu, Y., Fu, {\it et al.},
Astrophys. J. \textbf{933}, 122 (2022)
doi:10.3847/1538-4357/ac74b1

\bibitem{janiuk00}
A. Janiuk, {\it et al.},
Astrophys. J. L., \textbf{542}, L33 (2000)
doi:10.1086/312911

\bibitem{Carter:1971zc}
B.~Carter,
Phys. Rev. Lett. \textbf{26}, 331-333 (1971)
doi:10.1103/PhysRevLett.26.331

\bibitem{Robinson:1975bv}
D.~C.~Robinson,
Phys. Rev. Lett. \textbf{34}, 905-906 (1975)
doi:10.1103/PhysRevLett.34.905

\bibitem{Chrusciel:2012jk}
P.~T.~Chrusciel, J.~Lopes Costa and M.~Heusler,
Living Rev. Rel. \textbf{15}, 7 (2012)
doi:10.12942/lrr-2012-7
[arXiv:1205.6112 [gr-qc]].

\bibitem{Kerr:1963ud}
R.~P.~Kerr,
Phys. Rev. Lett. \textbf{11}, 237-238 (1963)
doi:10.1103/PhysRevLett.11.237

\bibitem{Remillard:2006fc}
R.~A.~Remillard and J.~E.~McClintock,
Ann. Rev. Astron. Astrophys. \textbf{44}, 49-92 (2006)
doi:10.1146/annurev.astro.44.051905.092532
[arXiv:astro-ph/0606352 [astro-ph]].

\bibitem{Kormendy:1995er}
J.~Kormendy and D.~Richstone,
Ann. Rev. Astron. Astrophys. \textbf{33}, 581 (1995)
doi:10.1146/annurev.aa.33.090195.003053

\bibitem{Miller:2003sc}
M.~C.~Miller and E.~J.~M.~Colbert,
Int. J. Mod. Phys. D \textbf{13}, 1-64 (2004)
doi:10.1142/S0218271804004426
[arXiv:astro-ph/0308402 [astro-ph]].

\bibitem{Penrose:1969pc}
R.~Penrose,
Riv. Nuovo Cim. \textbf{1}, 252-276 (1969)
doi:10.1023/A:1016578408204

\bibitem{Reynolds:2020jwt}
C.~S.~Reynolds,
Ann. Rev. Astron. Astrophys. \textbf{59}, 117-154 (2021)
doi:10.1146/annurev-astro-112420-035022
[arXiv:2011.08948 [astro-ph.HE]].

\bibitem{Woosley:2006fn}
S.~E.~Woosley and J.~S.~Bloom,
{\it The Supernova Gamma-Ray Burst Connection},
Ann. Rev. Astron. Astrophys. \textbf{44}, 507-556 (2006),
doi:10.1146/annurev.astro.43.072103.150558
[arXiv:astro-ph/0609142 [astro-ph]].

\bibitem{Yoon:2006fr}
S.~C.~Yoon, N.~Langer and C.~Norman,
{\it Single star progenitors of long gamma-ray bursts. 1. Model grids and redshift dependent GRB rate},
Astron. Astrophys. \textbf{460}, 199 (2006). doi:10.1051/0004-6361:20065912
[arXiv:astro-ph/0606637 [astro-ph]].

\bibitem{Fuller:2019sxi}
J.~Fuller and L.~Ma,
Astrophys. J. Lett. \textbf{881}, no.1, L1 (2019)
doi:10.3847/2041-8213/ab339b
[arXiv:1907.03714 [astro-ph.SR]].

\bibitem{King:1999aq}
A.~R.~King and U.~Kolb,
Mon. Not. Roy. Astron. Soc. \textbf{305}, 654 (1999)
doi:10.1046/j.1365-8711.1999.02482.x
[arXiv:astro-ph/9901296 [astro-ph]].

\bibitem{Valsecchi:2010cw}
F.~Valsecchi, E.~Glebbeek, W.~M.~Farr, T.~Fragos, B.~Willems, J.~A.~Orosz, J.~Liu and V.~Kalogera,
. Nature \textbf{468}, 77 (2010). doi:10.1038/nature09463
[arXiv:1010.4809 [astro-ph.SR]].

\bibitem{Wong:2011eg}
T.~W.~Wong, F.~Valsecchi, T.~Fragos and V.~Kalogera. \apj, \textbf{747}, 111 (2012)
doi:10.1088/0004-637X/747/2/111
[arXiv:1107.5585 [astro-ph.HE]].

\bibitem{Podsiadlowski:2002ww}
P.~Podsiadlowski, S.~Rappaport and Z.~Han,
Mon. Not. Roy. Astron. Soc. \textbf{341}, 385 (2003)
doi:10.1046/j.1365-8711.2003.06464.x
[arXiv:astro-ph/0207153 [astro-ph]].

\bibitem{Fragos:2014cva}
T.~Fragos and J.~E.~McClintock. \apj, \textbf{800}, 17 (2015). doi:10.1088/0004-637X/800/1/17
[arXiv:1408.2661 [astro-ph.HE]].

\bibitem{Qin:2018sxk}
Y.~Qin, P.~Marchant, T.~Fragos, G.~Meynet and V.~Kalogera. \apjl, \textbf{870}, L18 (2019). doi:10.3847/2041-8213/aaf97b
[arXiv:1810.13016 [astro-ph.SR]].

\bibitem{Zdziarski:2025ozs}
A.~A.~Zdziarski, G.~Marcel, A.~Veledina, A.~Olejak and D.~Lancova,
[arXiv:2506.00623 [astro-ph.HE]].

\bibitem{Barausse:2012fy}
E.~Barausse,
\mnras. \textbf{423}, 2533-2557 (2012)
doi:10.1111/j.1365-2966.2012.21057.x
[arXiv:1201.5888 [astro-ph.CO]].

\bibitem{Sesana:2014bea}
A.~Sesana, E.~Barausse, M.~Dotti and E.~M.~Rossi,
\apj, \textbf{794}, 104 (2014)
doi:10.1088/0004-637X/794/2/104
[arXiv:1402.7088 [astro-ph.CO]].

\bibitem{Bardeen:1972fi}
J.~M.~Bardeen, W.~H.~Press and S.~A.~Teukolsky,
Astrophys. J. \textbf{178}, 347 (1972)
doi:10.1086/151796

\bibitem{Steiner:2010kd}
J.~F.~Steiner, J.~E.~McClintock, R.~A.~Remillard, L.~Gou, S.~Yamada and R.~Narayan,
Astrophys. J. Lett. \textbf{718}, L117-L121 (2010)
doi:10.1088/2041-8205/718/2/L117
[arXiv:1006.5729 [astro-ph.HE]].

\bibitem{Penna:2010hu}
R.~F.~Penna, J.~C.~McKinney, R.~Narayan, A.~Tchekhovskoy, R.~Shafee and J.~E.~McClintock,
\mnras, \textbf{408}, 752 (2010).
doi:10.1111/j.1365-2966.2010.17170.x
[arXiv:1003.0966 [astro-ph.HE]].

\bibitem{Zdziarski:2020arg}
A.~A.~Zdziarski, B.~De Marco, M.~Szanecki, A.~Niedzwiecki and A.~Markowitz,
Astrophys. J. \textbf{906}, 69 (2021)
doi:10.3847/1538-4357/abca9c
[arXiv:2006.12829 [astro-ph.HE]].

\bibitem{Wang-Ji:2017oly}
J.~Wang-Ji, J.~A.~Garc\'\i{}a, J.~F.~Steiner, J.~A.~Tomsick, F.~A.~Harrison, C.~Bambi, P.~O.~Petrucci, J.~Ferreira, S.~Chakravorty and M.~Clavel,
Astrophys. J. \textbf{855}, 61 (2018)
doi:10.3847/1538-4357/aaa974
[arXiv:1712.02571 [astro-ph.HE]].

\bibitem{Liu:2023ovm}
H.~Liu, J.~Jiang, Z.~Zhang, C.~Bambi, A.~C.~Fabian, J.~A.~Garcia, A.~Ingram, E.~Kara, J.~F.~Steiner and J.~A.~Tomsick, \textit{et al.}
Astrophys. J. \textbf{951}, 145 (2023)
doi:10.3847/1538-4357/acd8b9
[arXiv:2303.10593 [astro-ph.HE]].

\bibitem{Zdziarski:2021jrw}
A.~A.~Zdziarski, M.~A.~Dzie{\l}ak, B.~De Marco, M.~Szanecki and A.~Nied{\'z}wiecki,
Astrophys. J. Lett. \textbf{909}, L9 (2021)
doi:10.3847/2041-8213/abe7ef
[arXiv:2101.04482 [astro-ph.HE]].

\bibitem{Zdziarski:2021nem}
A.~A.~Zdziarski, E.~Jourdain, P.~Lubi{\'n}ski, M.~Szanecki, A.~Nied{\'z}wiecki, A.~Veledina, J.~Poutanen, M.~A.~Dzielak and J.~P.~Roques,
Astrophys. J. Lett. \textbf{914}, L5 (2021)
doi:10.3847/2041-8213/ac0147
[arXiv:2104.04316 [astro-ph.HE]].

\bibitem{Will:2014kxa}
C.~M.~Will,
Living Rev. Rel. \textbf{17}, 4 (2014)
doi:10.12942/lrr-2014-4
[arXiv:1403.7377 [gr-qc]].

\bibitem{Bambi:2015kza}
C.~Bambi,
Rev. Mod. Phys. \textbf{89}, no.2, 025001 (2017)
doi:10.1103/RevModPhys.89.025001
[arXiv:1509.03884 [gr-qc]].

\bibitem{Karas:2022}
V. Karas, M. Zaja{\v{c}}ek, D. Kunneriath and M. {Dov{\v{c}}iak},
Advances in Sp. Research, \textbf{69}, 448-466 (2022),
doi:10.1016/j.asr.2021.09.012,
[arXiv:2110.11136 [astro-ph.HE]]

\bibitem{Tripathi:2018lhx}
A.~Tripathi, S.~Nampalliwar, A.~B.~Abdikamalov, \textit{et al.}
Astrophys. J. \textbf{875}, 56 (2019)
doi:10.3847/1538-4357/ab0e7e
[arXiv:1811.08148 [gr-qc]].

\bibitem{Tripathi:2020yts}
A.~Tripathi, Y.~Zhang, A.~B.~Abdikamalov, \textit{et al.}
Astrophys. J. \textbf{913}, 79 (2021)
doi:10.3847/1538-4357/abf6cd
[arXiv:2012.10669 [astro-ph.HE]].

\bibitem{LIGOScientific:2016lio}
B.~P.~Abbott \textit{et al.} [LIGO Scientific and Virgo],
Phys. Rev. Lett. \textbf{116}, 221101 (2016)
[erratum: Phys. Rev. Lett. \textbf{121}, 129902 (2018)]
doi:10.1103/PhysRevLett.116.221101
[arXiv:1602.03841 [gr-qc]].

\bibitem{Vagnozzi:2022moj}
S.~Vagnozzi, R.~Roy, Y.~D.~Tsai, \textit{et al.}
Class. Quant. Grav. \textbf{40}, 165007 (2023)
doi:10.1088/1361-6382/acd97b
[arXiv:2205.07787 [gr-qc]].

\bibitem{Gou:2009}
L.~Gou, J.~E.~McClintock, J.~Liu, R.~Narayan, J.~F.~Steiner, R.~A.~Remillard, J.~A.~Orosz, S.~W.~Davis, K.~Ebisawa, E.~M.~Schlegel,
\textit{The Astrophysical Journal} \textbf{701}, no. 2, 1076 (2009)
doi:10.1088/0004-637X/701/2/1076.

\bibitem{Gou:2011extreme}
L.~Gou, J.~E.~McClintock, M.~J.~Reid, J.~A.~Orosz, J.~F.~Steiner, R.~Narayan, J.~Xiang, R.~A.~Remillard, K.~A.~Arnaud, and S.~W.~Davis,  
Astrophys. J. \textbf{742}, no.2, 85 (2011)  
doi:10.1088/0004-637X/742/2/85

\bibitem{svoboda_first_2024}
J.~Svoboda, M.~Dov{\v c}iak, J.~F. Steiner, F.~Muleri, A.~Ingram, A.~Yilmaz,
  N.~Rodriguez~Cavero, L.~Marra, J.~Poutanen, A.~Veledina, et al., Astrophys. J. \textbf{960}, 3 (2024). doi:10.3847/1538-4357/ad0842

\bibitem{marra_ixpe_2024}
L.~Marra, M.~Brigitte, N.~Rodriguez~Cavero, S.~Chun, J.~F. Steiner, M.~Dov{\v
  c}iak, M.~Nowak, S.~Bianchi, F.~Capitanio, A.~Ingram, et al., Astron. Astrophys. \textbf{684}, A95 (2024). doi:10.1051/0004-6361/202348277

\bibitem{steiner_ixpe-led_2024}
J.~F. Steiner, E.~Nathan, K.~Hu, H.~Krawczynski, M.~Dov{\v c}iak, A.~Veledina,
  F.~Muleri, J.~Svoboda, K.~Alabarta, M.~Parra, et al., Astrophys. J. Lett.
  \textbf{969}, L30 (2024). doi:10.3847/2041-8213/ad58e4


\bibitem{Ingram2019_reltrans} A.~Ingram, G.~Mastroserio, T.~Dauser, et al. \mnras, \textbf{488}, 324 (2019). doi:10.1093/mnras/stz1720

\bibitem{Mastroserio2021} G.~Mastroserio, A.~Ingram, J.~Wang, et al. \mnras, \textbf{507}, 55 (2021). doi:10.1093/mnras/stab2056

\bibitem{Mastroserio2020} G.~Mastroserio, A.~Ingram \& M.~van der Klis. \mnras, \textbf{498}, 4971 (2020). doi:10.1093/mnras/staa2735

\bibitem{Cackett2014} E.~M.~Cackett, , A.~Zoghbi, C.~Reynolds et al. \mnras, \textbf{438}, 2980 (2014). doi:10.1093/mnras/stt2424

\bibitem{Mastroserio2018} G.~Mastroserio, A.~Ingram \& M.~van der Klis. \mnras, \textbf{475}, 4027 (2018). doi:10.1093/mnras/sty075

\bibitem{Paolillo2025} M.~Paolillo \& I.~Papadakis. arXiv preprint, arXiv:2506.23899 (2025)

\bibitem{Bambi:2020jpe}
C.~Bambi, L.~W.~Brenneman, T.~Dauser, et al.
Space Sci. Rev. \textbf{217}, 65 (2021)
doi:10.1007/s11214-021-00841-8

\bibitem{Draghis:2023vzj}
P.~A.~Draghis, J.~M.~Miller, E.~Costantini, L.~C.~Gallo, M.~Reynolds, J.~A.~Tomsick and A.~Zoghbi,
Astrophys. J. \textbf{969}, 40 (2024)
doi:10.3847/1538-4357/ad43ea
[arXiv:2311.16225 [astro-ph.HE]].

\bibitem{Dauser:2013xv}
T.~Dauser, J.~Garcia, J.~Wilms, M.~Bock, L.~W.~Brenneman, M.~Falanga, K.~Fukumura and C.~S.~Reynolds,
Mon. Not. Roy. Astron. Soc. \textbf{430}, 1694 (2013)
doi:10.1093/mnras/sts710
[arXiv:1301.4922 [astro-ph.HE]].

\bibitem{Garcia:2013lxa}
J.~Garc\'\i{}a, T.~Dauser, A.~Lohfink, T.~R.~Kallman, J.~Steiner, J.~E.~McClintock, L.~Brenneman, J.~Wilms, W.~Eikmann and C.~S.~Reynolds, \textit{et al.}
Astrophys. J. \textbf{782}, 76 (2014)
doi:10.1088/0004-637X/782/2/76
[arXiv:1312.3231 [astro-ph.HE]].

\bibitem{2023ApJ...955...53F} Feng, Y., Yuan, Y.-F., \& Zhang, S.-N.\ 2023, \apj, 955, 53. doi:10.3847/1538-4357/acedff

\bibitem{Feng:2025} Feng, Y., Yuan, Y.-F., \& Zhang, S.-N.\ 2025, \apj, submitted.

\bibitem{Podgorny:2023egq}
J.~Podgorn\'y, M.~Dov\v{c}iak, R.~Goosmann, F.~Marin, G.~Matt, A.~R\'o\.za\'nska and V.~Karas,
Mon. Not. Roy. Astron. Soc. \textbf{524}, no.3, 3853-3876 (2023)
doi:10.1093/mnras/stad2169
[arXiv:2307.08819 [astro-ph.HE]].

\bibitem{schnittman_x-ray_2010}
J.~D. Schnittman and J.~H. Krolik, Astrophys. J. \textbf{712}, 908 (2010). doi:10.1088/0004-637X/712/2/908

\bibitem{tamborra_moca:_2018}
F.~Tamborra, G.~Matt, S.~Bianchi, and M.~Dov{\v c}iak, Astron. Astrophys. \textbf{619}, A105 (2018). doi:10.1051/0004-6361/201732023

\bibitem{zhang_constraining_2019}
W.~Zhang, M.~Dov{\v c}iak, and M.~Bursa, Astrophys. J.
  \textbf{875}, 148 (2019). doi:10.3847/1538-4357/ab1261

\bibitem{zhang_investigating_2022}
W.~Zhang, M.~Dov{\v c}iak, M.~Bursa, V.~Karas, G.~Matt, and F.~Ursini, Mon. Not. Roy. Astron. Soc. \textbf{515}, 2882 (2022). doi:10.1093/mnras/stac1937

\bibitem{ursini_prospects_2022}
F.~Ursini, G.~Matt, S.~Bianchi, A.~Marinucci, M.~Dov{\v c}iak, and W.~Zhang, Mon. Not. Roy. Astron. Soc. \textbf{510}, 3674 (2022). doi:10.1093/mnras/stab3745

\bibitem{poutanen_polarized_2023}
J.~Poutanen, A.~Veledina, and A.~M. Beloborodov, Astrophys. J. Lett. \textbf{949}, L10 (2023). doi:10.3847/2041-8213/acd33e

\bibitem{dovciak_thermal_2008}
M.~Dov{\v c}iak, F.~Muleri, R.~W. Goosmann, V.~Karas, and G.~Matt, Mon. Not. Roy. Astron. Soc. \textbf{391}, 32 (2008).

\bibitem{li_inferring_2009}
L.-X. Li, R.~Narayan, and J.~E. McClintock, Astrophys. J.
  \textbf{691}, 847 (2009). doi:10.1088/0004-637X/691/1/847

\bibitem{schnittman_x-ray_2009}
J.~D. Schnittman and J.~H. Krolik, The Astrophys. J.\textbf{701}, 1175 (2009). doi:10.1088/0004-637X/701/2/1175

\bibitem{taverna_towards_2020}
R.~Taverna, W.~Zhang, M.~Dov{\v c}iak, S.~Bianchi, M.~Bursa, V.~Karas, and
  G.~Matt, Mon. Not. Roy. Astron. Soc. \textbf{493}, 4960
  (2020). doi:10.1093/mnras/staa598

\bibitem{taverna_spectral_2021}
R.~Taverna, L.~Marra, S.~Bianchi, M.~Dov{\v c}iak, R.~Goosmann, F.~Marin,
  G.~Matt, and W.~Zhang, Mon. Not. Roy. Astron. Soc.
  \textbf{501}, 3393 (2021). doi:10.1093/mnras/staa3859

\bibitem{mikusincova_x-ray_2023}
R.~Mikusincova, M.~Dov{\v c}iak, M.~Bursa, N.~D. Lalla, G.~Matt, J.~Svoboda,
  R.~Taverna, and W.~Zhang, Mon. Not. Roy. Astron. Soc.
  \textbf{519}, 6138 (2023). doi:10.1093/mnras/stad077

\bibitem{weisskopf_imaging_2022}
M.~C. Weisskopf, P.~Soffitta, L.~Baldini, B.~D. Ramsey, S.~L. O'Dell, R.~W.
  Romani, G.~Matt, W.~D. Deininger, W.~H. Baumgartner, R.~Bellazzini, et al., J. Astron. Telesc. Instrum. Syst.
  \textbf{8}, 026002 (2022). doi:10.1117/1.JATIS.8.2.026002

\bibitem{krawczynski_polarized_2022}
H.~Krawczynski, F.~Muleri, M.~Dov{\v c}iak, A.~Veledina, N.~Rodriguez~Cavero, J.~Svoboda, A.~Ingram, G.~Matt, J.~A. Garcia, V.~Loktev, et al., Science \textbf{378}, 650
  (2022). doi:10.1126/science.add5399

\bibitem{Abdulghani:2024}
Y.~Abdulghani, A.~M.~Lohfink, and J.~Chauhan. \mnras, \textbf{530}, no.1, 424--445 (2024),  
doi:10.1093/mnras/stae767.

\bibitem{zhu2012eye}
Y.~Zhu, S.~W.~Davis, R.~Narayan, A.~K.~Kulkarni, R.~F.~Penna, J.~E.~McClintock. \mnras, \textbf{424}, no. 4, 2504--2521 (2012)
doi:10.1111/j.1365-2966.2012.21181.x.

\bibitem{kulkarni2011measuring}
A.~K.~Kulkarni, R.~F.~Penna, R.~V.~Shcherbakov, J.~F.~Steiner, R.~Narayan, A.~Sk{a}dowski, Y.~Zhu, J.~E.~McClintock, S.~W.~Davis, J.~C.~McKinney \mnras, \textbf{414}, no. 2, 1183--1194 (2011)
doi:10.1111/j.1365-2966.2011.18446.x

\bibitem{Wielgus2022MNRAS.514..780W} M.~Wielgus, D.~Lan{\v{c}}ov{\'a}, O.~Straub, et al. \mnras, \textbf{514}, 1, 780 (2022). doi:10.1093/mnras/stac1317

\bibitem{Lancova2023AN....34430023L} D.~Lan{\v{c}}ov{\'a}, A.~Yilmaz, M.~Wielgus, et al. Astronomische Nachrichten, \textbf{344}, 4 (2023), e20230023. doi:10.1002/asna.20230023

\bibitem{Mummery2024MNRAS.531..366M} A.~Mummery, A.~Ingram, S.~Davis, et al. \mnras, \textbf{531}, 1, 366 (2024). doi:10.1093/mnras/stae1160

\bibitem{mummery2024plunging}
A.~Mummery, J.~Jiang, A.~Fabian. \mnras, \textbf{533}, no. 1, L83--L90 (2024),  
doi:10.1093/mnrasl/slae056.

\bibitem{Mummery2025arXiv250513119M} A.~Mummery, J.~Jiang, A.~Ingram, et al. arXiv:2505.13119 (2025). doi:10.48550/arXiv.2505.13119

\bibitem{Lightman1974ApJ...187L...1L} A.~P.~Lightman \& D.~M.~Eardley. \apjl, \textbf{187}, L1 (1974). doi:10.1086/181377
\bibitem[Begelman \& Pringle(2007)]{Begelman2007MNRAS.375.1070B} M.~C.~Begelman \& J.~E.~ Pringle. \mnras, \textbf{375}, 3, 1070 (2007). doi:10.1111/j.1365-2966.2006.11372.x

\bibitem[Lan{\v{c}}ov{\'a} et al.(2019)]{Lancova2019ApJ...884L..37L} D.~Lan{\v{c}}ov{\'a}, D.~Abarca, W.~ Klu{\'z}niak, et al. \apjl, 884, 2, L37 (2019). doi:10.3847/2041-8213/ab48f5

\bibitem{2024ApJ...966..226K} O. Kop{\'a}{\v{c}}ek, V. Karas,
Astrophys. J. \textbf{966}, id. 226 (2024). doi:10.3847/1538-4357/ad3932

\bibitem{2025MNRAS.537.1963M} A. Mummery. \mnras, {\bf 537}, 1963 (2025). doi:10.1093/mnras/staf060

\bibitem[\protect\citeauthoryear{Motta et al.}{2014}]{Motta2014} S.~Motta, T.~Belloni, L.~Stella, T.~Mu{\~n}oz-Darias \& R.~Fender et al. \mnras, \textbf{437}, 2554-2565 (2014). doi:10.1093/mnras/stt2068

\bibitem[\protect\citeauthoryear{Ingram \& Motta}{2014}]{Ingram2014} A.~Ingram \& S.~Motta. \mnras, \textbf{444}, 2065-2070 (2014). doi:10.1093/mnras/stu1585

\bibitem[\protect\citeauthoryear{Motta \& Belloni}{2024}]{Motta2024}  S.~Motta \& T.~Belloni. \aap, \textbf{684}, A209 (2024). doi:10.1051/0004-6361/202347331

\bibitem[Karas et al.(2023)]{2023CoSka..53d.175K} V.~Karas, K.~{Klimovi{\v{c}}ov{\'a}}, D.~{Lan{\v{c}}ov{\'a}}, M.~{{\v{S}}tolc}, J.~{Svoboda}, G,~{T{\"o}r{\"o}k}, M.~{Matuszkov{\'a}}, E.~{{\v{S}}r{\'a}mkov{\'a}}, R.~{{\v{S}}pr{\v{n}}a} \& M.~{Urbanec}. Contributions of the Astronomical Observatory Skalnate Pleso, \textbf{53}, 4, 175 (2023). doi:10.31577/caosp.2023.53.4.175

\bibitem[\protect\citeauthoryear{Ingram, Done, \& Fragile}{2009}]{Ingram2009} A.~Ingram, C.~Done, P.~Fragile. \mnras, \textbf{397}, L101 (2009). doi:10.1111/j.1745-3933.2009.00693.x

\bibitem[\protect\citeauthoryear{Ingram \& Done}{2012}]{Ingram2012} A.~Ingram \& C.~Done. \mnras, \textbf{427}, 934 (2012). doi:10.1111/j.1365-2966.2012.21907.x

\bibitem[\protect\citeauthoryear{Ingram et al.}{2016}]{Ingram2016} A.~Ingram, M.~van der Klis, M.~Middleton, C.~Done, D.~Altamirano, L.~Heil, P.~Uttley, et al. \mnras, 461, 1967 (2016). doi:10.1093/mnras/stw1245

\bibitem[\protect\citeauthoryear{Nathan et al.}{2022}]{Nathan2022} E.~Nathan, A.~Ingram, J.~Homan, D.~Huppenkothen, P.~Uttley, M.~van der Klis,  S.~Motta, et al. \mnras, \textbf{511}, 255 (2022). doi:10.1093/mnras/stab3803

\bibitem[\protect\citeauthoryear{Ingram et al.}{2017}]{Ingram2017} A.~Ingram, M.~van der Klis, M.~ Middleton, D.~Altamirano, P.~Uttley. \mnras, \textbf{64}, 2979 (2017). doi:10.1093/mnras/stw2581

\bibitem[\protect\citeauthoryear{Ingram}{2019}]{Ingram2019} A.~Ingram. \mnras, \textbf{89}, 3927 (2019). doi:10.1093/mnras/stz2409

\bibitem[\protect\citeauthoryear{Ingram \& Maccarone}{2017}]{Ingram2017a} A.~Ingram \& T.~Maccarone. \mnras, \textbf{471}, 4206 (2017). doi:10.1093/mnras/stx1881

\bibitem[\protect\citeauthoryear{Ingram et al.}{2015}]{Ingram2015} A.~Ingram, T.~Maccarone, J.~Poutanen, H.~Krawczynski. \apj, \textbf{807}, 53 (2015). doi:10.1088/0004-637X/807/1/53

\bibitem[\protect\citeauthoryear{Zhao et al.}{2024}]{Zhao2024} Q.~C.~Zhao, L.~Tao, H.~C.~Li, S.~N.~Zhang, H.~Feng, M.~Y.~Ge, L.~Ji, et al. \apjl, \textbf{961}, L42 (2024). doi:10.3847/2041-8213/ad1e6c

\bibitem{Krawczynski:2012ac}
H.~Krawczynski,
Astrophys. J. \textbf{754}, 133 (2012)
doi:10.1088/0004-637X/754/2/133
[arXiv:1205.7063 [gr-qc]].

\bibitem{Liu:2015ibq}
D.~Liu, Z.~Li, Y.~Cheng and C.~Bambi,
Eur. Phys. J. C \textbf{75}, 383 (2015)
doi:10.1140/epjc/s10052-015-3600-9
[arXiv:1504.06788 [gr-qc]].

\bibitem{Zhou:2019fcg}
M.~Zhou, A.~B.~Abdikamalov, D.~Ayzenberg, {\it et al.},
Phys. Rev. D \textbf{99}, 104031 (2019)
doi:10.1103/PhysRevD.99.104031
[arXiv:1903.09782 [gr-qc]].

\bibitem{Tripathi:2020qco}
A.~Tripathi, M.~Zhou, A.~B.~Abdikamalov, {\it et al.},
Astrophys. J. \textbf{897}, 84 (2020)
doi:10.3847/1538-4357/ab9600
[arXiv:2001.08391 [gr-qc]].

\bibitem{Bambi:2012pa}
C.~Bambi,
JCAP \textbf{09}, 014 (2012)
doi:10.1088/1475-7516/2012/09/014
[arXiv:1205.6348 [gr-qc]].

\bibitem{Bambi:2013fea}
C.~Bambi,
Eur. Phys. J. C \textbf{75}, 162 (2015)
doi:10.1140/epjc/s10052-015-3396-7
[arXiv:1312.2228 [gr-qc]].

\bibitem{Bambi:2016sac}
C.~Bambi, A.~Cardenas-Avendano, T.~Dauser, J.~A.~Garcia and S.~Nampalliwar,
Astrophys. J. \textbf{842}, 76 (2017)
doi:10.3847/1538-4357/aa74c0
[arXiv:1607.00596 [gr-qc]].

\bibitem{Abdikamalov:2019yrr}
A.~B.~Abdikamalov, D.~Ayzenberg, C.~Bambi, {\it et al.},
Astrophys. J. \textbf{878}, 91 (2019)
doi:10.3847/1538-4357/ab1f89
[arXiv:1902.09665 [gr-qc]].

\bibitem{Tripathi:2020dni}
A.~Tripathi, A.~B.~Abdikamalov, D.~Ayzenberg, {\it et al.},
Astrophys. J. \textbf{907}, 31 (2021)
doi:10.3847/1538-4357/abccbd
[arXiv:2010.13474 [astro-ph.HE]].

\bibitem{Zhang:2021ymo}
Z.~Zhang, H.~Liu, A.~B.~Abdikamalov, {\it et al.},
Astrophys. J. \textbf{924}, 72 (2022)
doi:10.3847/1538-4357/ac350e
[arXiv:2106.03086 [astro-ph.HE]].

\bibitem[Belloni et al.(2005)]{Belloni:2005} T.~Belloni, J.~Homan, P.~Casella, et al. \aap, \textbf{440}, 1, 207 (2005). doi:10.1051/0004-6361:20042457

\bibitem[Merloni et al.(2003)]{Merloni:2003} A.~Merloni, S.~Heinz and T.~di Matteo, T. \mnras, \textbf{345}, 4, 1057 (2003). doi:10.1046/j.1365-2966.2003.07017.x

\bibitem[Kang et al.(2025)]{Kang:2025} J.~L.~Kang, C.~Done, S.~Hagen, et al.\, arXiv:2504.08067 (2025). doi:10.48550/arXiv.2504.08067

\bibitem[Hagen et al.(2024)]{Hagen:2024} S.~Hagen, C.~Done, J.~D.~Silverman, et al.\, \mnras, \textbf{534}, 3, 2803 (2024). doi:10.1093/mnras/stae2272




\bibitem{2014ARA&A..52..529Y}
F. {Yuan}, {Narayan}, R. \araa, \textbf{52}, 529-588 (2014)
doi:10.1146/annurev-astro-082812-141003

\bibitem{2007A&ARv..15....1D}
C. {Done}, {Gierli{\'n}ski}, M., {\it et al.},
Astron. Astrophys. Rev. \textbf{15}, 1-66 (2007)
doi:10.1007/s00159-007-0006-1

\bibitem{1997ApJ...489..865E}
A. {Esin}, {McClintock}, J., {\it et al.} \apj, \textbf{489}, 865-889 (1997)
doi:10.1086/304829

\bibitem{2019Natur.565..198K}
E. {Kara}, {Steiner}, J., {\it et al.},
Nature \textbf{565}, 198-201 (2019)
doi:10.1038/s41586-018-0803-x

\bibitem{2011MNRAS.414L..60U}
P. {Uttley}, {Wilkinson}, T., {\it et al.},
\mnras, \textbf{414}, L60-L64 (2011)
doi:10.1111/j.1745-3933.2011.01056.x

\bibitem{2006MNRAS.367..801A}
P.~{Ar{\'e}valo}, P~{Uttley}. \mnras, \textbf{367}, 801-814 (2006)
doi:10.1111/j.1365-2966.2006.09989.x

\bibitem{2025MNRAS.536.3284U} P.~Uttley \& J.~Malzac. \mnras, \textbf{536}, 3284 (2025). doi:10.1093/mnras/stae2514

\bibitem{2018ApJ...858...82Y}
B. {You}, {Bursa}, M., {\it et al.}. \apj, \textbf{858}, 82 (2018)
doi:10.3847/1538-4357/aabd33

\bibitem{2020ApJ...897...27Y} B. {You}, {{\.Z}ycki}, P., {\it et al.},
 \apj, \textbf{897}, 27 (2020). doi:10.3847/1538-4357/ab9838

\bibitem{2025arXiv250203995Z} Y. {Zhan}, {You}, B., {\it et al.}, arXiv:2502.03995 (2025). doi:10.48550/arXiv.2502.03995

\bibitem[Novikov \& Thorne(1973)]{1973blho.conf..343N} I.~D.~Novikov \& K.~S.~Thorne. Astrophysics of black holes., 343 (1973)

\bibitem[Garc{\'\i}a et al.(2013)]{2013ApJ...768..146G} J.~Garc{\'\i}a, T.~Dauser, C.~S.~Reynolds, et al. \apj, \textbf{768}, 2, 146 (2013). doi:10.1088/0004-637X/768/2/146

\bibitem[Dauser et al.(2013)]{2013MNRAS.430.1694D} T.~Dauser, J.~Garc{\'\i}a, J.~Wilms, et al. \mnras, \textbf{430}, 3, 1694 (2013). doi:10.1093/mnras/sts710

\bibitem{matteo_bachetti_2024_10813181}
M.~Bachetti, D.~Huppenkothen, U.~Khan, {\it et al.},
StingraySoftware/stingray: Stingray Version 2.0. Zenodo (2024).
doi:10.5281/zenodo.10813181

\bibitem{2019ApJ...881...39H}
D. {Huppenkothen}, M.~{Bachetti}, A.~L.~Stevens, {\it et al.},
\apj, \textbf{881}, 39 (2019)
doi:10.3847/1538-4357/ab258d

\bibitem{bachettiStingrayFastModern2024}
M. Bachetti,  D.~Huppenkothen, A.~L.~Stevens, {\it et al.},
Journal of Open Source Software, \textbf{9}, 7389 (2024)
doi:10.21105/joss.07389

\bibitem{Uttley2014} P.~Uttley,  E.~M.~Cackett, A.~C.~Fabian, et al. \aapr, \textbf{22}, 72 (2014). doi:10.1007/s00159-014-0072-0

\bibitem{Kara2016a} E.~Kara, W.~N.~Alston, A.~C.~Fabian, {\it et al.} \mnras, \textbf{462}, 511 (2016). doi:10.1093/mnras/stw1695

\bibitem{Veledina2023} A. Veledina, F. Muleri, M. Dov{\v{c}}iak, J. Poutanen, A. Ratheesh, F. Capitanio, G. Matt, P. Soffitta, A. Tennant, M. Negro, et al. \apjl, \textbf{958}, L16 (2023). doi:10.3847/2041-8213/ad0781

\bibitem{Ingram2024} A. Ingram, N. Bollemeijer, A. Veledina, M. Dov{\v{c}}iak, J. Poutanen, E. Egron, T.D. Russell, S.A. Trushkin, M. Negro, A. Ratheesh, et al. \apj, \textbf{968}, 76 (2024). doi:10.3847/1538-4357/ad3faf

\bibitem{Podgorny2024} J. Podgorn{\`y}, F. Marin, F \& M. Dov{\v{c}}iak. \mnras, \textbf{527}, 1114-1134 (2024). doi:10.1093/mnras/stad3266

\bibitem{Wang2022} J.~Wang, E.~Kara, M.~Lucchini, et al. \apj, \textbf{930}, 18 (2022). doi:10.3847/1538-4357/ac6262

\bibitem{Cao2022} Z.~Cao, M.~Lucchini, S.~Markoff, et al. \mnras, \textbf{509}, 2517 (2022). doi:10.1093/mnras/stab3080

\bibitem{Liu2023} H. Liu, C. Bambi, J. Jiang, et al. \apj, \textbf{950}, 5 (2023). doi:10.3847/1538-4357/acca17

\bibitem{Yang2023} Z.-X. Yang, L. Zhang, S. Zhang~N., et al. \mnras, \textbf{521}, 3570 (2023). doi:10.1093/mnras/stad795

\bibitem{Li2025} Y. Li, Z. Yan, C. Gao, et al. 2024, arXiv:2407.08421. doi:10.48550/arXiv.2407.08421


\bibitem{2024ApJ...973...59S} Q.-C. Shui, S. Zhang, J.-Q. Peng, et al. \apj, \textbf{973}, 59 (2024). doi:10.3847/1538-4357/ad676a

\bibitem{1997ApJ...482L.155Z} S.-N. Zhang, W. Cui, \& W. Chen. \apjl, \textbf{482}, L155 (1997). doi:10.1086/310705

\bibitem{1995ApJ...445..780S} T. Shimura \& F. Takahara. \apj, \textbf{445}, 780 (1995). doi:10.1086/175740

\bibitem{2005ApJ...621..372D} S.~W. Davis, O.~M. Blaes W., I. Hubeny, et al. \apj, \textbf{621}, 372 (2005). doi:10.1086/427278

\bibitem{2006ApJ...636L.113S} R. Shafee, J. McClintock, E. Narayan, et al. \apjl, \textbf{636}, L113 (2006). doi:10.1086/498938

\bibitem{1975ApJ...202..788C} C. Cunningham. \apj, \textbf{202}, 788 (1975). doi:10.1086/154033

\bibitem{1998RSPSA.454..903H} N.~E. Huang, Z. Shen, S.~R. Long, et al. Proceedings of the Royal Society of London Series A, \textbf{454}, 903 (1998). doi:10.1098/rspa.1998.0193

\bibitem{2023ApJ...957...84S} Q.~C. Shui, S. Zhang, S.~N. Zhang, et al. \apj, \textbf{957}, 84 (2023). doi:10.3847/1538-4357/acfc42

\bibitem{Marinucci2020} A. Marinucci, S. Bianchi, V. Braito, {\it et al.} \mnras, \textbf{496}, 3412 (2020). doi:10.1093/mnras/staa1683

\bibitem{Markowitz2003} A. Markowitz, R. Edelson, \& S. Vaughan, \apj, \textbf{598}, 935 (2003). doi:10.1086/379103

\bibitem{Fabian2004} A. Fabian, Miniutti C., Gallo G., L., {\it et al.}\, \mnras, {\bf 353}, 1071 (2004). doi:10.1111/j.1365-2966.2004.08036.x

\bibitem{Miniutti2004} G. Miniutti, A. Fabian, \& J. Miller, \mnras, {\bf 351}, 466 (2004). doi:10.1111/j.1365-2966.2004.07794.x

\bibitem{Shu2010} X.~W. Shu, T. Yaqoob, K.~D. Murphy, {\it et al.}\, \apj, {\bf 713}, 1256 (2010). doi:10.1088/0004-637X/713/2/1256

\bibitem{Liang2022} W.~C. Liang, X.~W. Shu, J.~X. Wang, {\it et al.}\, Journal of High Energy Astrophysics, {\bf 33}, 20 (2022). doi:10.1016/j.jheap.2022.01.002

\bibitem{Madau2024} P. Madau \& F. Haardt\ 2024, \apjl, {\bf 976}, L24 (2024). doi:10.3847/2041-8213/ad90e1

\bibitem{Suh2024} H. Suh, J. Scharw{\"a}chter, E.~P. Farina, {\it et al.} Nature Astronomy, \textbf{9}, 271-279 (2025). doi:10.1038/s41550-024-02402-9

\bibitem{Abramowicz1988} M.~A. Abramowicz, B. Czerny, J.~P. Lasota B., {\it et al.}, \apj, {\bf 332}, 646 (1988). doi:10.1086/166683

\bibitem{Ohsuga2009} K. Ohsuga, S. Mineshige, M. Mori, {\it et al.}, \pasj, {\bf 61}, L7 (2009). doi:10.1093/pasj/61.3.L7

\bibitem{Jiang2014} Y.-F. Jiang, J.~M. Stone, \& S.~W. Davis. \apj, {\bf 796}, 106 (2014). doi:10.1088/0004-637X/796/2/106

\bibitem{Thomsen2019} L.~L. Thomsen, L. Dai, J. L., E. Ramirez-Ruiz, E. Kara, C. Reynolds. \apjl, {\bf 884}, L21 (2019). doi:10.3847/2041-8213/ab4518

\bibitem{Thomsen2022} L.~L. Thomsen, L, Dai, E. Kara, {\it et al.}\, \apj, {\bf 925}, 151 (2022). doi:10.3847/1538-4357/ac3df3

\bibitem{Zhang2024} Z. Zhang, L.~L. Thomsen, L. Dai, C. Reynolds, J.~A. Garc{\'\i}a, E. Kara, R. Connors, {\it et al.}, \apj, {\bf 977}, 157 (2024). doi:10.3847/1538-4357/ad86c0


\bibitem{Kara2016b} E. Kara, J.~M. Miller, C. Reynolds, {\it et al.}\, \nat, {\bf 535}, 388 (2016). doi:10.1038/nature18007

\bibitem{Wang2021} J. Wang, G. Mastroserio, E. Kara, {\it et al.}\, \apjl, {\bf 910}, L3 (2021). doi:10.3847/2041-8213/abec79

\bibitem{Gianolli2023} V. Gianolli, D. Kim, S. Bianchi, {\it et al.}\, \mnras, {\bf 523}, 4468 (2023). doi:10.1093/mnras/stad1697

\bibitem{bardeen_lense-thirring_1975} J. Bardeen \& J.~A.Petterson. \apjl, {\bf 195}, L65 (1975). doi:10.1086/181711

\bibitem{zhang_predictions_2015} W.~D. Zhang, W.~F. Yu, V. Karas, Dov{\v c}iak V., M.\, \apj, {\bf 807}, 89 (2015). doi:10.1088/0004-637X/807/1/89

\bibitem{zhang_probing_2019} W.~D. Zhang, W.~F. Yu, V. Karas, M. Dov{\v c}iak. \apj, {\bf 884}, 72 (2019). doi:10.3847/1538-4357/ab3e3e

\bibitem{Gutierrez2020} E.~M. Guti{\'e}rrez, R. Nemmen,  \& F. Cafardo, F. \apjl, {\bf 891(2)}, L36 (2020). doi:10.3847/2041-8213/ab7998

\bibitem{2006NewAR..50..796P} B.~M. Peterson \& M.~C. Bentz. \nar, {\bf 50}, 796 (2006). doi:10.1016/j.newar.2006.06.062

\bibitem[McHardy(2010)]{2010LNP...794..203M} I. McHardy.\ Lecture Notes in Berlin Springer Verlag Physics, 203 (2010). doi:10.1007/978-3-540-76937-8\_8

\bibitem[Gonz{\'a}lez-Mart{\'\i}n \& Vaughan(2012)]{2012A&A...544A..80G} O. Gonz{\'a}lez-Mart{\'\i}n, \& S. Vaughan, \aap, {\bf 544}, A80 (2012). doi:10.1051/0004-6361/201

\bibitem[Markoff et al.(2015)]{2015ApJ...812L..25M} S. Markoff, M. Nowak, E. Gallo, et al. \apjl, {\bf 812}, L25 (2015). doi:10.1088/2041-8205/812/2/L25

\bibitem[Scaringi et al.(2015)]{2015SciA....1E0686S} S. Scaringi, T. Maccarone, E. Kording, et al. Science Advances, {\bf 1}, e1500686 (2015). doi:10.1126/sciadv.1500686

\bibitem[Uttley et al.(2002)]{2002MNRAS.332..231U} P. Uttley, I. McHardy, \& E. Papadakis. \mnras, {\bf 332}, 231 (2002). doi:10.1046/j.1365-8711.2002.05298.x

\bibitem[Markowitz et al.(2003)]{2003ApJ...593...96M} A. Markowitz, R. Edelson, S. Vaughan, et al. \apj, {\bf 593}, 96 (2003). doi:10.1086/375330

\bibitem[Vaughan et al.(2003)]{2003MNRAS.345.1271V} S. Vaughan, R. Edelson, R. Warwick, et al. \mnras, {\bf 345}, 1271 (2003). doi:10.1046/j.1365-2966.2003.07042.x

\bibitem[Vaughan et al.(2011)]{2011MNRAS.413.2489V} S. Vaughan, P. Uttley, K. Pounds, et al. \mnras, {\bf 413}, 2489 (2011). doi:10.1111/j.1365-2966.2011.18319.x

\bibitem[McHardy et al.(2006)]{2006Natur.444..730M} I. McHardy, E. Koerding, C. Knigge, et al. \nat, {\bf 444}, 730 (2006). doi:10.1038/nature05389

\bibitem[Strohmayer(2001)]{2001ApJ...552L..49S} T. Strohmayer. \apjl, \textbf{552}, L49 (2001). doi:10.1086/320258

\bibitem[Stella et al.(1999)]{1999ApJ...524L..63S} L. Stella, M. Vietri, \& S. Morsink. \apjl, {\bf 524}, L63 (1999). doi:10.1086/312291

\bibitem[Abramowicz \& Klu{\'z}niak(2001)]{2001A&A...374L..19A} M. Abramowicz \& W. Klu{\'z}niak. \aap, {\bf 374}, L19 (2001). doi:10.1051/0004-6361:20010791

\bibitem[Rezzolla et al.(2003)]{2003MNRAS.344L..37R} L. Rezzolla, S. Yoshida, T. Maccarone, et al. \mnras, {\bf 344}, L37 (2003). doi:10.1046/j.1365-8711.2003.07018.x

\bibitem[Li \& Narayan(2004)]{2004ApJ...601..414L} L.-X. Li \& R. Narayan. \apj, {\bf 601}, 414 (2004). doi:10.1086/380446

\bibitem[Tagger \& Pellat(1999)]{1999A&A...349.1003T} M. Tagger \& R. Pellat. \aap, {\bf 349}, 1003 (1999). doi:10.48550/arXiv.astro-ph/9907267

\bibitem[Li et al.(2003)]{2003ApJ...593..980L} L.-X. Li, J. Goodman, \& R. Narayan. \apj, {\bf 593}, 980 (2003). doi:10.1086/376695

\bibitem[Gierli{\'n}ski et al.(2008)]{2008Natur.455..369G} M. Gierli{\'n}ski, M. Middleton, M. Ward, et al. \nat, {\bf 455}, 369 (2008). doi:10.1038/nature07277

\bibitem[Pan et al.(2016)]{2016ApJ...819L..19P} H.-W. Pan, W. Yuan, S. Yao, et al. \apjl, {\bf 819}, L19 (2016). doi:10.3847/2041-8205/819/2/L19

\bibitem[Zhang et al.(2020)]{2020AcASn..61....2Z} P. Zhang, J.~Z. Yan, \& Q.~Z. Liu. Acta Astronomica Sinica, {\bf 61}, 2 (2020).

\bibitem[Zhang et al.(2017)]{2017ApJ...849....9Z} P. Zhang, P.~F. Zhang, J.~Z. Yan, et al. \apj, {\bf 849}, 9 (2017). doi:10.3847/1538-4357/aa8d6e

\bibitem[Gupta et al.(2018)]{2018A&A...616L...6G} A. Gupta, A. Tripathi, P. Wiita, et al. \aap, {\bf 616}, L6 (2018). doi:10.1051/0004-6361/201833629

\bibitem[Alston et al.(2015)]{2015MNRAS.449..467A} W. Alston, M. Parker, J. Markevi{\v{c}}i{\={u}}t{\.{e}}, et al. \mnras, {\bf 449}, 467 (2015). doi:10.1093/mnras/stv351

\bibitem[Lin et al.(2013)]{2013ApJ...776L..10L} D. Lin, J. Irwin, O. Godet, et al. \apjl, {\bf 776}, L10 (2013). doi:10.1088/2041-8205/776/1/L10

\bibitem[Zhang et al.(2023)]{2023ApJ...946...52Z} H. Zhang, S. Yang, \& B. Dai. \apj, {\bf 946}, 52 (2023). doi:10.3847/1538-4357/acbe37


\bibitem{Marin2024} F. Marin, V. Gianolli, A. Ingram, et al. Galaxies, {\bf 12}, 35 (2024). doi:10.3390/galaxies12040035


\bibitem{Marinucci2022} A. Marinucci, F. Muleri, M. Dov{\v c}iak, et al. \mnras, {\bf 516}, 5907 (2022). doi:10.1093/mnras/stac2634

\bibitem{Tagliacozzo2023} D. Tagliacozzo, A. Marinucci, F. Ursini, et al. \mnras, {\bf 525}, 4735 (2023). doi:10.1093/mnras/stad2627

\bibitem{Ingram2023} A. Ingram, M. Ewing, A. Marinucci, et al. \mnras, {\bf 525}, 5437 (2023). doi:10.1093/mnras/stad2625

\bibitem{Gianolli2024} V. Gianolli~Bianchi E., Kammoun S., E., et al. \aap, {\bf 691}, A29 (2024). doi:10.1051/0004-6361/202451645

\bibitem[Sun et al.(2013)]{Sun2013} L. Sun, X. Shu, \& T. Wang. \apj, {\bf 768}, 2, 167 (2013). doi:10.1088/0004-637X/768/2/167

\bibitem[Miniutti et al.(2019)]{Miniutti2019} G. Miniutti, R. Saxton, M. Giustini, et al. \nat, {\bf 573}, 7774, 381 (2019). doi:10.1038/s41586-019-1556-x

\bibitem[Nicholl et al.(2024)]{2024Natur.634..804N} M. Nicholl, D. Pasham~Mummery R., A., et al.\ 2024, \nat, 634, 8035, 804. doi:10.1038/s41586-024-08023-6

\bibitem[Giustini et al.(2020)]{Giustini2020} M. Giustini, G. Miniutti, \& R. Saxton. \aap, {\bf 636}, L2 (2020). doi:10.1051/0004-6361/202037610

\bibitem[Arcodia et al.(2021)]{Arcodia2021} R. Arcodia, A. Merloni, K. Nandra, et al. \nat, {\bf 592}, 7856, 704 (2021). doi:10.1038/s41586-021-03394-6

\bibitem{Quintin2023} E. Quintin, N. Webb, S. Guillot, et al. Astronomy \& Astrophysics, {\bf 675}, p.A152 (2023). doi:10.1051/0004-6361/202346440

\bibitem[Arcodia et al.(2022)]{Arcodia2022} R. Arcodia, G. Miniutti, G. Ponti, et al. \aap, {\bf 662}, A49 (2022). doi:10.1051/0004-6361/202243259

\bibitem[Bykov et al.(2025)]{Bykov2025} S. Bykov, M. Gilfanov, R. Sunyaev, et al. \mnras, {\bf 540}, 1, 30 (2025). doi:10.1093/mnras/staf686

\bibitem[Zhou et al.(2024)]{Zhou2024a} C. Zhou, L. Huang, K. Guo, et al. Phys.Rev. D, {\bf 109}, 10, 103031 (2024). doi:10.1103/PhysRevD.109.103031

\bibitem[Franchini et al.(2023)]{Franchini2023} A. Franchini, M. Bonetti, A. Lupi, et al. \aap, {\bf 675}, A100 (2023). doi:10.1051/0004-6361/202346565

\bibitem[Linial \& Metzger(2023)]{Linial2023} I. Linial \& B. Metzger. \apj, {\bf 957}, 1, 34 (2023). doi:10.3847/1538-4357/acf65b

\bibitem[Tagawa \& Haiman(2023)]{Tagawa2023} H. Tagawa \& Z. Haiman. \mnras, {\bf 526}, 1, 69 (2023). doi:10.1093/mnras/stad2616

\bibitem{2021ApJ...917...43S} P. {Sukov{\'a}}, M. {Zaja{\v{c}}ek}, V. {Witzany},  and V. {Karas}, \aj, \textbf{917}, id.43 (2021)

\bibitem[Zhou et al.(2024)]{Zhou2024b} C. Zhou, B. Zhong, Y. Zeng, et al. Phys.Rev. D, {\bf 110}, 8, 083019 (2024). doi:10.1103/PhysRevD.110.083019

\bibitem[Xian et al.(2021)]{Xian2021} J. Xian, F. Zhang, L. Dou, et al. \apjl, {\bf 921}, 2, L32 (2021). doi:10.3847/2041-8213/ac31aa

\bibitem[Zhou et al.(2025)]{Zhou2025a} C. Zhou, Y. Zeng, \& Z. Pan. \apj, {\bf 985}, 2, 242 (2025). doi:10.3847/1538-4357/adcee2

\bibitem[Zhou et al.(2025)]{Zhou2025b} C. Zhou, Z. Pan, \& N. Jiang. arXiv:2504.11078 (2025). doi:10.48550/arXiv.2504.11078

\bibitem{Hills1975} J. Hills. \nat, {\bf 254}, 295–298 (1975). doi:10.1038/254295a0

\bibitem{Rees1988} M. Rees. \nat, {\bf 333}, 523–528 (1988). doi:10.1038/333523a0

\bibitem{Gezari2021} S. Gezari. \araa, {\bf 59}, 21–58 (2021). doi:10.1146/annurev-astro-111720-030029

\bibitem{Papaloizou1995} J. Papaloizou \& D. Lin. \apj, {\bf 438}, 841 (1995). doi:10.1086/175127

\bibitem{Fragile2007} P. Fragile, O. Blaes, P. Anninos M., J. Salmonson. \apj, {\bf 668}, 417–429 (2007). doi:10.1086/521092

\bibitem{Stone2012} N. Stone \& A. Loeb. \prl, {\bf 108(6)}, 061302 (2012). doi:10.1103/PhysRevLett.108.061302

\bibitem{Franchini2016} A. Franchini, G. Lodato, S. Facchini. \mnras, {\bf 455}, 1946–1956 (2016). doi:10.1093/mnras/stv2417

\bibitem{Zhang2025} W. Zhang, X. Shu, L. Sun, et al. Nature Astronomy, 1-8 (2025). doi:10.1038/s41550-025-02502-0

\bibitem{Lei2013} W.-H. Lei, B. Zhang, H. Gao. \apj, {\bf 762}, 98 (2013). doi:10.1088/0004-637X/762/2/98

\bibitem{2024SciA...10J8898P} D.~R. {Pasham}, F. {Tombesi}, P. {Sukov{\'a}}, et al. Science Advances, \textbf{10(13)} (2024), eadj8898. doi:10.1126/sciadv.adj8898

\bibitem{Liska2018} M. Liska, C. Hesp, A. Tchekhovskoy, A. Ingram, van der M. Klis, S. Markoff. \mnras, {\bf 474}, L81–L85 (2018). doi:10.1093/mnrasl/slx174

\bibitem{Ryu2023} T. Ryu, J. Krolik, \& T. Piran. \apjl, {\bf 946}, L33 (2023). doi:10.3847/2041-8213/acc390








\end{thebibliography}










\end{multicols}
\end{document}